\documentstyle[12pt]{article}
\newlength{\pubnumber} 
\newcommand\pubblock[2]{\begin{flushright}\parbox{\pubnumber}
{\begin{flushleft}#1\\ #2\\ \end{flushleft}}\end{flushright}}
\catcode`\@=11
\@addtoreset{equation}{section}

\def\section{\@startsection{section}{1}{\z@}{3.5ex plus 1ex minus .2ex}
 {2.3ex plus .2ex}{\large\bf}}
\def\subsection{\@startsection{subsection}{2}{\z@}{2.3ex plus .2ex}
 {2.3ex plus .2ex}{\bf}}
\newcommand\Appendix[1]{\def\thesection{Appendix \Alph{section}}
 \section{\label{#1}}\def\thesection{\Alph{section}}}
\def\beqq{\begin{equation}}
\def\beq#1{\begin{equation}\label{#1}}
\def\eeq{\end{equation}}
\def\beqqa{\begin{eqnarray}}
\def\beqa#1{\begin{eqnarray}\label{#1}}
\def\eeqa{\end{eqnarray}}
\def\bZ{{\bf Z}}
\def\eptilde{\tilde\epsilon}
\def\epsdag{{\epsilon^\dagger}}
\def\ket#1{\vert #1\rangle}
\def\id{1{\hbox{\kern-4pt $1$}}}
\def\slp{\not{\hbox{\kern-2.3pt $p$}}}
\def\usl{\not{\hbox{\kern-2.0pt $u$}}}
\def\wsl{\not{\hbox{\kern-3.5pt $w$}}}
\def\a{\mbox{\boldmath $\alpha$}}
\def\b{\mbox{\boldmath $\beta$}}
\def\c{\mbox{\boldmath $\gamma$}}

\begin{document}

\begin{titlepage}
\samepage{
\pubblock{CLNS 91/1121}{May 1992}
\vfill
\begin{center}
{\Large \bf Low-Lying States of the Six-Dimensional\\
Fractional Superstring\\}
\vfill
{\large Philip C. Argyres\footnote{E-mail address:
pca@strange.tn.cornell.edu; pca@crnlnuc.bitnet}, Edwin
Lyman\footnote{E-mail address: lyman@beauty.tn.cornell.edu}
and S.-H. Henry Tye\\}
\vspace{.25in}
{\it Newman Laboratory of Nuclear Studies\\
Cornell University\\
Ithaca, N.Y.  14853-5001\\}
\end{center}
\vfill
\begin{abstract}
The $K=4$ fractional superstring Fock space is constructed in terms
of $\bZ_4$ parafermions and free bosons.  The bosonization of the
$\bZ_4$ parafermion theory and the generalized commutation relations
satisfied by the modes of various parafermion fields are reviewed.
In this preliminary analysis, we describe a Fock space which is simply
a tensor product of $\bZ_4$ parafermion and free boson Fock spaces.
It is larger than the Lorentz-covariant Fock space indicated by the
fractional superstring partition function.  We derive the form of the
fractional superconformal algebra that may be used as the constraint
algebra for the physical states of the FSS.  Issues concerning the
associativity, modings and braiding properties of the fractional
superconformal algebra are also discussed.  The use of the constraint
algebra to obtain physical state conditions on the spectrum is
illustrated by an application to the massless fermions and bosons of
the $K=4$ fractional superstring.  However, we fail to generalize these
considerations to the massive states.  This means that the appropriate
constraint algebra on the fractional superstring Fock space remains to
be found.  Some possible ways of doing this are discussed.
\end{abstract}
\vfill}
\end{titlepage}

\setcounter{footnote}{0}
\section{\label{sone}Introduction}

String theory \cite{GSW} is the only known theory with the potential
for describing all matter and forces in nature in a unified way.  In
particular, the superstring and the closely related heterotic string
entail many structures, including gravity, Yang-Mills fields and chiral
fermions, that are central to our present understanding of the world.
However, their critical space-time dimension is ten, and though there
are numerous proposed mechanisms to reduce the number of observable
dimensions, there is no known compelling reason why the superstring
theory should have only four large space-time dimensions.  While it is
important to search for dynamical and/or symmetry reasons explaining how
our world could be realized in the heterotic/superstring framework, we
would like to ask instead if other string theories with lower critical
dimensions exist, for they could provide more natural descriptions of
the world.

Recently strong evidence has been presented for the existence of such
string theories \cite{AT}.  Since string theories are characterized by
the local symmetries of a two-dimensional conformal field theory on the
string world-sheet, it is natural to try to construct string theories
with smaller critical space-time dimensions by changing the world-sheet
symmetry.  It is well-known that fractional-spin fields exist in
two-dimensional theories.  One can imagine new local symmetries on the
world-sheet which involve fractional-spin currents (replacing the
spin-$3/2$ supercurrent of the superstring) and which lead to string
propagation in space-times with dimensions less than 10.  In ref.~\cite{AT},
string theories, called fractional superstrings (FSS), with
spin $4/3$ and $6/5$ currents on the world-sheet were found to have
potentially interesting phenomenologies in $6$ and $4$ critical space-time
dimensions, respectively.  In this paper, we discuss in detail
the spectrum and physical state conditions of the $6$-dimensional
FSS.  Although the results obtained are unsatisfactory (or, at best,
incomplete), we believe the analysis presented below illuminates some of
the main issues that are involved in understanding the FSS.

The basic idea behind the FSS is to replace the world-sheet supersymmetry
of the superstring theory with a world-sheet fractional supersymmetry
parametrized by an integer $K\geq 2$.  Such a fractional supersymmetry
relates world-sheet bosons not to fermions but rather to world-sheet
parafermions.  The world-sheet fractional superpartner of the space-time
coordinate $X^\mu$ is a field $\epsilon^\mu$ of spin $2/(K+2)$.  This
field is the so-called ``energy operator'' of the $\bZ_K$ parafermion
theory \cite{ZFpara}.  The fractional supersymmetry is generated by a
generalization of the supercurrent, a new chiral current $G$
\cite{KZ,KMQ,BNY,Rav,ALT} whose conformal dimension is $(K+4)/(K+2)$.
This new current transforms $X^\mu$ to the fractional-spin field
$\epsilon^\mu$.

By demanding that the FSS have only transverse propagating modes, a
generalization of an argument of Brink and Nielson \cite{BrN} implies
that the critical dimensions of such string theories should be \cite{AT}
\beqq  D=2+16/K~.  \eeq
The case $K=2$ ($D=10$) corresponds to the superstring.  The new theories
are those with $K>2$; for $K=4$, $8$ and $16$ we have the integer critical
dimensions $D=6$, $4$ and $3$, respectively.

In this paper we will concentrate exclusively on the simplest case after
the ($K=2$) superstring, the $K=4$ FSS.  The reasons for this are twofold.
The first is that the complexity of these theories increases considerably
with increasing $K$.  Although the world-sheet fractional supersymmetry
algebra is non-local, the $\bZ_4$ parafermion fields that appear in the
$K=4$ FSS can be simply represented by free bosons, which enables the
calculations to be simplified tremendously.  This is not the case in the
$K=8$ and $K=16$ theories \cite{AGT1,AGT2}.  Furthermore, a close examination
shows that the appropriate world-sheet fractional supersymmetry algebra for
the $K=8$ theory contains two spin-$13/5$ currents in addition to the
spin-$6/5$ current \cite{AGT2}, which further complicate the analysis.
The second reason for our emphasis on the $K=4$ FSS is because it is
potentially the most interesting one from the phenomenological point of view.
As argued in \cite{DT,ADT}, the requirements of quantum mechanics, Lorentz
invariance and locality suggest that the $K=4$ FSS may be automatically
compactified from six to four space-time dimensions.  Furthermore, as argued
in \cite{DT}, the compactification from the critical dimension $6$ to the
natural dimension $4$ offers the possibility of the construction of heterotic
type $K=4$ FSS models that have chiral space-time fermions.  This is
encouraging, since the $K=4$ FSS, because of its relative simplicity, affords
the best prospect for detailed examination in the near future.

Let us highlight some of the similarities and differences between the
$K=4$ FSS and the superstring.  In the superstring, the world-sheet
superpartner of the space-time coordinate boson $X^{\mu}$ is a Majorana
fermion $\psi^\mu$.  $\psi$ is the primary field of dimension $1/2$ in the
Ising model, the $c=1/2$ minimal unitary conformal field theory (CFT)
\cite{BPZ}.  This theory has two other primary fields, the identity $\id$
and the spin field $\sigma$, which play specific roles in the construction
of the superstring Fock space.  All Neveu-Schwarz (space-time bosonic)
states can be generated by the action of the modes of the Majorana fermion
field  on the identity, and all states in the Ramond sector (space-time
fermions) can be generated by the action of the modes of the Majorana fermion
on the spin field.

In the $K=4$ string, on the other hand, the world-sheet fractional
superpartner of the space-time coordinate boson is the dimension-$1/3$
energy operator $\epsilon^\mu$, a primary field in the $\bZ_4$ parafermion
theory.  This CFT has central charge $c=1$ and an infinite number of primary
fields.  In Sect.~\ref{stwo} we develop a Fock space description of the
$\bZ_4$ parafermion theory, and show how the states are divided into
sectors that close under the action of the energy operator modes.  Different
sectors will be seen to correspond to space-time bosons and fermions, in
analogy to the Neveu-Schwarz and Ramond sectors of the superstring.  The
analysis of the $\bZ_4$ parafermion theory is made easier by the fact
that the $\bZ_4$ parafermion theory can be realized as a free boson $\phi$
compactified on a circle of radius $2$ (where the boson is normalized
by $\langle\rho(z)\rho(0)\rangle = -{2\over3}{\rm ln}z$) \cite{Yang}.
Nevertheless a number of technical issues having to do with the modings of
fractional-spin operators on the world-sheet will have to be unraveled.
In this paper we consider the simplest case, where the Fock space of the
open $K=4$ FSS is described as the $D$-fold tensor product of the $\bZ_4$
parafermion Fock spaces and free coordinate boson Fock spaces, with $D$ the
dimension of space-time.

At a fundamental level, the superstring is described by its world-sheet
gauge invariance---two-dimensional superconformal supergravity.  Classically,
the gauge fixing constraints in the superconformal gauge consist of the
vanishing of the energy-momentum tensor $T(z)$ and the supercurrent
$G_s(z)=\psi^\mu\partial X_\mu$.  Upon quantization, these constraints
are realized weakly on the space of physical states of the superstring:
 \beqq
 \langle\phi|T(z)\ket{\chi}=\langle\phi|G_s(z)\ket{\chi}=0
 \eeq
for all physical states $\phi$ and $\chi$.  By writing $T$ and $G_s$ in
modes, these conditions can be factorized into the familiar physical state
conditions, in which the positive modes of $T$ and $G_s$ annihilate physical
states.  We have no understanding at present of the classical world-sheet
gauge invariance of the $K=4$ FSS.  However, there does exist a natural
analog of the supercurrent $G_s$, the dimension $4/3$ fractional supercurrent
$G$ of the form
 \beqq
  G(z)\sim\epsilon^\mu\partial X_\mu+\ldots~.
 \eeq
The precise expression for $G(z)$ will be explored in Sect.~\ref{sfour},
where we will also discuss some aspects of the algebra satisfied by $G$.
A basic assumption that will be made in this paper is that there exists
some sort of ``fractional superconformal gauge'' whose classical constraint
equation is the vanishing of $G$ (and $T$, of course---we still require
two-dimensional reparametrization and Weyl invariance).  One goal in this
paper is to formulate and solve, at the massless levels, the resulting FSS
physical state conditions.

In Sect.~\ref{sfive}, we begin our exploration of the FSS constraint
algebra by solving the physical state conditions for the massless propagating
modes of the open $K=4$ FSS in the space-time fermion sector.  We show how
space-time spinors are naturally described in the FSS Fock space, and that
the massless state is a Weyl fermion.  We find that there exists a natural
left-right pairing of massive states in the fermionic sector necessary for
Lorentz invariance.  The existence of this pairing for all mass levels follows
from a counting argument relying on the Euler pentagonal number theorem.

We turn to an examination of the physical states in the bosonic sector in
Sect.~\ref{ssix}.  The open FSS massless propagating modes turn out to
describe a massless vector particle.  A novel feature compared to the
superstring is the existence of normal-ordering constant (``intercept'')
for the fractional supercurrent physical state condition in the bosonic
sector.

Sect.~6 describes two problems that appear upon solving the physical
state conditions at higher mass levels of the open FSS Fock space.
These are, firstly, that the tensor product Fock space description of
states is Lorentz non-covariant above the massless level, and, secondly,
that the physical state consitions do not remove nearly enough states
to agree with the counting of states in the FSS partition function.  We
do not have a clear resolution of these problems; however, they suggest
that the tensor product Fock space described in this paper is larger than
the Lorentz-covariant space of states implied by the  FSS partition
function.  One can either search for a direct way of reducing the
tensor-product Fock space to one in which the physical state conditions
have a Lorentz-covariant action, or modify/supplement the physical
state conditions to reduce the number of propagating degrees of freedom
and cancel the Lorentz non-covariant terms in the equations of motion.

In Sect.~7 we remark on two features of the $K=4$ FSS which may provide
useful hints for finding a complete and correct description of the FSS
space of states.  First, we discuss the possible equivalence of the
$K=4$ FSS with the spin-$4/3$ string introduced in ref.~\cite{ALT}.
The critical central charge of the spin-$4/3$ string, unlike the $K=4$
string, can be calculated by constructing extra towers of null states
using the algebra of physical state conditions.  A representation of the
spin-$4/3$ algebra would then give a non-tensor-product realization of
the FSS space of states.  However, the representation theory of this algebra
is not well understood, and thus a direct comparison of its spectrum of
physical states with the $K=4$ FSS partition function cannot be made yet.
Second, we show that a chiral closed $K=4$ FSS must have extra massless
scalars in its spectrum if gravitational anomalies are to cancel.

We have collected some of the more technical or tangential discussions
in a series of appendices.  Various chiral associative solutions to the
$\bZ_4$ parafermion theory are constructed using cocycles in \ref{appA}.
A review of the derivation of the generalized commutation relations satisfied
by fractional-spin fields, based on examples taken from the $\bZ_4$
parafermion theory is presented in \ref{appB}.  \ref{appC} discusses the
representation theory and associativity constraints of the fractional
superconformal algebra.  In \ref{sthree} we review the construction of the
modular invariant partition function for the closed $K=4$ FSS.  This makes
the analog of the GSO projection \cite{GSO} in the FSS apparent, and aids
in the identification of bosonic and fermionic states in the FSS Fock space.
Finally, in \ref{appD} we discuss the physical state conditions and null
states of the spin-4/3 string.

\setcounter{footnote}{0}
\section{\label{stwo}The $K=4$ FSS Fock Space}

Each space-time dimension of the $K=4$ FSS corresponds to a free coordinate
boson $X^\mu$ and a $\bZ_4$ parafermion theory on the string world-sheet.
In this section we will construct a free-field representation of the $\bZ_4$
parafermion theory, first pointed out by Yang \cite{Yang}.  We use this
representation to compute the operator product expansions (OPEs), mode
expansions, generalized commutation relations, and characters of primary
fields in the parafermion theory.  In this way we will be able to build up
a Fock space realization of the space of states for each dimension of the FSS.
We assume that the total Fock space is a tensor product of $D$ of these
individual Fock spaces.  This tensor product structure will be the subject
matter of later sections.

\subsection{Review of $\bZ_4$ parafermions}

The operator content of the chiral $\bZ_4$ parafermion theory can be
realized by the $SU(2)_4/U(1)$ coset model \cite{ZFpara}.  The chiral
$SU(2)_4$ WZW theory \cite{KZ} has central charge $c_{\rm WZW}=2$ and
consists of holomorphic primary fields $\Phi^j_m(z)$ of conformal
dimension $j(j+1)/6$.  The indices $j,m~\in{\bf Z}/2$ label $SU(2)$
representations where $0\leq j\leq 2$ and $|m|\leq j$ with $j-m\in \bZ$.
When we factor a $U(1)$ subgroup out of $SU(2)_4$, we correspondingly
factor the primary fields as
\beq{primfac}
 \Phi^j_m(z)~=~\phi^j_m(z)\,\exp\left\{i{m\over2}\,\varphi(z)\right\}~.
\eeq
Here $\varphi$ is the free $U(1)$ boson normalized so that
$\langle\varphi(z)\varphi(w)\rangle=-2{\rm ln}(z-w)$. The $\phi^j_m(z)$
are Virasoro primary fields in the $\bZ_4$ parafermion theory with
conformal dimensions:
\beq{pfdims}
 \Delta(\phi^j_m)~=~{j(j+1)\over6}~-~{m^2\over4}
 \quad\quad\quad{\rm for}~~|m|\leq j~.
\eeq
The central charge of the $\bZ_4$ parafermion theory is then
$c=c_{\rm WZW}-c_{\varphi}=1$.  This is an indication that the $\bZ_4$
parafermion can be realized by a free boson.  The definition of the
$\phi^j_m$ fields can be consistently extended to the case where
$|m|>j$ by the rules
\beq{phid} \phi^j_m~=~\phi^j_{m+4}~=~\phi^{2-j}_{m-2}~. \eeq
With these identifications, an independent set of fields can be taken
to be $\phi^0_0$, $\phi^0_{\pm1}$, $\phi^{1/2}_{\pm1/2}$, $\phi^1_0$,
$\phi^1_1$, $\phi^{3/2}_{\pm1/2}$ and $\phi^2_0$.

The fusion rules of the parafermion fields follow from those of the
$SU(2)_4$ theory:
\beq{fusionrules}
 [\phi^{j_1}_{m_1}]\otimes[\phi^{j_2}_{m_2}]~\sim
 \sum_{j=|j_1-j_2|}^r[\phi^j_{m_1+m_2}]
\eeq
where $r={\rm min}\{j_1+j_2\, ,\, 4-j_1-j_2\}$. The sectors $[\phi^j_m]$
include the primary fields $\phi^j_m$ and a tower of higher-dimension fields
(with dimensions differing by integers) defined as in (\ref{primfac}) from
current algebra descendants of the $\Phi^j_m$.  Indeed, the $SU(2)_4$
currents factorize as well:
\beqa{KMcurralg}
    J^+ &=&2\psi_1\,e^{i\varphi/2}\nonumber\\
    J^0 &=&i\,\partial\varphi \\
    J^- &=&2\psi_{-1}\,e^{-i\varphi/2}\nonumber
\eeqa
where the parafermion currents $\psi_\ell\equiv\phi^0_\ell=\phi^2_{\ell-2}$
and $\psi_{-\ell}\equiv\psi_{4-\ell}$ have conformal dimensions $\ell(4-
\ell)/4$ in accordance with (\ref{pfdims}).  It follows from the fusion
rules (\ref{fusionrules}) that the current blocks $[\psi_\ell]$,
$\ell=0,1,2,3$ form a closed algebra, namely, the $\bZ_4$ parafermion
current algebra.  Note that $\psi_1$ acting on a field $\phi^j_m$ increases
the $m$ quantum number by one but does not change the $SU(2)$ spin $j$.

Another special field---the one that will play the central role in the
construction of the FSS Fock space to follow---is the energy operator
$\epsilon\equiv \phi^1_0$ of dimension $1/3$.  Operating on a field
$\phi^j_m$ with $\epsilon$ preserves the $m$ quantum number but yields
sectors with $j$ quantum numbers $j-1$, $j$, and $j+1$,  when permitted
by the fusion rule (\ref{fusionrules}).

Before plunging into the detailed construction of the parafermion Fock
space, we can already make some important remarks concerning the roles
the various parafermion fields can be expected to play in the FSS.
The energy operator $\epsilon$, as the fractional superpartner of the
coordinate boson field $X$, will play a role analogous to the one the
Majorana fermion field $\psi$ plays in the superstring.  There, the
Neveu-Schwarz sector is built from the action of $\psi$ on the identity.
{}From the fusion rules (\ref{fusionrules}) we see that $\epsilon$ acting
on the identity can create all parafermions with $m$ quantum number zero.
Thus, we expect that the set of fields $\{[\phi^j_0]\}$ will be the analog
of the Neveu-Schwarz sector in the FSS.  Just as the Ramond sector is
generated by the action of $\psi$ on another field, we should look for
other sectors in the parafermion theory that close under the action of
$\epsilon$.  Because of the identifications in (\ref{phid}), there are
only three such other sectors: $\{[\phi^j_{\pm1/2}]\}$ and $\{[\phi^j_1]\}$.
These are formed by the action of $\epsilon$ on the so-called ``spin''
fields of the parafermion theory, $\sigma_{\pm1} \equiv \phi^{1/2}_{\pm1/2}$
and $\sigma_2 \equiv \phi^1_1$, respectively.  The arguments of \ref{sthree}
indicate that the $m=1$ sector plays the role of the Ramond sector in the
FSS, while the $m=\pm1/2$ sectors do not enter into the FSS Fock space at
all.  The fusion rules (\ref{fusionrules}) show that it is consistent to
project out the half-odd integral spin states since the integral spin states
do not close on them; \ref{sthree} shows that this projection is required
by modular invariance.

\subsection{Bosonization of $\bZ_4$ parafermions}

The spectrum of the $c=1$ $\bZ_4$ parafermion theory is the same as that
of the $\bZ_2$ orbifold of a boson on a circle of a certain radius
\cite{Yang}.  In line with the above discussion, we will ignore the
parafermion fields with spin $j\in\bZ+{1\over2}$ which correspond to the
twist fields in the orbifold theory.  Therefore we will just be interested
in a free boson taking values on a circle.  This bosonization will allow us
to easily construct the Fock space of the parafermion theory and to derive a
simple form for the parafermion characters as sums over the winding modes
of the compactification.

Consider a chiral boson $\rho(z)$ satisfying
\beq{rhorho}
 \langle\,\rho(z)~\rho(w)\,\rangle~=~-\,{2\over3}\,{\rm ln}(z-w)~.
\eeq
With this choice of normalization, the energy-momentum tensor is given by
\beqq
 T_{\rho}(z)~=~-\,{3\over4}:(\partial\rho(z))^2:~.
\eeq
If we take this boson to be compactified on a circle of radius $2$ [in the
units implied by (\ref{rhorho})], so that $\rho=\rho+4\pi n$, $n\in{\bf Z}$,
the Virasoro primary fields include the dimension $1$ field $i\partial\rho(z)$,
higher dimension fields built from it, and the infinite set of fields
\beqq
 \epsilon^{(a)}(z)~=~:\!{\rm e}^{ia\rho(z)}\! :~,\quad a\in{\bf Z}/2~,
\eeq
with conformal dimensions
\beqq
 \Delta(\epsilon^{(a)})~=~{{a^2}\over 3}~.
\eeq
Comparing to the conformal dimensions of the $\phi^j_m$ parafermion fields
(\ref{pfdims}), we make the identifications
\beqa{pfid}
 \epsilon^{(0)}(z)&=&\phi^0_0(z)~~~(=\id)~,~~~~~\Delta=0~,\nonumber\\
 \epsilon^{(\pm1/2)}(z)&=&\phi^1_1(z)~~~(=\sigma_2)~,~~~~\,\Delta=1/12,
  \nonumber\\
 \epsilon^{(\pm1)}(z)&=&\phi^1_0(z)~~~(=\epsilon)~,~~~~~~\Delta=1/3~,\\
 \epsilon^{(\pm3/2)}(z)&=&\phi^0_{\pm1}(z)~(=\psi_{\pm1})~,~~~\Delta=3/4~,
  \nonumber\\
 i\partial\rho(z)&=&\phi^2_0(z)~~~(=\psi_2)~,~~~~\,\Delta=1~,\nonumber
\eeqa
where we have also given in parentheses the corresponding spin-field,
energy operator, or parafermion current symbol.  The identification of
fields in the free boson theory with $\bZ_4$ parafermion fields is
discussed in more detail in \ref{appA}.

The OPEs of these fields are easily computed to be, for $a+b\neq0$,
\beq{epope}
 \epsilon^{(a)}(z)~\epsilon^{(b)}(w)=(z-w)^{2ab/3}~
 \left[\epsilon^{(a+b)}(w)+{a\over a+b}\,(z-w)\,
 \partial\epsilon^{(a+b)}+\ldots\right]~,
\eeq
or, if $a+b=0$,
\beq{epope1}
 \epsilon^{(a)}(z)~\epsilon^{(-a)}(w)=(z-w)^{-2{a^2}/3}~
 \bigl[\id+a\,(z-w)\,i\partial\rho(w)+\ldots\bigr]~;
\eeq
and
\beq{epope2}
 i\partial\rho(z)~\epsilon^{(a)}(w)={2a\over3}\,{1\over(z-w)}\,
 \epsilon^{(a)}(w)+{1\over a}\,\partial\epsilon^{(a)}(w)+\ldots
\eeq

A special feature of the free boson representation is its $\bZ_3$ symmetry.
We can associate a $\bZ_3$ charge $q$ with the $\epsilon^{(a)}$ fields by
the rule
\beq{z3} q~\equiv~-2a~({\rm mod}~3)~,~a\in\bZ/2~. \eeq
Thus the parafermion sectors $[\phi^j_m]$ with $j=0$ or $2$ have $\bZ_3$
charge $q=0$, while the $j=1$ sectors have $q=\pm1$. {}From the OPEs it
follows that this charge is additive under fusion. We will see below that
its occurrence makes for important technical simplifications.  Note that
the $\bZ_3$ symmetry of this free boson representation has nothing to do
with the $\bZ_4$ symmetry of the parafermion current algebra---its existence
is ``accidental'' in the sense that there do not in general exist $\bZ_N$
symmetries of the whole spectrum of other $\bZ_K$ parafermion theories.

The Fock space of a free boson compactified on a circle is usually built
up by the action of the modes of the boson field.  The (chiral) mode
expansion of $\rho(z)$ is
\beqq
 \rho(z)=r_0-is_0~{\rm ln}z+i\sum_{n\neq0}{1\over n}s_nz^{-n}~,
\eeq
where the modes satisfy the commutation relations following from (\ref{rhorho})
\beqq
 [r_0,s_0]={2i\over3}~,\quad\quad [s_n,s_m]={2n\over3}\delta_{n+m}~.
\eeq
The Fock space is built with the $\rho(z)$ creation operators from the
vacuum $\ket{0}$ satisfying $s_m\ket{0}=0$ for $m\geq0$.  Because $\rho(z)$
is compactified on a circle of radius $2$,
\beqq r_0=r_0+4\pi n~,\quad\quad n\in\bZ~. \eeq
In the zero-mode sector the allowed $s_0$ eigenstates are therefore
\beqq
 \ket{a}\equiv{\rm e}^{iar_0}\ket{0}~,\quad\quad a\in\bZ/2~,
\eeq
satisfying
\beqq s_0\ket{a}={2a\over3}\ket{a}~. \eeq
Clearly, these states $\ket{a}$ are created from the vacuum by the vertex
operators $\epsilon^{(a)}(z)$:
\beq{aket} \ket{a}=\epsilon^{(a)}(0)\ket{0}~. \eeq
The rest of the Fock space is built up by acting on the $\ket{a}$ states
with the $s_m$ modes with $m<0$.

{}From this description of the Fock space, and the identifications
(\ref{pfid}) of the parafermion operators in each sector, we can easily
derive an expression for the character ${\cal Z}^j_m$ of the parafermion
sector $[\phi^j_m]$.  Indeed, each parafermion sector $[\phi^j_m]$ consists
of the primary field $\phi^j_m$ and its parafermion and Virasoro descendants
whose dimensions differ by integers.  In the bosonic Fock space, these
sectors correspond to the ``momentum'' state $\ket{a}$ identified in
(\ref{pfid}), its associated winding states $\ket{a+3n}$, and all their
descendants created by the action of the $s_m$ modes.  The counting of
the $s_m$ descendants simply contributes a factor of the free boson partition
function $\eta(q)^{-1}$ to the parafermion character, where
\beq{dede} \eta(q)=q^{1/24}\prod_{n=1}^\infty(1-q^n) \eeq
is the Dedekind $\eta$-function.  Here $q=e^{2\pi i\tau}$, where $\tau$
is the complex modulus on the torus. Since the character is defined to be
${\rm Tr}q^{L_0}$ over each sector, the factor of $\eta(q)^{-1}$ must be
multiplied by $q^\Delta$ where $\Delta$ is the dimension of each winding
state $\ket{a+3n}$ appearing in $[\phi^j_m]$.  Thus, we find the characters
\beqa{pfchars}
 {\cal Z}^0_0+{\cal Z}^2_0 &=& {1\over\eta}\left(\sum_{n=-\infty}^\infty
  q^{3n^2} \right)~,\nonumber\\
 {\cal Z}^1_1 &=& {q^{1/12}\over\eta}\left(\sum_{n=-\infty}^\infty q^{3n^2+n}
  \right)~,\\
 {\cal Z}^1_0 &=& {q^{1/3}\over\eta}\left(\sum_{n=-\infty}^\infty q^{3n^2+2n}
  \right)~,\nonumber\\
 {\cal Z}^0_1~=~{\cal Z}^0_{-1} &=& {q^{3/4}\over\eta}\left(\sum_{n=0}^\infty
  q^{3n^2+3n} \right)~.\nonumber
\eeqa
These characters will form the basis for our discussion of modular invariance
of the closed $K=4$ FSS in \ref{sthree}.

\subsection{Mode expansions and commutation relations}

Though the description of the $\bZ_4$ parafermion Fock space given above
is complete, it is not expressed in the right language for our purposes.
We will need to define the mode expansions of the $\epsilon^{(a)}$ fields
and derive the commutation relations these modes satisfy in order to state
and solve the FSS physical state conditions.  Because the parafermion fields
have fractional spin, the usual contour deformation argument \cite{BPZ} for
deriving commutation relations of modes from the OPE of the fields must be
modified.  For fields that satisfy abelian braid relations, as the
$\epsilon^{(a)}$ do, Zamolodchikov and Fateev \cite{ZFpara} have invented
the necessary modifications.  Their construction is reviewed in \ref{appB}.
The result is that abelian braided fields satisfy generalized commutation
relations (GCRs) which involve infinite sums of terms in place of the usual
commutator or anticommutator.

First, though, we must introduce the mode expansions of the $\epsilon^{(a)}$.
{}From the OPEs (\ref{epope})--(\ref{epope2}) we see that in general
\beq{gope}
 \epsilon^{(a)}(z)\chi_q(0)=\sum_{n\in\bZ}z^{n-aq/3}\,\chi^{(n)}_{q+q_a}(0)~,
\eeq
where $\chi_q$ represents any field with $\bZ_3$ charge $q$, and $q_a$ is
the $\bZ_3$ charge of $\epsilon^{(a)}$.  The $\chi^{(n)}_{q+q_a}$ fields
are all the primary and descendant fields that appear on the right-hand
side of the OPE.  From \ref{gope}, the dimension of $\chi^{(n)}_{q+q_a}$
is $\Delta(\chi^{(n)}_{q+q_a}) = \Delta(\chi_q)+n+a(a-q)/3$, where we have
used $\Delta(\epsilon^{(a)})=a^2/3$.  Following the usual convention that
the subscript on a mode operator is the negative of its dimension, we define
the $\epsilon^{(a)}$ modes by
\beqq
 \chi^{(n)}_{q+q_a}(0)\equiv \epsilon^{(a)}_{-n+a(q-a)/3}\chi_q(0)~.
\eeq
Thus the general OPE (\ref{gope}) can be written
\beq{epmode}
 \epsilon^{(a)}(z)~\chi_q(0)=\sum_{n}z^{n-aq/3}
 \epsilon^{(a)}_{-n+a(q-a)/3}~\chi_q(0)~,
\eeq
or
\beq{edompe}
 \epsilon^{(a)}_{n+a(q-a)/3}~\chi_q(0)~=~\oint_{\gamma}{{\rm dz}\over2\pi i}
 \ z^{n-1+aq/3}\epsilon^{(a)}(z)~\chi_q(0)~,
\eeq
where $\gamma$ is a contour encircling the $\chi_q(0)$ insertion once.
Because the OPE (\ref{gope}) describes abelian braiding (there is only
one cut on the right hand side) it was possible to choose the integrand
in (\ref{edompe}) so that the contour of integration closes.  In the case
of non-abelian braiding this is not in general possible.

{}From the OPEs of the fields we can now derive their GCRs following the
discussion in \ref{appB}.  However, as explained there, there are actually
an infinite number of these relations which can be obtained from a single
OPE depending on how many terms on the right hand side of the OPE are
included.  Of course, these GCRs are all consistent, but those encoding
more terms of an OPE contain more information.  We will find in the following
analysis that only the first few terms of the $\bZ_4$ parafermion OPEs are
sufficient to determine all the mode relations in the parafermion Fock space.

We can choose in deriving these relations to select only the first term on
the right-hand side of the $\epsilon^{(a)}\epsilon^{(b)}$ OPEs (\ref{epope})
and (\ref{epope1}) by multiplying them by the factor $(z-w)^{\alpha}$ with
$\alpha = -1-2ab/3$. Then, following the procedure outlined in \ref{appB},
we obtain the GCR
\beqa{epgcr1}
 &\displaystyle{ \sum^{\infty}_{\ell=0}c^{(\alpha)}_\ell\left[
 \epsilon^{(a)}_{n-\ell-1+{a(q-2b-a)\over3}}
 \epsilon^{(b)}_{m+\ell+1+{b(q-b)\over3}}+
 \epsilon^{(b)}_{m-\ell+{b(q-2a-b)\over3}}
 \epsilon^{(a)}_{n+\ell+{a(q-a)\over3}}\right] }&\nonumber\\
 &=\epsilon^{(a+b)}_{n+m+{(a+b)(q-a-b)\over3}}~.&
\eeqa
The $c_\ell^{(\alpha)}$ are the binomial coefficients defined by
$(1-x)^{\alpha} = \sum_{\ell=0}^{\infty} c_\ell^{(\alpha)}x^\ell$.
The above expressions are understood to be valid only when acting on a
state with $\bZ_3$ charge $q$.  Note that for $a+b=0$, the right hand
side is $\epsilon^{(0)}_{n+m} = \delta_{n+m}$. {}From the $\rho\epsilon$
OPE (\ref {epope}), a standard commutator is obtained:
\beq{sigcr}
 \left[s_n, \epsilon^{(a)}_{m+a(q-a)/3}\right]~=~
 {{2a} \over 3}~\epsilon^{(a)}_{n+m+a(q-a)/3}~.
\eeq
As we will see, it turns out that this relation is insufficient for our
analysis.  We will also need the GCR corresponding to keeping one more
term in the $\rho\epsilon$ OPE,
\beqa{siggcr}
 &&\sum^{\infty}_{\ell=0}~\left[s_{n-\ell-1}~\epsilon^{(a)}_{m+\ell+1+
 a(q-a)/3}~+~\epsilon^{(a)}_{m-\ell+a(q-a)/3}~s_{n+\ell}\right]\nonumber\\
 &&~~~~~~~=\left\{ {2an\over3}-{1\over a}\left(n+m+{aq\over3}
 \right)\right\}~\epsilon^{(a)}_{n+m+a(q-a)/3}~.
\eeqa
Note that (\ref{sigcr}) can be derived from (\ref{siggcr}).

These generalized commutators may seem complicated and difficult to use
due to the infinite summation.  However, when acting on any given state
only a finite number of terms from the infinite sum contributes.  This occurs
because, as the summation index $\ell$ gets larger, the dimension of the
right-most operators in the GCRs become more and more negative, so that
eventually they annihilate any given state.  For example, the vacuum
satisfies the properties
\beqa{epvac}
 \epsilon^{(a)}_{n-a^2/3}\ket{0}&=&0~,~~~n>0~,\nonumber\\
 s_n\ket{0}&=&0~,~~~n\geq0~,
\eeqa
following from the definition of the modes.  Note that the definition of
the ``momentum'' states (\ref{aket}) implies
\beq{epaket} \epsilon^{(a)}_{-a^2/3}\ket{0}=\ket{a}~. \eeq
Starting from (\ref{epvac}) and (\ref{epaket}) the GCRs
(\ref{epgcr1})--(\ref{siggcr}) are sufficient to build up the whole Fock
space of the $\bZ_4$ parafermion theory using $\epsilon^{(a)}$ modes.

To translate between the description of the Fock space in terms of the
$s_m$ modes presented in the last subsection, and the $\epsilon^{(a)}$
modes being discussed here, we need to derive a relation of the form
\beq{eptrans} \epsilon^{(a)}_r\ket{b}=P^{(a)}_r(s)\ket{a+b}~, \eeq
where $P^{(a)}_r(s)$ is a polynomial in the boson creation modes $s_m$
with $m<0$.  Then, using  the $s_m$--$\epsilon^{(a)}_r$ commutation relation
(\ref{sigcr}), an arbitrary $\epsilon$-state can be translated to the $s_m$
Fock space description:
\beqq
 \prod_{j=1}^N\epsilon^{(a_j)}_{r_j}\ket{a_0}
 =P(s)\left|\textstyle\sum_{j=0}^N \displaystyle a_j\right\rangle~,
\eeq
for some definite polynomial $P$.

We will now use the GCRs to derive a recursive formula for the polynomial
$P^{(a)}_r(s)$ in (\ref{eptrans}).  This will also serve as an example of
the use of the GCRs.  Consider the GCR (\ref{epgcr1}) with $m=-1$ acting
on the vacuum $\ket{0}$, so that $q=0$.  Using the vacuum properties
(\ref{epvac}) and (\ref{epaket}), we obtain
\beqa{epket2}
 \epsilon^{(a)}_{n-a(a+2b)/3}\ket{b} &=& 0~,\quad n>0~,\nonumber\\
 \epsilon^{(a)}_{-a(a+2b)/3}\ket{b} &=& \ket{a+b}~.
\eeqa
Now consider the $s_m$--$\epsilon^{(a)}_r$ GCR (\ref{siggcr}) acting on the
state $\ket{b}$.  By the $\bZ_3$ charge assignments (\ref{z3}) $q=-2b$, and
if we choose $n=0$ and $m=-k<0$, we obtain
\beqqa
 &\displaystyle{\left(\sum_{\ell=0}^\infty s_{-\ell-1}\,
 \epsilon^{(a)}_{\ell+1-k-a(a+2b)/3}\right)\ket{b}
 +\epsilon^{(a)}_{-k-a(a+2b)/3}s_0\ket{b}~= }&\nonumber\\
 &\displaystyle{\left({k\over a}+{2b\over3}\right)
 \epsilon^{(a)}_{-k-a(a+2b)/3}\ket{b}~. }&
\eeqa
Using (\ref{epket2}) and the fact that $\ket{b}$ is an $s_0$ eigenvector
$s_0\ket{b}=(2b/3)\ket{b}$, this implies
\beq{sep2}
 \sum_{\ell=0}^{k-1} s_{-\ell-1}\,\epsilon^{(a)}_{\ell+1-k-a(a+2b)/3}
 \ket{b}={k\over a}\,\epsilon^{(a)}_{-k-a(a+2b)/3}\ket{b}~.
\eeq
If we define the polynomial $P^{(a)}_k(s)$ by
\beq{Pdef}
 \epsilon^{(a)}_{-k-a(a+2b)/3}\ket{b}=P^{(a)}_k(s)\ket{a+b}~,
\eeq
then (\ref{sep2}) becomes the recursion relation
\beq{Prec}
 P^{(a)}_k=\left({a\over k}\right)\sum_{\ell=0}^{k-1}
 s_{-\ell-1}P^{(a)}_{k-1-\ell}~.
\eeq
The initial condition for this recursion, $P^{(a)}_0=1$, is provided by
(\ref{epket2}).  The first few solutions are $P^{(a)}_1 = as_{-1}$ and
$P^{(a)}_2 = {a\over2}(s_{-2} + as_{-1}s_{-1})$.

To conclude this discussion of the $\bZ_4$ parafermion Fock space, we
write down all the low-lying states, since they will be useful for
building the low-lying FSS physical states.  Consider first the states
in the parafermion sectors with $SU(2)$ quantum number $m=0$. This is the
sector analogous to the superstring Neveu-Schwarz sector---it consists
of all states generated from the vacuum by modes of the energy operators
$\epsilon^{(\pm1)}$.  The complete set of independent states for the
lowest levels of the $m=0$ sector can be written as follows:
\beqa{epfock}
 &\ket{0}&{\rm level}~0~,\nonumber\\
 &\epsilon^{(1)}_{-1/3}\ket{0}~,~
 \epsilon^{(-1)}_{-1/3}\ket{0}&{\rm level}~{1\over3}~,\nonumber\\
 &s_{-1}\ket{0}&{\rm level}~1~,\\
 &\epsilon^{(2)}_{-4/3}\ket{0}~,~\epsilon^{(-2)}_{-4/3}\ket{0}~,~
 s_{-1}\epsilon^{(1)}_{-1/3}\ket{0}~,~s_{-1}\epsilon^{(-1)}_{-1/3}\ket{0}
 &{\rm level}~{4\over3}~,\nonumber\\
 &\ldots&\ldots\nonumber
\eeqa
The counting of states here agrees with that implied by the characters
${\cal Z}^0_0+{\cal Z}^2_0$ and ${\cal Z}^1_0$ obtained previously. Note
that the GCR's given in eq.~(\ref{epgcr1}) have been used to eliminate
states that are dependent on the ones listed.  Indeed, it is easy to solve
the recursion relations (\ref{Pdef}) and (\ref{Prec}) to the first
few levels to obtain the identities
\beq{2.41}
 \epsilon^{(\pm1)}_{-2/3}~\epsilon^{(\mp1)}_{-1/3}\ket{0}~
 =~\pm s_{-1}\ket{0}~,
\eeq
\beq{seprel}
 s_{-1}~\epsilon^{(\pm1)}_{-1/3}\ket{0}~
 =~\pm\epsilon^{(\pm1)}_{-4/3}\ket{0}~,
\eeq
and
\beq{2.43}
 \epsilon^{(\pm1)}_{-1}\epsilon^{(\pm1)}_{-1/3}\ket{0}~
 =~\epsilon^{(\pm2)}_{-4/3}\ket{0}~.
\eeq

Similarly, the analogs of the Ramond sector are the $m=\pm1$ sectors built
up by $\epsilon^{(\pm1)}$ modes acting on the $\epsilon^{(\pm1/2)}$ spin
fields.  The complete set of independent states for the lowest levels are
\beqa{2.44}
 & \epsilon^{(1/2)}_{-1/12}\ket{0}
 ~,~\epsilon^{(-1/2)}_{-1/12}\ket{0}
 & \quad\quad{\rm level}~{1\over12}~,\nonumber\\
 & \epsilon^{(3/2)}_{-3/4}\ket{0}
 ~,~\epsilon^{(-3/2)}_{-3/4}\ket{0}
 & \quad\quad{\rm level}~{1\over12}+{2\over3}~,\\
 & s_{-1}\epsilon^{(1/2)}_{-1/12}\ket{0}
 ~,~s_{-1}\epsilon^{(-1/2)}_{-1/12}\ket{0}
 & \quad\quad{\rm level}~{1\over12}+1~,\nonumber\\
 & \ldots & \quad\quad\ldots\nonumber
\eeqa
These states can be written in other ways using the identities
following from the GCRs:
\beq{2.45}
 \epsilon^{(\mp1)}_0\,\epsilon^{(\pm1/2)}_{-1/12}\ket{0}~=~
 \epsilon^{(\mp1/2)}_{-1/12}\ket{0}~,
\eeq
\beqq
 \epsilon^{(\pm1)}_{-2/3}\,\epsilon^{(\pm1/2)}_{-1/12}\ket{0}~=~
 \epsilon^{(\pm3/2)}_{-3/4}\ket{0}~,
\eeq
\beqq
 \epsilon^{(\mp1)}_{-1/3}\,\epsilon^{(\pm3/2)}_{-3/4}\ket{0}~=~
 \mp s_{-1}\,\epsilon^{(\pm1/2)}_{-1/12}\ket{0}~,
\eeq
and
\beq{2.48}
 \epsilon^{(\mp1)}_{-1}\,\epsilon^{(\pm1/2)}_{-1/12}\ket{0}~=~
 \mp s_{-1}\,\epsilon^{(\mp1/2)}_{-1/12}\ket{0}~.
\eeq

Note also that if we had only used the standard commutator (\ref{sigcr})
as the defining commutation relation, rather than the GCR (\ref{siggcr}),
we would not have found the correct counting of states.  For example,
the relation (\ref{seprel}) would have been absent.

\subsection{The coordinate boson field}

To complete the description of one dimension's worth of the FSS Fock space,
we must tensor the parafermion theory with a free boson $X$, which will
have the interpretation of a space-time coordinate field.  We will only
consider the left-moving (holomorphic) part of this boson on the world-sheet.
We set its normalization by
\beqq \langle\,X(z)\,X(w)\,\rangle~=~-\,{\rm ln}(z-w)~, \eeq
from which follows the energy-momentum tensor
\beqq T_{X}(z)~=~-\,{1\over2}:(\partial X(z))^2:~.  \eeq
The primary fields are 
\beqq V_p(z)~\equiv~:\!{\rm e}^{ipX(z)}\!: \eeq
of dimension $\Delta(V_p)=p^2/2$.  Because $X$ is not compactified, the
momentum $p$ can take on any real value. The mode expansion of $X$ is
\beq{Xmode}
 X(z)~=~x_0-i\alpha_0\,{\rm ln}(z)+
 i\sum_{n\neq0}~{1\over n}\,\alpha_n z^{-n}~,
\eeq
where the modes satisfy the standard commutation relations
\beqq
 [x_0,\alpha_0]=i~,\quad\quad\quad [\alpha_n,\alpha_m]=n\delta_{n+m}~.
\eeq
The Fock space is built up by these modes from the vacuum $\ket{0}$
satisfying $\alpha_n\ket{0}=0$ for $n\geq0$. The highest weight states
\beqq \ket{p}~\equiv~V_p(0)\ket{0}~=~{\rm e}^{ipx_0}\ket{0} \eeq
satisfy $\alpha_0\ket{p}=p\ket{p}$ and $\alpha_n\ket{p}=0$ for $n>0$.

Upon tensoring the $\bZ_4$ parafermion theory with the coordinate boson,
we obtain a CFT with central charge $c_0=2$, and energy-momentum tensor
\beqq T(z)~=~T_X(z)+T_{\rho}(z)~.  \eeq
Defining the Virasoro modes in the usual way by $T(z)=\sum z^{-n-2}L_n$,
we find
\beq{dimvir}
 L_n~=~{1\over2}\sum_{\ell} :\!\alpha_\ell\alpha_{n-\ell}\!:
 + {3\over4}\sum_\ell :\!s_\ell s_{n-\ell}\!:~.
\eeq
Note that $T(z)$ is not the energy-momentum tensor for the full FSS, but
instead corresponds to only one dimension of the FSS.  We take the full
$K=4$ FSS Fock space to be the tensor product of $D$ copies of this $c_0=2$
CFT, where $D$ is the number of space-time dimensions.  The full $L_n$'s
will then obey a Virasoro algebra with central charge $c=Dc_0$.  In
Sect.~\ref{sfour} we will derive expressions for the fractional supercurrent
$G(z)$ in the $c_0=2$ and $c=Dc_0$ Fock spaces.

The $\bZ_4$ parafermion fields $\epsilon^{(a)}(z)$ and the coordinate boson
field $\partial X(z)$ are primary fields with respect to $T(z)$ with
conformal dimensions $a^2/3$ and $1$, respectively.  Their modes satisfy
the usual comutation relations with the $L_n$:
\beqa{2.57}
 {}[L_n,\alpha_m]&=&-m\alpha_{n+m}~,\nonumber\\
 {}[L_n,\epsilon^{(a)}_r]&=&\left[\left({a^2\over3}-1\right)n-r\right]
 \epsilon^{(a)}_{n+r}~.
\eeqa

\setcounter{footnote}{0}
\section{\label{sfour}The fractional superconformal algebra}

In this section we construct the fractional supercurrent, $G(z)$, of the
$K=4$ FSS.  This current is the analog of the dimension-$3/2$ supercurrent
of the superstring.  We argue that it is a dimension-$4/3$ chiral primary
field in the CFT describing the FSS, and find its explicit form in the
free boson representation introduced in Sect.~\ref{stwo}.  This current
$G(z)$ and the energy-momentum tensor $T(z)$ together generate the
fractional superconformal algebra. By analogy with the superconformal gauge
of the superstring, in which the energy-momentum tensor and supercurrent
generate the physical state conditions, we discuss the physical state
conditions that follow from the fractional superconformal mode algebra.

It is important to note that the real justification for singling out the
dimension-$4/3$ field as the fractional supercurrent, and for taking its
modes as generators of FSS physical state conditions, rests in showing that
a sensible spectrum results.  In Sects.~\ref{sfive} and \ref{ssix} we will
succeed in doing this for the massless states in the spectrum, but will fail
for the massive states.

\subsection{Constructing the fractional supercurrent}

We start by constructing the fractional supercurrent in the CFT
corresponding to a single dimension of the $K=4$ FSS.  Later we will
tensor together $D$ components to get an expression for the full
fractional supercurrent.

Recall the form of the supercurrent in a single
dimension of the usual superstring:
\beq{supcur} G_s~=~\psi\,\partial X~.  \eeq
In terms of a $\bZ_2$ parafermion description, the Majorana
fermion field $\psi$ is the Virasoro primary field in the
$j=1$, $m=0$ parafermion sector $[\phi^1_0]$.  By analogy
with this construction, we might naively expect the fractional
supercurrent to have the form $G\sim\phi^1_0\,\partial X$,
where $\phi^1_0$ stands for a Virasoro primary in the $[\phi^1_0]$
sector of the $\bZ_4$ parafermion theory.  However, as we have seen
in Sect.~\ref{stwo}, the $\bZ_4$ parafermion sectors (unlike the
$\bZ_2$ case) contain an infinite number of primary fields.  In
particular the $[\phi^1_0]$ sector consists of a tower of primary fields
$\epsilon^{(3n\pm1)}$ of dimensions $(3n\pm1)^2/3$ for $n\in\bZ$.
The lowest-dimension fields are the so-called energy operators
$\epsilon^{(\pm1)}$ with dimension $1/3$, implying the
fractional supercurrent has dimension $4/3$.  Note, however, that
there are also Virasoro primary fields $\epsilon^{(\pm2)}$ in the
$[\phi^1_0]$ sector with dimension $4/3$.
In general, if we demand that the algebra generated by the fractional
supercurrent and the energy-momentum tensor close, we will find
that both the $\epsilon^{(\pm1)}$ and $\epsilon^{(\pm2)}$ fields
will have to be included in the definition of $G$.

Another difference between the fractional supercurrent and the
usual supercurrent (\ref{supcur}) stems from the fact that the
$[\phi^1_0]$ sector appears with multiplicity two in the free
boson representation of the $\bZ_4$ parafermion theory described
in Sect.~\ref{stwo}.  This means that the fractional supercurrent
$G$ will be naturally split into two currents, $G^{+}(z)$ and
$G^{-}(z)$.  Indeed, demanding that two spin-$4/3$ currents
and the energy-momentum tensor built from a boson field $X$
and $\bZ_4$ parafermion fields form a closed
operator product algebra, one discovers the expressions
\beqa{curform}
  G^{+}(z)&=&{1\over\sqrt2}\epsilon^{(1)}(z)\,i\partial X(z)
  +{1\over2}\epsilon^{(-2)}(z)~,\nonumber\\
  G^{-}(z)&=&{1\over\sqrt2}\epsilon^{(-1)}(z)\,i \partial X(z)
  +{1\over2}\epsilon^{(2)}(z)~,\\
  T(z)&=&-\,{1\over2}:\partial X(z)\partial X(z):
  -\,{3\over4}:\partial\rho(z)\partial\rho(z):~,\nonumber
\eeqa
which satisfy the algebra
\beqa{ftalg}
  G^{+}(z)G^{+}(w)&= & {1\over(z-w)^{4/3}}
   \left\{G^{-}(w)+{1\over2}(z-w)\partial G^{-}(w)\right\},\nonumber\\
  G^{-}(z)G^{-}(w)&= & {1\over(z-w)^{4/3}}
   \left\{G^{+}(w)+{1\over2}(z-w)\partial G^{+}(w)\right\},\\
  G^{+}(z)G^{-}(w)&= & {(1/2)\over(z-w)^{8/3}}\left\{{3\over2}+
   2(z-w)^2 T(w)\right\}.\nonumber
\eeqa
In the above OPEs, only the singular terms ({\it i.e.}\ those with
negative powers of $z-w$) have
been included.  The normalization of the right hand side of
the $G^{+}G^{-}$ OPE does not completely fix the normalization
of the $G^\pm$ currents separately. This extra freedom
was used make the structure constants appearing in the
$G^{+}G^{+}$ and $G^{-}G^{-}$ OPEs equal.  We
have also fixed the trivial
symmetry which takes all $\epsilon^{(a)}(z)$ to $x^a\epsilon^{(a)}(z)$
for $x$ some complex number, corresponding to a shift in the
origin of the $\rho(z)$ boson.

This split algebra (\ref{ftalg}) is a special case of the spin-$4/3$
algebra studied by Zamolodchikov and Fateev \cite{FZft}.  An
important property of the split algebra is that its currents,
$G^{+}$ and $G^{-}$, inherit definite $\bZ_3$ charges, $q=1$ and $-1$
respectively, from the parafermion representation.  This is
reflected in the fact that $G^\pm$
satisfy abelian braiding relations.
In terms of their OPEs, this means that only one kind
of cut appears on the right hand side.  For example, the
terms in the $G^{+} G^{+}$ OPE are all proportional to
$(z-w)^{n+2/3}$ where $n$ in an integer.  Because of this, we
will be able to derive generalized commutation relations
satisfied by the current modes, following the discussion
in \ref{appB}.

The full fractional superconformal current $G(z)$ of the FSS
is defined in the $D$-fold tensor product space of coordinate
boson plus $\bZ_4$ parafermion theories.  Let us rename the
currents in (\ref{curform}) associated with the $\mu$th
dimension $G^{(\mu)\pm}$ and $T^{(\mu)}$.
Then the algebra generated by the fields
\beqa{tensor}
 G(z)&=&\sum_{\mu=0}^{D-1}\left(G^{(\mu)+}(z)
 +G^{(\mu)-}(z)\right)~,\nonumber\\
 T(z)&=&\sum_{\mu=0}^{D-1}{T}^{(\mu)}(z)~,
\eeqa
is the full fractional superconformal algebra:
\beqa{truefss}
 T(z)T(w)&=&{D\over(z-w)^4}+{2T(w)\over(z-w)^2}
  +{\partial T(w)\over(z-w)}~,\nonumber\\
 T(z)G(w)&=&{{4\over3}G(w)\over(z-w)^2}
  +{\partial G(w)\over(z-w)}~,\\
 G(z)G(w)&=&{1\over(z-w)^{8/3}}\left\{{3D\over2}+
   2(z-w)^2 T(w)\right\}\nonumber\\
  &&\mbox{}+{1\over(z-w)^{4/3}}
   \left\{G(w)+{1\over2}(z-w)\partial G(w)\right\}~.\nonumber
\eeqa
It is this non-local, non-abelianly braided algebra which we
will take as the generator of the physical state conditions for
the $K=4$ FSS.  It will play a role in the FSS analogous to that
played by the superVirasoro algebra in the superstring.  The
representation theory of the fractional superconformal algebras
(\ref{ftalg}) and (\ref{truefss}) is similar to
that of the superVirasoro algebra in that both have a discrete
series of minimal unitary representations.  However, the
fractional superconformal algebras have some qualitatively
new properties arising from their non-local nature (the cuts
in their OPEs).  The representation theory of the fractional
superconformal algebras is discussed further in \ref{appC}.

\subsection{Fractional superconformal mode algebra}

We now turn to the mode expansions and generalized
commutators following from the fractional superconformal
algebra.  We noted above that the split algebra (\ref{ftalg}) has a
$\bZ_3$-symmetry and is abelianly braided.  Therefore the
arguments of \ref{appB} can be directly applied in deriving
the GCRs following from the split algebra.  The mode expansions
for the full superconformal current can be built from the split
algebra pieces.

The Fock space of split algebra representations fall into
sectors ${\cal H}_q$ labelled by their $\bZ_3$ charge.  The currents
$G^{+}$ and $G^{-}$ have $\bZ_3$ charges $q=+1$ and $q=-1$, respectively,
and act on the Fock space sectors according to the rules
\beqq
  G^{+}: {\cal H}_q\rightarrow{\cal H}_{q+1}~,\quad
  G^{-} : {\cal H}_q\rightarrow{\cal H}_{q-1}~,
\eeq
where the $\bZ_3$ charge is defined mod 3.  With these
actions, the mode expansions of $G^{+}$ and $G^{-}$ are
defined as
  \beqa{Jmode}
  G^{+}(z)\chi_q(0)&=&\sum_n z^{n-q/3}
   G^{+}_{-1-n-(1-q)/3}\chi_q(0)~,\nonumber\\
  G^{-}(z)\chi_q(0)&=&\sum_n z^{n+q/3}
   G^{-}_{-1-n-(1+q)/3}\chi_q(0)~,
  \eeqa
where $\chi_q$ is an arbitrary state in ${\cal H}_q$.  These
mode expansions can be inverted to give
\beqa{edomJ}
  G^{+}_{n-(1-q)/3}\chi_q(0)&=&\oint_\gamma{{\rm d}z
   \over2\pi i} z^{n+q/3}G^{+}(z)\chi_q(0)~,\nonumber\\
  G^{-}_{n-(1+q)/3}\chi_q(0)&=&\oint_\gamma{{\rm d}z
   \over2\pi i} z^{n-q/3}G^{-}(z)\chi_q(0)~.
\eeqa
Here, $\gamma$ is a contour encircling the origin once,
where $\chi_q(0)$ is inserted.

{}From the argument reviewed in \ref{appB}, the GCRs for
the current modes of the split algebra (\ref{ftalg}) can be
derived.  As explained in \ref{appB}, there are many GCRs that
can be derived from a single OPE, depending on how many terms
on the right hand side of the OPE one wishes to include.  We will
include only the singular terms, shown in eq.~(\ref{ftalg}).  With
this choice, the split algebra GCRs become \cite{FZft}
\beqa{Jalg}
 &\displaystyle{\sum_{\ell=0}^\infty c^{(-2/3)}_\ell\left[
  G^{+}_{{q\over3}+n-\ell}G^{+}_{{2+q\over3}+m+\ell}-
  G^{+}_{{q\over3}+m-\ell}G^{+}_{{2+q\over3}+n+\ell}\right]
  ~= } & \nonumber\\
  &\displaystyle{ {1\over2}(n-m)G^{-}_{{2+2q\over3}+n+m}
  ~, } &\nonumber\\
 &\displaystyle{ \sum_{\ell=0}^\infty c^{(-2/3)}_\ell\left[
  G^{-}_{-{q\over3}+n-\ell}G^{-}_{{2-q\over3}+m+\ell}-
  G^{-}_{-{q\over3}+m-\ell}G^{-}_{{2-q\over3}+n+\ell}\right]
  ~= } &\\
  &\displaystyle{ {1\over2}(n-m)G^{+}_{{2-2q\over3}+n+m}
  ~, } &\nonumber\\
 &\displaystyle{ \sum_{\ell=0}^\infty c^{(-1/3)}_\ell\left[
  G^{+}_{{1+q\over3}+n-\ell}G^{-}_{-{1+q\over3}+m+\ell}+
  G^{-}_{-{2+q\over3}+m-\ell}G^{+}_{{2+q\over3}+n+\ell}\right]
  ~= } &\nonumber\\
 &\displaystyle{ L_{n+m}+{3\over8} \left(n+1+{q\over3}\right)
  \left(n+{q\over3}\right)\delta_{n+m}~, } &\nonumber
\eeqa
where these expressions are understood to be acting on a
state in ${\cal H}_q$.
For completeness, we also write down the Virasoro algebra and
the standard commutators following from the fact that $G^\pm$
are dimension $4/3$ primary fields:
\beqa{LJalg}
 \left[L_m,L_n\right]&=&
 (m-n)L_{m+n}+
 {1\over6}(m^3-m)\delta_{m+n}~,\nonumber\\
 \left[L_m,G^\pm_r\right]&=&
 \left({1\over3}m-r\right)G^\pm_{m+r}~,
\eeqa
where the moding $r$ is the one appropriate to whichever
$\bZ_3$ sector the $G^\pm$ currents are acting on.
The Virasoro algebra has central charge $c_0=2$,
corresponding to one dimension of the full $K=4$ FSS.

It is useful to have expressions for the $G^\pm$
modes in terms of the $\bZ_4$ parafermion and coordinate
boson modes that we used in Sect.~\ref{stwo} to construct
the Fock space of FSS states.  Using the explicit form for the
currents (\ref{curform}) and the parafermion and boson
mode expansions (\ref{epmode}) and (\ref{Xmode}),
we express the current modes as
\beqa{curmode}
 G^{+}_{n-(1-q)/3}&=&{1\over\sqrt2}\sum_{\ell\in\bZ}~
  \alpha_{-\ell}\epsilon^{(1)}_{n+\ell-(1-q)/3}~+~{1\over2}
  \epsilon^{(-2)}_{n-(1-q)/3}~,\nonumber\\
 G^{-}_{n-(1+q)/3}&=&{1\over\sqrt2}\sum_{\ell\in\bZ}~
  \alpha_{-\ell}\epsilon^{(-1)}_{n+\ell-(1+q)/3}~+~{1\over2}
  \epsilon^{(2)}_{n-(1+q)/3}~.
\eeqa
The analogous formula for the energy-momentum modes
$L_n$ has already been given in eq.~(\ref{dimvir}).
{}From these expressions and the parafermion and boson
commutation relations derived in Sect.~\ref{stwo}, one can
verify the current commutators (\ref{Jalg}) and (\ref{LJalg}).

The $G(z)$ current of the full fractional superconformal
algebra (\ref{truefss}) is a sum of one copy of $G^{+}$ and $G^{-}$
for each space-time dimension.  The full Fock space can be
decomposed into $\bZ_3$ sectors for each dimension, which we denote
by ${\cal H}_{\{q_\mu\}}$ where $\mu=0,1,\ldots,D-1$, and
$D$ is the space-time dimension.  Since the action of $G$ mixes
these sectors,
\beqq
 G:{\cal H}_{\{q_0,\ldots,q_{D-1}\}}\rightarrow\bigoplus^{D-1}_{\mu=0}
 \left({\cal H}_{\{q_0,\ldots,q_\mu-1,\ldots,q_{D-1}\}}\oplus
 {\cal H}_{\{q_0,\ldots,q_\mu+1,\ldots,q_{D-1}\}}\right)~,
\eeq
it will have no definite moding when acting on states in these
sectors.  However, we can define mode operators for $G(z)$
acting between two specific sectors:
\beqq
 G^{(\mu)\pm}_r:
 {\cal H}_{\{q_0,\ldots,q_\mu,\ldots,q_{D-1}\}}\rightarrow
 {\cal H}_{\{q_0,\ldots,q_\mu\pm1,\ldots,q_{D-1}\}}~.
\eeq
The moding $r$ is determined (up to an integer part) by the
initial and final sectors.  In other words, because of its
nonabelian braiding, the moding of the currents depends not
only on the Fock space sector upon which it is acting, but
also on the sector it maps to.
The mode expansion of $G(z)$ can thus be written
\beqq
 G(z)=\sum_{\mu=0}^{D-1}\sum_{n\in\bZ}\left(
 z^{n-q_\mu/3}G^{(\mu)+}_{-1-n-(1-q_\mu)/3}+
 z^{n+q_\mu/3}G^{(\mu)-}_{-1-n-(1+q_\mu)/3}\right)~,
\eeq
when acting on an arbitrary state in ${\cal H}_{\{q_\mu\}}$.
{}From our construction of the fractional supercurrent from
the split algebra currents (\ref{tensor}),
it is clear that the modes $G^{(\mu)\pm}$ are just $G^\pm$
for the $\mu$th $\bZ_4$ parafermion plus coordinate boson
CFT in the full $D$-fold tensor product CFT of the FSS.

When we apply the physical state conditions, we will be
interested in acting with a given moding of the full current
$G(z)$ on states which are not in a definite Fock space sector.
We can define such a mode operator as follows.  Write the Fock
space states as a $3^D$-dimensional column vector with elements
labeled by the set $\{q_\mu\}$ of their $\bZ_3$
quantum numbers in each space-time dimension.
Then the action of $G(z)$ on this state can be represented by a
$3^D\times3^D$ matrix of operators with $G^{(\mu)\pm}$ being its
various (off-diagonal) elements.  We can think of the coordinate
boson and $\bZ_4$ parafermion modes as matrices in a similar way.
Thus, if, for example, there were only one space-time dimension,
the $\epsilon^{(1)}_{n-(1-q)/3}$ mode would be represented by a
$3\times3$ matrix with one non-zero element, namely
the mode itself, in column $\{q\}$ and row $\{q+1\}$.
For $D$ dimensions, $\epsilon^{(1),\mu}_{n-(1-q)/3}$ would
be the tensor product of $D$ $3\times3$ matrices
\beqq
 \epsilon^{(1),\mu}_{n-(1-q)/3}~=~\id\otimes\ldots\otimes
 \epsilon^{(1)}_{n-(1-q)/3}\otimes\ldots\otimes\id~,
\eeq
where the $\epsilon^{(1)}$--matrix is in the $\mu$th position.
Similar comments apply to the general parafermion mode
$\epsilon^{(a),\mu}_r$ and the coordinate boson modes
$\alpha^\mu_n$.  With this convention, we can write the
general formula for the mode $G_r$ in terms of boson and
parafermion modes from the expressions (\ref{curmode}):
\beq{Gcurmode}
 G_r~=~\sum_{\mu=0}^{D-1}\left({1\over\sqrt2}\sum_{\ell\in\bZ}
 (\alpha_{-\ell})_{\mu} \left[\epsilon^{(1),\mu}_{r+\ell}
 +\epsilon^{(-1),\mu}_{r+\ell}\right]~+~{1\over2}
 \left[\epsilon^{(-2),(\mu)}_r+\epsilon^{(2),(\mu)}_r\right]\right)~,
\eeq
for an arbitrary moding $r\in\bZ/3$.
The matrix notation we have introduced to write (\ref{Gcurmode})
may be more easily utilized by noting that it is formally
equivalent to adopting the convention
that when a parafermion mode operator has the wrong moding
to act on a given sector, it vanishes.

Note that we have embellished
the boson and parafermion modes with an extra superscript
$\mu$ denoting which of the $D$ tensored copies they act in.
In the case of the coordinate boson modes $\alpha^\mu$, we
interpret this superscript as a space-time
Lorentz covariant index.  Lorentz invariance of the fractional
supercurrent then implies that the index on the
$\epsilon^{(\pm1),\mu}$ fields is also Lorentz covariant,
and that a factor of the Minkowski metric should be understood
in the first term of (\ref{Gcurmode}).  By the same token,
the index on the $\epsilon^{(\pm2),(\mu)}$ fields can {\it not\/}
be Lorentz covariant.  This should not be surprising, since,
by the parafermion GCR (\ref{epgcr1}) any $\epsilon^{(\pm2)}$
mode can be written in terms of two $\epsilon^{(\pm1)}$ modes;
for example
\beqa{4.19}
 \epsilon^{(2),(\mu)}_{n+m+(2q-1)/3}&=&
 \sum^{\infty}_{\ell=0}c^{(-5/3)}_\ell\,\Bigl[
 \epsilon^{(1),\mu}_{n-1-\ell+q/3}~
 \epsilon^{(1),\mu}_{m+1+\ell+(q-1)/3}\nonumber\\
 &&\quad\quad\mbox{}+\epsilon^{(1),\mu}_{m-1-\ell+q/3}~
 \epsilon^{(1),\mu}_{n+1+\ell+(q-1)/3}\Bigr]\,.
\eeqa
Replacing the $\sum_\mu\epsilon^{(\pm2),(\mu)}_r$ terms
in (\ref{Gcurmode}) with the above expression (and including
a factor of the Minkowski metric to tie the indices together),
we recover a space-time Lorentz covariant interpretation of the
index $\mu$.

The $G_r$ modes (\ref{Gcurmode}) of the full fractional supercurrent
satisfy no simple GCRs such as those satisfied by the
component modes $G^{(\mu)\pm}_r$ in (\ref{Jalg}).  The reason for
this is that for $\mu\neq\nu$, $G^{(\mu)\pm}_r$ and $G^{(\nu)\pm}_s$
satisfy simple (anti)commutation relations which can not
be combined with the GCRs in (\ref{Jalg}) to form expressions
involving only the full $G_r$ modes.  This is simply a
reflection of the fact that the $G^{(\mu)\pm}(z)$ currents
satisfy abelian braid relations, whereas the full current
$G(z)$, does not.

The energy-momentum tensor and its mode expansion
for the full $D$-dimensional FSS is built up in a similar
manner from pieces acting between Fock space
sectors.  However, since the $L_n$ moding is always integral
(equivalently, the $L_n$ modes act diagonally on the Fock
space sectors), there is never any need to keep track of the sector
indices $q$ and $\mu$.  So, in the matrix notation, these modes satisfy
the usual commutators
\beqa{LGalg}
 \left[L_m,L_n\right]&=&(m-n)L_{m+n}+
 {D\over6}(m^3-m)\delta_{m+n}~,\nonumber\\
 \left[L_m,G_r\right]&=&\left({m\over3}-r\right)G_{m+r}~.
\eeqa

\subsection{Physical state conditions}

In the usual superstring, the physical state conditions are
constraints following from gauge-fixing the local world-sheet
symmetry.  Classically these constraints in the superconformal
gauge are given by the vanishing of the energy-momentum tensor
and superconformal current:  $T(z)=G_s(z)=0$.
The full local world-sheet symmetry of the $K=4$ FSS is unknown,
though it should include reparametrization and Weyl invariance.
We will assume that some analog of the superconformal gauge
exists in the FSS, giving rise to an algebra of constraints
generated by the vanishing of $T(z)$ and the fractional
superconformal current $G(z)$.  In other words, we assume that
the fractional superconformal algebra is the quantum version of
some classical constraint algebra.  Thus, although we do not
know of any classical local symmetry on the world-sheet that
gives rise to a spin-4/3 current as a constraint upon
gauge-fixing, we nevertheless assume the weak physical state
conditions
  \beq{qcon}
  \langle\psi\vert :\!T(z)\!:\ket{\phi}=
   \langle\psi\vert :\!G(z)\!:\ket{\phi}=0~,
  \eeq
for any physical states $\ket{\phi}$ and $\ket{\psi}$.  The
normal ordering symbols are there to remind us that there may
be normal ordering constants in the quantum definition of the
currents.

The energy-momentum constraint is ``factorized'' by
expressing $T(z)$ in terms of its mode operators,
  \beqq
  T(z)=\sum_n L_n z^{-n-2}~,
  \eeq
and using the mode algebra following from the constraint
algebra OPEs
  \beq{Vir}
  [L_m,L_n]=(m-n)L_{m+n}+{D\over6}(m^3-m)\delta_{m+n}~,
  \eeq
as well as the hermiticity conditions
  \beqq
  (L_n)^{\dagger}=L_{-n}~,
  \eeq
to factorize the quantum constraints into the usual physical
state conditions
  \beqa{Vpsc}
  L_0\ket{\phi}&=&v\ket{\phi}~,\nonumber\\
  L_n\ket{\phi}&=&0~,\quad n\geq0~.
  \eeqa
Here $v$ is the intercept, a normal ordering constant in the
definition of $T$.  This is a consistent set of constraints
because the positive modes form a closed subalgebra of the
Virasoro algebra generated by the two modes $L_1$ and $L_2$.

Let us mimic this discussion in the case of the fractional
superconformal constraint.  In our matrix notation, between
two physical states, $G(z)$ will have the mode expansion
\beqq
  \langle\psi\vert :\!G(z)\!:\ket{\phi}=
 \sum_{r\in\bZ/3}z^{r-4/3} \langle\psi\vert :\!G_r\!:\ket{\phi}~.
\eeq
{}From the explicit formula (\ref{Gcurmode}) for the current
modes and the hermiticity properties of the $\bZ_4$ parafermion
field modes
\beqq
 (\epsilon^{(a)}_r)^\dagger=\epsilon^{(-a)}_{-r}~,
\eeq
it follows that the fractional superconformal current
satisfies the hermiticity condition
\beqq
 (G_r)^\dagger=G_{-r}~.
\eeq
This makes it plausible to take as physical state conditions
factorizing the $G(z)$ quantum constraint (\ref{qcon})
\beqa{Gpsc}
  G_0\ket{\phi}&=&\beta\ket{\phi}~,\nonumber\\
  G_r\ket{\phi}&=&0~,\quad r>0~,
\eeqa
where $\beta$ is an undetermined normal-ordering constant.

{}From the GCRs for the fractional supercurrent components (\ref{Jalg}),
it is not hard to see that the physical state conditions (\ref{Gpsc}) are
consistent with themselves and with the Virasoro conditions
(\ref{Vpsc}).  In particular, from (\ref{Jalg}) and (\ref{LGalg}),
it is not possible to derive an identity of the form
\beqq
 G_{-r}G_s\ket{\phi}=G_{s-r}\ket{\phi}~,
\eeq
for $r>s>0$, or similar relations with $L_{-r}$ replacing
the $G_{-r}$ mode.  Because of the infinite sums that
appear in the GCR algebra, it is unclear in what sense, if
any, the positive modes of $G$ can be said to form a
closed subalgebra of the constraint algebra.  However, from
the $L_m$--$G_r$ commutator (\ref{LGalg}), we can generate
all the conditions (\ref{Vpsc}) and (\ref{Gpsc})
from the set $\{L_0,L_1,G_0,G_{1/3},G_{2/3}\}$.

\setcounter{footnote}{0}
\section{\label{sfive} Low-lying states of the fermionic sector}

In this section we construct the full space of states in the
fermionic sector of a $D$-dimensional open $K=4$ FSS.  The ground
state of this sector forms a representation of the $D$-dimensional
Clifford algebra.  We
then solve for the subset of states at the massless
level that satisfy the physical state conditions.  They obey
Lorentz covariant equations of motion as well as a sufficient
number of constraints to eliminate all unphysical degrees of
freedom.  The number of propagating
modes at these levels can be halved by a chirality projection
analogous to the GSO projection \cite{GSO} in the Ramond sector of the
superstring.  We derive, using a counting argument involving
the Euler pentagonal number theorem, the explicit form of the
chirality operator at all levels of the fermionic sector.

\subsection{Fermionic ground state and Clifford algebra}

In Sect.~\ref{stwo} and \ref{sthree} we have argued that the
fermionic sector of the FSS Fock space consists of all states
obtained by successive applications of the dimension-$1/3$
$\epsilon^{(\pm1),\mu}(z)$ parafermion fields and the
coordinate boson fields $\partial X^\mu(z)$ on the ground
state:
\beq{Fgrnd}
 \ket{\alpha,p}\equiv\left(\prod_{\mu=0}^{D-1}
 \epsilon^{(\pm1/2),\mu}\right):\!{\rm e}^{ip\cdot X}\!:\ket{0}~.
\eeq
Here $\epsilon^{(\pm1/2),\mu}(z)$ are the two dimension-$1/12$
parafermion spin fields associated with the $\mu$th space-time
dimension.  At a given momentum $p$, the $\alpha$ index of the
ground state labels its $2^D$-fold degeneracy.  We will show that
the zero-modes of the $\epsilon^{(\pm1),\mu}$ fields naturally
form the $D$-dimensional Clifford algebra when acting on
(\ref{Fgrnd}).

Let us start by considering the CFT corresponding to a single
space-time dimension.  For the sake of notational simplicity,
we rename the dimension-$1/3$ parafermion fields
\beqq
 \epsilon\equiv\epsilon^{(+1)}~,\quad\quad
 \epsdag\equiv\epsilon^{(-1)}~,
\eeq
and the spin states
\beqqa
 \ket{+}&\equiv&\epsilon^{(+1/2)}_{-1/12}\ket{0}~,\nonumber\\
 \ket{-}&\equiv&\epsilon^{(-1/2)}_{-1/12}\ket{0}~.
\eeqa
Note that the states $\ket{\pm}$ have $\bZ_3$ charge $q=\mp1$.

{}From the discussion of Sect.~\ref{stwo}, the allowed modings $r$
of $\epsilon$ and $\epsdag$  when acting on $\ket{\pm}$ are
either $r=n$ or $r=n-2/3$, where $n \in \bZ$.  The integral modings
map between states with $\bZ_3$ charge $q=+1$ and $q=-1$,
whereas the other modings map between $q=\pm1$ and $q=0$
states.  In particular, by eq.~(\ref{2.45}) we find that
\beqa{epact}
 \epsilon_0\ket{+}&=&\ket{-}~,\nonumber\\
 \epsdag_0\ket{-}&=&\ket{+}~.
\eeqa
Fig.~1 summarizes the allowed modings of the
integral winding-number fields $\epsilon^{(a)}$, $a\in\bZ$,
on the different $\bZ_3$ sectors.

\begin{figure}
 \vspace{3.0in}
 \caption{
 The action and modings of the parafermion
 fields $\epsilon^{(a)}$ on the Fock space sectors of $\bZ_3$
 charge $q$.  The parafermion field winding number (superscript)
 is understood mod 3, and its moding (subscript) mod 1.
 }
\end{figure}

We can think of $\ket{\pm}$ as basis vectors in a two-dimensional
space of ground states $\ket{\alpha}$, $\alpha\in\{1,2\}$:
\beqq
 \ket{1}\equiv\left({\ket{+}}\atop{0}\right)~,\quad\quad
 \ket{2}\equiv\left({0}\atop{\ket{-}}\right)~.
\eeq
Acting on this space we can define the operators $\eptilde_n$ in the
following way:
\beq{eptilde}
 \eptilde_n~\equiv~\pmatrix{0&\epsdag_n\cr \epsilon_n & 0\cr}
 ~,\quad\quad n\in\bZ~.
\eeq
(In terms of the matrix notation for the parafermion modes described
in Sect.~\ref{sfour}, (\ref{eptilde}) is simply the statement
$\eptilde_n=\epsilon_n+\epsdag_n$.)  From (\ref{epact}) we learn
\beqq
 \eptilde_0~\eptilde_0\ket{\alpha}=\ket{\alpha}~,
\eeq
which can be rewritten as the one-dimensional Clifford algebra
\beqq
\left\{ \eptilde_0,\eptilde_0\right\}\ket{\alpha}~=~2\ket{\alpha}~.
\eeq
Thus we can identify $\eptilde_0$ with the gamma matrix of the
Clifford algebra when acting on the ground state $\ket{\alpha}$.

Now we turn to the tensor product theory.  In the $D$-dimensional
tensor product theory, the (reducible) ground state of the fermionic
sector is represented by
\beq{redgr}
\ket{\alpha}~=~\bigotimes_{\mu=0}^{D-1}~\ket{\alpha^\mu}~,
\eeq
where $\ket{\alpha^\mu}$ is the fermionic ground state of the
$\mu$th component of the tensor product theory and $\alpha =
\{\alpha^0,\ldots,\alpha^{D-1}\}$.  (Note that $\mu$ in these
expressions is not a Lorentz index but simply a dimensional
label.)  $\ket{\alpha}$ spans a $2^D$-dimensional vector space.
Ultimately we will reduce this space to obtain an irreducible
spinor representation of the Lorentz group.

We define the $\eptilde^\mu_n$ modes acting on this space in
the obvious way:
\beqq
 \eptilde^\mu_n\equiv\id\otimes\ldots\otimes\eptilde_n
 \otimes\ldots\otimes\id~,
\eeq
where $\eptilde_n$ is in the $\mu$th position in the tensor
product.  If we now consider the algebra of the ${\eptilde}^{\mu}_0$
modes, we find
\beqq
\left\{ \eptilde^\mu_0,\eptilde^\mu_0
\right\}\ket{\alpha}~=~2\ket{\alpha}~.
\eeq
The modes $\eptilde^\mu_0$ and $\eptilde^\nu_0$ can be chosen
to anticommute for $\mu\neq\nu$ by an appropriate choice of
Klein factors \cite{SW}.  Combining the $\mu=\nu$ and $\mu\neq\nu$
cases we find the $D$-dimensional Clifford algebra
\beqq
\left\{ \eptilde^\mu_0,\eptilde^\nu_0
\right\}\ket{\alpha}~=~2g^{\mu\nu}\ket{\alpha}~,
\eeq
so again we have, on the ground state,
\beqq
\eptilde^\mu_0\ket{\alpha}~=~\gamma^\mu\ket{\alpha}~.
\eeq
Here we take $g^{\mu\nu}$ to be the Minkowski metric.
We can define $\ket{\alpha}$, the ground state of the
fermionic sector in the tensor product theory, so that it
furnishes an irreducible spinor representation of the Lorentz
algebra SO$(D-1,1)$.  The dimension of this representation is
$2^{D/2}$.  All of the states in this representation can be
constructed from linear combinations of the states $\ket{\alpha}$
defined by eq.~(\ref{redgr}).  This state must be multiplied by a
momentum $p$ vertex operator made from the coordinate boson
$X^\mu$ zero modes, to form the full fermionic sector
ground state $\ket{\alpha,p}$.

It should be clear that so far this discussion closely
parallels that of the Ramond sector of superstring theory.
However, we must remark upon an unusual feature of the FSS Fock
space.  Because the $\eptilde^\mu$ modes satisfy GCRs instead
of simple commutation or anticommutation relations, the
Lorentz-covariant meaning of their space-time index $\mu$ is
unclear.  We have shown from the structure of the GCRs that the
$\eptilde_0$ modes satisfy the Clifford algebra when acting on
the ground state.  However, this will not be true in general when they
act upon excited states in the Fock space.  Thus we
are not free to replace the $\eptilde_0$ modes with gamma matrices
unless they are acting on the ground state.  More generally, by taking
a tensor product structure for the FSS Fock space, we have only
ensured a permutation symmetry among the different dimensions,
but not necessarily the rotational symmetry of the Lorentz group.

\subsection{Massless physical states}

We now wish to obtain the set of physical states at the
massless level.  These states are a subset of the states spanned by the
level $0$ state $\ket{\alpha,p}$.  As discussed in
Sect.~\ref{sfour}, physical states should
satisfy the conditions (\ref{Vpsc}) and (\ref{Gpsc}):
\beqa{pscF}
 L_0\ket{\psi_{\rm phys}}&=&v\ket{\psi_{\rm phys}}~,\nonumber\\
 G_0\ket{\psi_{\rm phys}}&=&\beta\ket{\psi_{\rm phys}}~,\nonumber\\
 L_n \ket{\psi_{\rm phys}}&=&0~, \quad 0<n\in\bZ~,\\
 G_r \ket{\psi_{\rm phys}}&=&0~, \quad 0<r\in\bZ/3~.\nonumber
\eeqa
Here $v$ and $\beta$ are the as-yet-undetermined normal-ordering
constants (``intercepts'') of the fermionic sector.

Let us impose the physical state conditions (\ref{pscF}) on
level $0$ states.   Consider the general state at level $0$:
\beqq
\ket{ \psi_0 }~=~\ket{\alpha,p}u^\alpha(p)~,
\eeq
where $u^\alpha(p)$ is a spinor polarization.  The $L_r$
and $G_r$ modes for $r>0$ automatically annihilate $\ket{\psi_0}$,
since it is the ground state of the fermionic sector.  Thus the
corresponding physical state conditions are identically satisfied.

Because $\ket{\alpha,p}$ has conformal dimension ${p^2\over2}
+{D\over12}$ for the $D$ space-time dimensional FSS Fock space,
the $L_0$ physical state condition is equvalent to
\beqq
 (L_0-v)\ket{\psi_0}=0~\Longrightarrow~
 {p^2\over2}+{D\over12}~=~v~,
\eeq
fixing the mass of the physical state in terms of the intercept $v$.

As follows easily from the expansion of the fractional supercurrent
modes (\ref{Gcurmode}), the action of the $G_0$ mode on
$\ket{\psi_0}$ is
\beqqa
 G_0\ket{\psi_0}
 &=&{1\over\sqrt2}\left(\alpha_0\cdot\eptilde_0\right)
 \ket{\alpha,p}u^\alpha\nonumber\\
 &=&\ket{\alpha,p}\slp u~.
\eeqa
Since $\eptilde_0$ acts on the ground state, we
have replaced it with a $\gamma$-matrix.  Thus the $G_0$
physical state condition is equivalent to
\beq{Gzer}
 (G_0-\beta)\ket{\psi_0}=0~\Longrightarrow~
 {1\over\sqrt2}{\slp}~u(p)=\beta u(p)~.
\eeq
Comparing the $L_0$ and $G_0$ conditions shows that the
two intercepts are related by
\beqq
 \beta^2=v-{D\over12}~.
\eeq

Since the $G_0$ mode, when acting on the ground
state, is the simple product of an $X^\mu$ mode and
a parafermion energy operator mode, it is hard to see
why the normal-ordering constant $\beta$ should
appear at all.  (When the modes of the $\epsilon^{(\pm2)}$
fields in the fractional supercurrent contribute, as can happen,
we will see, in the bosonic sector, then $\beta$ could
naturally be expected to be non-zero.)  Setting $\beta=0$
implies that $v=D/12$ and that the level zero physical
states are massless.  Indeed, this value of the intercept is
the critical value, because a tower of extra null states appears
in the FSS spectrum when $v=D/12$.  In particular, the fractional
superconformal algebra (\ref{Jalg}) implies when acting on physical
states with every space-time component in the ground state sector
({\it i.e.}\/ with $\bZ_3$ charge $q=\pm1$), that
\beqq
 G_0G_0~=~L_0-{D\over12}~.
\eeq
Thus if $\ket{\psi}$ is a physical state with intercept $v=D/12$,
then $\ket{\chi}=G_0\ket{\psi}$ also obeys the physical state
conditions with the same intercept.  $\ket{\chi}$ is thus both physical
and spurious, and therefore null.  This whole argument is precisely
analogous to the familiar argument for the Ramond ground state
of the usual superstring.\footnote{In the superstring, the intercept
$v$ in the Ramond sector is usually set to zero by redefining the
Virasoro mode $L_0$ to $L_0-D/16$ so that
the Ramond ground state automatically has $L_0$ eigenvalue zero.}

With the critical intercepts
\beqq
 v~=~{D\over12}~,\quad\quad\quad\beta~=~0~,
\eeq
the $G_0$ condition (\ref{Gzer}) implies that
$u^\alpha(p)$ satisfies the massless Dirac equation.
For space-time dimension $D=6$, the number of complex degrees
of freedom of the spinor representation is $2^{D/2}=8$.
Since the ground state is a massless spinor, and the space-time
dimension is even, we can impose a Weyl condition, reducing
this number to eight real degrees of freedom, of which only four
propagate, since $u^\alpha(p)$ is the solution to a Dirac equation.
This matches the counting of massless fermion propagating degrees
of freedom found from the study of the partition function in
\ref{sthree}.  The Weyl condition on the massless states is the
analog of the GSO projection \cite{GSO} in the Ramond sector of
the superstring.

\subsection{Fermionic sector Fock space}

In order to consider higher-mass states in the fermionic
sector, we will now derive a basis of states for all integral
levels of the fermionic sector Fock space.
We build the fermionic sector of the FSS Fock space
by the action of the coordinate boson modes $\alpha^\mu_n$
and the parafermion energy operator modes $\epsilon^{(\pm1)}_r$
on the ground state spinor $\ket{\alpha,p}$.
In Sect.~\ref{stwo} we showed that in the single space-time
component theory, the action of the energy operator modes on
the spin field states $\ket{\pm}$ correspond to the parafermion
sectors $[\phi^1_{\pm1}]$ and $[\phi^0_{\pm1}]$.  Recall
that $[\phi^1_{\pm1}]$ consists of the set of fields $\epsilon^{(3n\pm1/2)}$
of conformal dimensions $\bZ+1/12$ and $\bZ_3$ charge $\pm1$,
whereas $[\phi^0_{\pm1}]$ consists of the $\epsilon^{(3n+3/2)}$
fields of dimension $\bZ+3/4$ and $\bZ_3$ charge zero.  We will
call $[\phi^1_{\pm1}]$ the ground state sector, since it includes the
ground state (\ref{Fgrnd}).  The states in $[\phi^0_{\pm1}]$ we
will refer to as ``projection sector'' states for reasons to be made
clear later.  From the moding rules summarized in Fig.~1, we see
that we can define an operator,
$\eptilde_{n-2/3}$, similar to the $\eptilde_n$ mode, but which
has fractional moding on the $q=\pm 1$ sector:
\beqa{epfrac}
 \eptilde_{n-2/3}\ket{ \chi_{+1} }&=&\epsdag_{n-2/3}
 \ket{\chi_{+1}}~,\nonumber\\
 \eptilde_{n-2/3} \ket{\chi_{-1}}&=&\epsilon_{n-2/3}
 \ket{\chi_{-1}}~.
\eeqa
Note that, unlike the $\eptilde_n$ modes, the $\eptilde_{n-2/3}$
modes map $q=\pm1$ states to $q=0$ states.

In eq.~(\ref{2.44}) we wrote down a basis of states in the parafermion
theory for the first three levels.  Adding in the coordinate boson
field, and tensoring $D$ copies together, we find all the states
in the fermionic sector for these levels:
\beqa{Flevel}
 &\ket{\alpha,p}&\quad\quad\quad{\rm level}~0~,\nonumber\\
 &\eptilde^\mu_{-2/3}\ket{\alpha,p}
 &\quad\quad\quad{\rm level}~2/3~,\\
 &\eptilde^\mu_{-1}\ket{\alpha,p}+\alpha^\mu_{-1}\ket{\alpha,p}
 &\quad\quad\quad{\rm level}~1~,\nonumber\\
 &\ldots&\quad\quad\quad\ldots\nonumber
\eeqa
The states at level $\ell$ have conformal dimension
${p^2\over2}+{D\over12}+\ell$.  Note that, using the identities
(\ref{2.45})--(\ref{2.48}), these states can be written in many
other equivalent forms.

Each of the $D$-fold tensor product components of the integer
level states in (\ref{Flevel}) are in the ground state sector.
Thus, in the string function notation of \ref{sthree}, these
states will contribute to a term $(c^2_2)^{D-2}$ in the (light-cone)
partition function.  Indeed, such a term (for $D=6$) certainly appears
in the $A$ block of the FSS partition function (\ref{ABxpr}).
The level-$2/3$ state in (\ref{Flevel}), on the other hand, has one
component in the projection sector.  It would therefore contribute
to a term $(c^2_2)^{D-3}(c^4_2)$ in the partition function; however,
no such term appears in (\ref{ABxpr}).
This is analogous to the GSO projection in
the Neveu-Schwarz sector of the superstring,  where whole levels
of states in the Fock space are projected out.  Of course, in the
superstring this does not happen in the Ramond sector since all
modes automatically have integral moding there.  In the FSS, due
to the non-local nature of the operator algebra (the cuts in the
$\epsilon^{(a)}$ OPEs), integer as well as $\pm1/3$-moded
operators appear in both the bosonic and fermionic sectors.  It is thus
natural to expect that a GSO-like projection removing whole levels
of states will occur in both sectors.

Since the partition function found in \ref{sthree} has
contributions from states only at integral levels, we will
concentrate solely on such states below.  Though there is
nothing preventing us from applying the physical state conditions
to, say, the level-$2/3$ states, presumably such states will
decouple from all scattering amplitudes.  We do not address the
issue of scattering amplitudes in this paper.

A basis of states at level 2 is:
\beqa{Rfock}
 &\alpha^\mu_{-2}\ket{\alpha,p}~,
 ~\alpha^\mu_{-1}\alpha^\nu_{-1}\ket{\alpha,p}~,
 ~\eptilde^\mu_{-1}\eptilde^\nu_{-1}\ket{\alpha,p}~,&\nonumber\\
 &\alpha^\mu_{-1}\eptilde^\nu_{-1}\ket{\alpha,p}~,
 ~\eptilde^\mu_{-2}\ket{\alpha,p}~,&\\
 &\eptilde^{(2)\mu}_{-2}\ket{\alpha,p}~,&\nonumber
\eeqa
for the states in the ground state sector, and
\beq{Rfockp}
 \left. \matrix{
 \eptilde^\mu_{-2/3}\eptilde^\nu_{-2/3}
  \eptilde^\lambda_{-2/3}\ket{\alpha,p}\cr
 \eptilde^\mu_{-2/3}\eptilde^\nu_{-2/3}
  \eptilde^{(2)\lambda}_{-2/3}\ket{\alpha,p}\cr
 \eptilde^\mu_{-2/3}\eptilde^{(2)\nu}_{-2/3}
  \eptilde^{(2)\lambda}_{-2/3}\ket{\alpha,p}\cr
 \eptilde^{(2)\mu}_{-2/3}\eptilde^{(2)\nu}_{-2/3}
  \eptilde^{(2)\lambda}_{-2/3}\ket{\alpha,p}\cr}
 \right\}\quad{\rm for}~~\mu\neq\nu\neq\lambda~,
\eeq
for the projection sector states.  Note that we have introduced
the new notations, paralleling (\ref{eptilde}) and (\ref{epfrac})
\beqq
 \eptilde^{(2)}_n~\equiv~\pmatrix{0&\epsilon^{(+2)}_n\cr
 \epsilon^{(-2)}_n & 0\cr}~,\quad\quad n\in\bZ~,
\eeq
and
\beqqa
 \eptilde^{(2)}_{n-2/3}\ket{\chi_{+1}}&=&
 \epsilon^{(+2)}_{n-2/3}\ket{\chi_{+1}}~,\nonumber\\
 \eptilde^{(2)}_{n-2/3}\ket{\chi_{-1}}&=&
 \epsilon^{(-2)}_{n-2/3}\ket{\chi_{-1}}~,
\eeqa
following from the mode actions of $\epsilon^{(\pm2)}$ on the
$q=\pm1$ sectors (see Fig.~1).  Note that each of the states in
(\ref{Rfockp}) have three of their $D$-fold tensor product components
in the projection sector.  They will therefore contribute to a term of
the form $(c^2_2)^{D-5}(c^4_2)^3$ in the partition function.
Such a term does indeed appear in the $A$ block (\ref{ABxpr}).

In a similar manner, a basis of states for any integer level can be
built up from the $\alpha^\mu_n$ and $\epsilon^{(a)}_r$ modes
acting on the ground state $\ket{\alpha,p}$.  For example, using
the mode algebra derived in Sect.~\ref{stwo}, a basis of states in
the one-component ground state sector can be written
\beq{Fbasis}
 \eptilde^{(3\ell+1)}_{-\ell(3\ell+1)}
 \eptilde_{-m_1}\ldots\eptilde_{-m_j}
 \alpha_{-n_1}\ldots\alpha_{-n_k}\ket{\alpha,p}~,
\eeq
where $\ell,m_i,n_i\in\bZ$, $m_i,n_i>0$,
and the sets $\{m_1,\ldots,m_j\}$
and $\{n_1,\ldots,n_k\}$ are ``dictionary'' ordered.
The $\eptilde^{(a)}$ notation is the obvious generalization
of the tilde notation we have used for other modes above.
Note that in this notation $\eptilde^{(a)}\equiv\eptilde^{(-a)}$.
The state in (\ref{Fbasis}) has winding number $\pm(3\ell+1/2)$
in the parafermion Fock space---in other words, it is a descendent
of the $\eptilde^{(3\ell+1)}_{-\ell(3\ell+1)}\ket{\alpha} \equiv
\epsilon^{(\pm(3\ell+1/2))}_{-\ell(3\ell+1)-1/12}\ket{0}$
state.\footnote{A basis can also be written with just
$\epsilon^{(\pm1)}$ modes without using the higher winding
number modes $\epsilon^{(a)}$ with $|a|>1$.  However, in this
case fractionally moded as well as integrally moded operators
are required in the ground state sector.}  Indeed, an alternative
basis of states, used in Sect.~\ref{stwo} to derive the parafermion
characters (\ref{pfchars}), is for the one-component ground state sector
\beq{Fbasis2}
 \eptilde^{(3\ell+1)}_{-\ell(3\ell+1)}s_{-m_1}\ldots s_{-m_j}
 \alpha_{-n_1}\ldots\alpha_{-n_k}\ket{\alpha,p}~.
\eeq
The recursion relation (\ref{Pdef}) and (\ref{Prec}) can be used
to express the (\ref{Fbasis}) basis elements in terms of the
(\ref{Fbasis2}) basis elements.

A basis of states for the single
component projection sector can similarly be written as
\beq{Fbasis3}
 \eptilde^{(3\ell+1)}_{-3\ell(\ell+1)-2/3}
 \eptilde_{-m_1}\ldots\eptilde_{-m_j}
 \alpha_{-n_1}\ldots\alpha_{-n_k}\ket{\alpha,p}~.
\eeq
The general fermionic state is then a linear combination of tensor
products of (\ref{Fbasis}) or (\ref{Fbasis3}) states for each
space-time dimension.
Note that the (\ref{Fbasis3}) basis actually overcounts states by a factor
of two.  For example, $\eptilde_{-2/3}\ket{\alpha}$ and
$\eptilde^{(2)}_{-2/3}\ket{\alpha}$ are actually the same state
by the rule (\ref{epket2}) derived in Sect.~\ref{stwo}.  Thus
(\ref{Rfockp}) has $2^3$ copies of each independent state.  This
overcounting could be rectified by restricting $\ell\geq0$ in
(\ref{Fbasis3}).  However, allowing $\ell$ to run over the negative
as well as positive integers (and thus allowing the doubling of states
in the projection sector) will turn out to be necessary for the
construction of a Lorentz-invariant GSO-like chiral projection in
the fermionic sector.

\subsection{Chiral projection}

{}From experience with the superstring we expect that we will have to
implement a GSO-like projection halving the number of degrees of freedom
at all fermionic mass levels in order to have space-time supersymmetry.
We will be able to do this if we can define an analog of the $(-1)^{F}$
operator in superstring theory, which we will call $(-1)^\epsilon$,
that will enable us to generalize the Weyl condition on the massless
states to all massive levels. A straightforward generalization from
the ($K=2$) superstring case leads one to guess the form
\beq{chiralguess}
 (-1)^{\epsilon}~{\buildrel ? \over =}~\gamma_{D+1}(-1)^{N({\epsilon})}~,
\eeq
where $N({\epsilon})$ is the number operator for the ${\eptilde}$ modes.
However, this prescription does not tell us what signs to assign the
winding modes.
We will deduce below, from general arguments, the form of the
$(-1)^{\epsilon}$ operator which is correct for all levels.  It will
turn out that the existence of such an operator consistent with
Lorentz invariance follows from a counting argument relying on the
Euler pentagonal number theorem.

Consider the general state in the ground state
sector of the fermionic Fock space---the $D$-fold tensor product
of the basis states given in eq.~(\ref{Fbasis}):
\beq{fullbas}
 \ket{\{\ell_\lambda\},\{m,\mu\},\{n,\nu\}}_g=
 \left(\prod_{\lambda=0}^{D-1}
 \eptilde^{(3\ell_\lambda+1),\lambda}
 _{-\ell_\lambda(3\ell_\lambda+1)}\right)
 \eptilde^{\mu_1}_{-m_1}\ldots\eptilde^{\mu_j}_{-m_j}
 \alpha^{\nu_1}_{-n_1}\ldots\alpha^{\nu_k}_{-n_k}\ket{\alpha,p}~.
\eeq
(We will consider states with components in the projection sector later.)
We can determine the prescription for counting the chirality of the
winding modes $\eptilde^{(3\ell+1),\lambda}$ from the requirement that there
be an equal number of states with $(-1)^{\epsilon}=+1$ eigenvalue
built on positive and negative chirality ($\gamma_{D+1}$ eigenvalue)
states, $\ket{\alpha,p}_\pm$, at each massive level of the Fock
space.  This is a necessary condition for the states at these
levels to provide massive representations of the Lorentz group.
For example, at the second mass level excluding the one
winding mode state, there are $2D^2+D$ states satisfying
$(-1)^{\epsilon} =+1$ built on the positive chirality ground
state (the states on the first line in eq.~(\ref{Rfock})), but
only $D^2+D$ states built on the negative chirality ground state
(the second line).  Thus we must assign negative chirality to the
winding mode operator $\eptilde^{(2)\mu}$ in the $D$ states on the
third line of (\ref{Rfock}) in order to have left-right pairing.
We can generalize this counting argument to arbitrary level
and winding mode as follows.

Recall that the ground state sector of the fermionic Fock
space is made up of a $D$-fold tensor product of the free
coordinate boson theory with partition function $\eta^{-1}$,
and the $[\phi^1_1]$ parafermion fields with character
${\cal Z}^1_1$ given in (\ref{pfchars}).  Thus the character of the
ground state sector is
\beq{Rchar}
 {\rm ch}_g(q)~=~2^{D/2-1}\left({{{\cal Z}^1_1}\over\eta}\right)^D
 ~=~2^{D/2-1}\left({q^{1/12}\over\eta^2}
 \sum_\ell q^{3\ell^2+\ell}\,\right)^D~.
\eeq
where the factor $2^{D/2-1}$ takes into account the dimension of
the ground state, as well as the GSO-like projection.  Note that
this corresponds to only the term $2^{D/2-1}(c^2_2)^{D-2}$
in the $A$ block partition function (\ref{ABxpr}) of the
fermionic sector, since we are not
considering the projected sector states yet.  The terms
in the expression (\ref{Rchar}) for ${\rm ch}_g$ could have
been read off directly from the basis of states (\ref{fullbas}).
The two factors of $\eta^{-1}$ per dimension come from the
boson modes $\alpha^\nu_{-n}$ and the parafermion energy
operator modes $\eptilde^\mu_{-m}$.  The summation is
the contribution of the winding modes $\eptilde^{(3\ell+1),\lambda}$.

Define now the chiral index
\beqq
 I~\equiv~2^{1-D/2}~{\rm Tr}~{\gamma}_{D+1}~q^{L_0}~,
\eeq
which gives at each level the difference in the number of even and
odd chirality states (with the ground state multiplicity normalized
to one). If this difference is zero for massive levels (as it must be
in order that the states furnish good representations of the Lorentz
group at all levels), then the massless ground state will be the only
state to contribute to the chiral index.  Thus we must choose the
chiralities of the winding modes so that the chiral index $I=1$.

The ground state sector character ${\rm ch}_g$ can be rewritten as
\beqq
{\rm ch}_g(q)=2^{D/2-1}
\prod_{n=1}^{\infty}(1-q^n)^{-D}~\prod_{m=1}^{\infty}
(1-q^m)^{-D}\left(\sum_\ell q^{3\ell^2+\ell}\right)^D~,
\eeq
where we have substituted the definition of the Dedekind $\eta$-function
$\eta(q)=q^{1/24}\prod_{m=1}^{\infty}(1-q^m)$.  Since all the states
contributing to this character (by hypothesis) satisfy the GSO-like
condition $(-1)^{\epsilon}=+1$, the chirality of a given state will
follow from the product of the chiralities of the parafermion fields.
In particular, since each energy operator mode $\eptilde^\mu$ was found
earlier to have chirality $-1$, we should flip the sign of the $q^m$
terms in the $\eta^{-1}(q)$ factor corresponding to the $\eptilde$ modes.
This reflects the fact that in the $(-1)^{\epsilon}=+1$ sector, states
built on the ground state $\ket{\alpha,p}$ with an even number of
$\eptilde$ excitations have even chirality, and states with an odd number
of $\eptilde$ excitations have odd chirality.  Thus, the chiral index $I$
is given by
\beqa{chiralind}
 I&=&\prod_{n=1}^{\infty}(1-q^n)^{-D}~
 \prod_{m=1}^{\infty}(1+q^m)^{-D}
 \left(\sum_{\ell}f(\ell)q^{3\ell^2+\ell}\right)^D~,
 \nonumber\\
 &=&\prod_{m=1}^{\infty}(1-q^{2m})^{-D}
 \left(\sum_{\ell}f(\ell)q^{3\ell^2+\ell}\right)^D~,
\eeqa
where $f(\ell)$ is the (unknown) chirality of the
$\eptilde^{(3\ell+1)}$ winding modes.

Recalling the Euler pentagonal number theorem \cite{Gins}
\beq{Euler}
 \prod_{n=1}^{\infty}(1-q^n)~=~
 \sum_{\ell}(-1)^\ell q^{(3\ell^2+\ell)/2}~,
\eeq
we see that if we make the choice
\beqq
 f(\ell)~=~(-1)^\ell~,
\eeq
then the chiral index is
\beqq
 I~=~\prod_{m=1}^{\infty}(1-q^{2m})^{-D}~
 \prod_{n=1}^{\infty}(1-q^{2n})^D~=~1~.
\eeq
Thus the correct $(-1)^{\epsilon}$ operator for the ground state
fermionic sector is given by
\beq{GSOlike}
(-1)^{\epsilon}~=~\gamma_{N+1}(-1)^{N({\epsilon})}
(-1)^{N(\ell)}~,
\eeq
where we formally define the winding mode number operator by
\beq{Ngrnd}
 N(\ell)~\ket{\{\ell_\lambda\},\{m,\mu\},\{n,\nu\}}_g~\equiv~
 \left(\sum_{\lambda=0}^{D-1}\ell_\lambda\right)~
 \ket{\{\ell_\lambda\},\{m,\mu\},\{n,\nu\}}_g
\eeq
on the basis of states of the ground state sector given in (\ref{fullbas}).

We now turn to the states in the projection sector.  As noted
earlier, all these states are massive.  Therefore, by the
argument outlined above, the chiral index must be zero
in the projection sector to be consistent with Lorentz
invariance.  In other words, at each mass level there should
be an equal number of left- and right-chirality states.
Recall that the full projection sector
is the tensor product of $D-3$ one-component
ground state sectors and
$3$ one-component projection sectors.  Using the one-component
bases of states, (\ref{Fbasis}) and (\ref{Fbasis3}), we can easily
write down a basis of states for the full projection sector:
\beqa{projbas}
 &\ket{\{\ell,\lambda\},\{m,\mu\},\{n,\nu\}}_p~=&\\
 &\displaystyle{ \left(\prod_{i=1}^{D-3}
 \eptilde^{(3\ell_i+1),\lambda_i}_{-\ell_i(3\ell_i+1)}\right)
 \left(\prod_{i=D-2}^D
 \eptilde^{(3\ell_i+1),\lambda_i}_{-3\ell_i(\ell_i+1)-2/3}\right)
 \eptilde^{\mu_1}_{-m_1}\ldots\eptilde^{\mu_j}_{-m_j}
 \alpha^{\nu_1}_{-n_1}\ldots\alpha^{\nu_k}_{-n_k}
 \ket{\alpha,p}~,}&\nonumber
\eeqa
where where together the $D$ $\lambda_i$s span the set
$\{0,1,\ldots,D-1\}$ of space-time indices.  The partition function
for the projection sector is given by a similar argument as above:
\beqqa
 {\rm ch}_p(q)&=&2^{D/2-1}\left(
 {{{\cal Z}^1_1}\over\eta}\right)^{D-3}
 \left({{{\cal Z}^0_1+{\cal Z}^0_{-1}}\over\eta}\right)^3\\
 &=&2^{D/2-1}\left({q^{1/12}\over\eta^2}
 \sum_\ell q^{3\ell^2+\ell}\,\right)^{D-3}\left({2q^{3/4}\over\eta^2}
 \sum_p q^{3p(p+1)}\,\right)^3~.\nonumber
\eeqa
This corresponds to the term $2^{D/2-1}(c^2_2)^{D-5}(2c^4_2)^3$
in the $A$ block partition function (\ref{ABxpr}).
The first summation comes from the winding modes in the
$D-3$ one-component ground state sectors, while the second
summation comes from the winding modes in the $3$
one-component projection sectors.  Computing the chiral
index using the definition of $(-1)^{\epsilon}$ derived above,
we find
\beq{chproj}
 I~=~8q^2\prod_{m=1}^\infty(1-q^{2m})^{-3}\,
 \left(\sum_{\ell=-\infty}^{+\infty}f(\ell)q^{3\ell(\ell+1)}\right)^3~,
\eeq
where we have used the fact, derived above, that $I=1$ for each
of the $D-3$ ground state sector components separately.  Thus the
summation in (\ref{chproj}) is over the winding numbers of the
operators $\eptilde^{(3\ell+1)}$ contributing to the three projection
sector components.  If we take the same prescription for the
chirality of these winding modes as we took in the ground
state sector, namely $f(\ell)=(-1)^\ell$,
it is easy to see that the chiral index (\ref{chproj}) vanishes
due to the symmetry $\ell\rightarrow-\ell-1$ of the sum, which
is the desired result.

To summarize, the correct form of the
$(-1)^{\epsilon}$ projection operator consistent with Lorentz invariance
is given by eq.~(\ref{GSOlike}) for the fermionic sector of the
FSS, where the winding mode number operator $N(\ell)$ is defined
by (\ref{Ngrnd}) on ground state sector states, and by
\beqq
 N(\ell)~\ket{\{\ell,\lambda\},\{m,\mu\},\{n,\nu\}}_p~\equiv~
 \left(\sum_{i=1}^D\ell_i\right)~
 \ket{\{\ell,\lambda\},\{m,\mu\},\{n,\nu\}}_p
\eeq
on projection sector states.  We emphasize that the existence of
this GSO-like projection compatible with Lorentz invariance was by
no means guaranteed.  Indeed, we have seen that it works only through
the use of the non-trivial identity (\ref{Euler}).

\setcounter{footnote}{0}
\section{\label{ssix}The bosonic sector}

In this section we construct the full space of states in the
bosonic sector of a $D$-dimensional open $K=4$ FSS.  We
then solve for the subset of states at low-lying mass
levels that satisfy the physical state conditions.
The ground state is found to be tachyonic, while the
first excited state is a massless vector particle.  The
partition function for the closed FSS derived in \ref{sthree}
implies that the tachyonic state is removed by an analog of
the GSO projection.

\subsection{Bosonic Fock space}

We build the bosonic sector of the FSS Fock space
by the action of the coordinate boson modes $\alpha^\mu_n$
and the parafermion energy operator modes $\epsilon^{(\pm1)}_r$
on the ground state
\beqq
 \ket{p}~\equiv~:\!{\rm e}^{ip\cdot X(0)}\!:\ket{0}~.
\eeq
In Sect.~\ref{stwo} we showed that in the single space-time
component theory, the action of the energy operator modes on
the identity operator (vacuum state) $\ket{0}$ correspond to the parafermion
sectors $[\phi^0_0]$, $[\phi^1_0]$ and $[\phi^2_0]$.  Recall
that $[\phi^0_0]$ and $[\phi^2_0]$ consist
of the set of fields $\epsilon^{(3n)}$ and $\partial\rho$
with integer conformal dimensions and $\bZ_3$ charge zero,
whereas $[\phi^1_0]$ consists of the $\epsilon^{(3n\pm1)}$
fields of dimension $\bZ+1/3$ and $\bZ_3$ charge $\pm1$.
The allowed modings and actions of the
integral winding-number fields $\epsilon^{(a)}$, $a\in\bZ$, on
different $\bZ_3$ sector fields in the bosonic sector
is also summarized by Fig.~1.

In eq.~(\ref{epfock}) we wrote down a basis of states in the parafermion
theory for the first four levels.  Adding in the coordinate boson
field, and tensoring $D$ copies together, we find the complete list
of states in the bosonic sector for these levels:
\beqa{Blevel}
 \ket{p}&&{\rm level}~0~,\nonumber\\
 \epsilon^{(\pm1)\mu}_{-1/3}\ket{p}&&~~1/3~,\nonumber\\
 \epsilon^{\mu_1}_{-1/3}\epsilon^{\mu_2}_{-1/3}\ket{p}&&~~2/3~,\\
 \alpha^\mu_{-1}\ket{p}
  +\epsilon^{(\pm1)\mu}_{-2/3}\epsilon^{(\mp1)\mu}_{-1/3}\ket{p}
  +\epsilon^{\mu_1}_{-1/3}\epsilon^{\mu_2}_{-1/3}
   \epsilon^{\mu_3}_{-1/3}\ket{p} &&~~~1~,\nonumber\\
 \epsilon^{(\pm1)\mu}_{-1/3}\alpha^\nu_{-1}\ket{p}
  +\epsilon^{(\pm1)\mu}_{-1}\epsilon^{(\pm1)\mu}_{-1/3}\ket{p}
  +\epsilon^{(\pm1)\mu}_{-4/3}\ket{p}
  +\epsilon^{\mu_1}_{-1/3}\epsilon^{\mu_2}_{-1/3}
   \epsilon^{\mu_3}_{-1/3}\epsilon^{\mu_4}_{-1/3}\ket{p}
  &&~~4/3~.\nonumber\\
 \ldots\qquad\qquad&&~\ldots\nonumber
\eeqa
The mode operators in the $\epsilon^{\mu_1}_{-1/3}\cdots
\epsilon^{\mu_n}_{-1/3}\ket{p}$ states in (\ref{Blevel}) are to be
understood to be either $\epsilon^{(+1)\mu_i}$ or $\epsilon^{(-1)\mu_i}$
modes, independently for each factor, and that in these terms
$\mu_i\neq\mu_j$.  Thus, for example,  $\epsilon^{\mu_1}_{-1/3}
\epsilon^{\mu_2}_{-1/3}\epsilon^{\mu_3}_{-1/3}\ket{p}$
represents ${8}{D\choose3}$ independent states for fixed $p^\mu$.
The states at level $\ell$ have conformal dimension
${p^2\over2}+\ell$.  Note that we have used the identities
(\ref{2.41})--(\ref{2.43}) to write these states
in terms of $\epsilon^{(\pm1)}$ modes only.  Using the machinery
developed in Sect.~\ref{stwo}, this can be done for the whole
bosonic Fock space.

We can now make a correspondence between these states and the light-cone
spectrum of the $D=6$ FSS found in \ref{sthree}.  Recall that the ground
state, corresponding to a term $(c^0_0)^{D-2}$ in the partition function
does not appear in the $A$ block of the FSS partition function (\ref{ABxpr}).
This is a reflection of a GSO-like projection in the FSS spectrum.  In
general, only the level $\bZ+{1\over3}$ states survive this projection.
Thus, the $2D$ states at level $1/3$ in (\ref{Blevel}) contribute to the
$(D-2)(c^0_0)^{D-3}(c^2_0)$ term in the partition function.  We will see
below how the physical state conditions reduce the multiplicity of this
state from $2D$ to $D-2$.  At level $4/3$, the first three states in
(\ref{Blevel}) contribute to the same term in the partition function
as the level $1/3$ state; the last state, however, contributes to
the $(c^2_0)^4$ term in (\ref{ABxpr}).  This is one of the ``internal
projection'' states in the bosonic sector.

\subsection{Physical state conditions and the massless vector particle}

Now we wish to impose the physical state conditions
\beqa{pscep}
 (L_0-v)\ket{\phi}&=&0~,\nonumber\\
 (G_0-\beta)\ket\phi&=&0~,\\
 L_n\ket\phi~=~G_{n/3}\ket\phi&=&0~,\qquad n>0~.\nonumber
\eeqa
We will determine the intercepts $v$ and $\beta$ by demanding
that extra sets of null states appear at their critical values.
Note that, unlike in the usual superstring case, both integer
and fractional moding of the currents are allowed in the
bosonic sector.

Let us start with the general level zero state
\beqq
 \ket{\phi_0}~=~\zeta(p)\ket{p}~,
\eeq
where $\zeta(p)$ is an arbitrary (scalar) wave function.
Clearly only the $L_0$ and $G_0$ conditions are potentially
non-vanishing on $\ket{\phi_0}$.  From the moding diagram,
Fig.~1, it is also easy to see that $G_0\ket{\phi_0}=0$.
Finally, since $L_0\ket{\phi_0}={p^2\over2}\ket{\phi_0}$,
the physical state conditions on $\ket{\phi_0}$ are
$v=p^2/2$ and $\beta=0$.
We will show below that $\beta\neq0$ at its critical value, and
thus that the tachyonic state must decouple.  Note that
this is different from what happens in the usual superstring,
where the physical state conditions by themselves do not
remove the tachyonic states. (The GSO projection is only
required at the string loop level.)  This new situation
is clearly related to the fact that in the FSS the bosonic sector
admits integral moding of the fractional supercurrent.

We now consider the general level $1/3$ state:
\beqq
 \ket{\phi_{1/3}}~=~\left(\zeta_\nu\epsilon^\nu_{-1/3}+
 \zeta^\dagger_\nu\epsdag^\nu_{-1/3}\right)\ket{p}~,
\eeq
where $\zeta_\mu$ and $\zeta^\dagger_\mu$ are polarization
vectors, and we have introduced the notation
\beqq
 \epsilon~=~\epsilon^{(+1)}~,\quad{\rm and}\quad\epsdag~=~\epsilon^{(-1)}~.
\eeq
Acting on this state, the non-trivial physical state conditions
are $L_0$, $G_0$ and $G_{1/3}$.  The $L_0$ condition is easily seen
to give
\beq{5.7}
 (L_0-v)\ket{\phi_{1/3}}~=~0\quad
 \Longrightarrow\quad v~=~{p^2\over2}+{1\over3}~.
\eeq

Using the mode expansion of the $G$ current (\ref{Gcurmode})
we can write the action of the $G_{1/3}$ mode on $\ket{\phi_{1/3}}$
in terms of component fields as
\beqq
 G_{1/3}\ket{\phi_{1/3}}~=~
 {1\over\sqrt2}\sum_n\alpha_{n,\mu}
 \left[\epsilon^\mu_{1/3-n}+\epsdag^\mu_{1/3-n}\right]
 \left(\zeta_\nu\epsilon^\nu_{-1/3}+
 \zeta^\dagger_\nu\epsdag^\nu_{-1/3}\right)\ket{p}~,
\eeq
since on this state there is no contribution from the
$\epsilon^{(\pm2)}$ terms in the current.  Now, the expression
\beq{epmuepnu}
\epsdag^\mu_{1/3}~\epsilon^{\nu}_{-1/3} \ket{p}
\eeq
is only well-defined, with respect to the moding rules given in
Fig.~1, for $\mu=\nu$.  For $\mu \neq \nu$, the moding
is inappropriate, since $\epsdag^\mu_{1/3}$ is acting on
the vacuum state with $\bZ_3$-charge $0$ for the $\mu$-th
component, and thus has the wrong moding.  However, recall
that in Sect.~\ref{sfour} we learned how to deal with this problem---the
lesson being that ``wrong'' moding operators should be formally
set to zero.  With this understanding, we can give a covariant
meaning to the expression (\ref{epmuepnu}):
\beqq
\epsdag^\mu_{1/3}~\epsilon^{\nu}_{-1/3} \ket{p}~=~g^{\mu\nu}\ket{p}~,
\eeq
where $g^{\mu\nu}$ is the Minkowski metric.  Here we
have used the one-component commutation relations for the
$\mu=\nu$ case.  Thus
\beqqa
 G_{1/3}\ket{\phi_{1/3}}&=&{1\over\sqrt2}
 \alpha_0\cdot(\epsilon_{1/3}+\epsdag_{1/3})
 (\zeta\cdot\epsilon_{-1/3}+\zeta^\dagger\cdot\epsdag_{-1/3})
 \ket{p}\nonumber\\
&=&{1\over\sqrt2}p\cdot(\zeta+\zeta^\dagger)\ket{p}~,
\eeqa
where all other terms vanish by either using the
single-component GCRs, eq.~(\ref{epgcr1})
for the cases where $\mu=\nu$, or the annihilation property
of ``wrong-moding'' operators for $\mu\neq\nu$.
The $G_{1/3}$ physical state condition then gives
\beq{5.12}
 G_{1/3}\ket{\phi_{1/3}}~=~0\quad\Longrightarrow\quad
 p\cdot(\zeta+\zeta^\dagger)~=~0~.
\eeq

Finally, let us compute the $G_0$ condition on $\ket{\phi_{1/3}}$.
{}From (\ref{Gcurmode}) we have
\beqq
 G_0\ket{\phi_{1/3}}=\sum_\mu\left\{
 {1\over\sqrt2}\sum_n \alpha_{n,\mu}
 \left(\epsilon^\mu_n+\epsdag^\mu_n\right)+
 {1\over2}\left(\epsilon^{(+2)\mu}_0+\epsilon^{(-2)\mu}_0
 \right)\right\}\ket{\phi_{1/3}}~.
\eeq
The first term vanishes since $\epsilon^\mu_n$ for $n\in\bZ$
has the ``wrong'' moding for all the components of $\ket{\phi_{1/3}}$
(see Fig.~1).  However, the second term, involving the
$\epsilon^{(\pm2)\mu}_0$ modes, does not vanish.  Recall that we
can write the $\epsilon^{(\pm2)}$ modes in terms of the $\epsilon$
and $\epsdag$ modes (\ref{4.19}), so that
\beq{twotoone}
 \sum_\mu\epsilon^{(\pm2)\mu}_0=g_{\mu\nu}
 \sum_{\ell=0}^\infty c_\ell^{(-5/3)}\left[
 \epsilon^{(\pm1)\mu}_{-{1\over3}-\ell}
 \epsilon^{(\pm1)\nu}_{{1\over3}+\ell}+
 \epsilon^{(\pm1)\mu}_{-{4\over3}-\ell}
 \epsilon^{(\pm1)\nu}_{{4\over3}+\ell}\right]~,
\eeq
when acting on a state with $\bZ_3$ charge $q=\pm1$.  (By Fig.~1,
$\epsilon^{(\pm2)}_0$ does not act on states with $q=0$.)  Now,
by simple dimensional considerations, all $\epsilon^{(a)}_n$ modes
annihilate $\ket{\phi_{1/3}}$ if $n>{1/3}$.  This drastically
reduces the sum in (\ref{twotoone}), so that we have
\beqqa
 G_0\ket{\phi_{1/3}}&=&{1\over2}\left(
 \epsilon_{-1/3}\cdot\epsilon_{1/3}+
 \epsdag_{-1/3}\cdot\epsdag_{1/3}\right)\left(
 \zeta\cdot\epsilon_{-1/3}+
 \zeta^\dagger\cdot\epsdag_{-1/3}\right)\ket{p}\nonumber\\
 &=&{1\over2}\left(\zeta^\dagger\cdot\epsilon_{-1/3}+
 \zeta\cdot\epsdag_{-1/3}\right)\ket{p}~,
\eeqa
where we have used the $\bZ_4$ parafermion identities
\beqq
 \epsilon^{(\pm1)}_{1/3}\epsilon^{(\mp1)}_{-1/3}\ket{0}~=~\ket{0}~,
\eeq
which follow from (\ref{epket2}).  Therefore the
$G_0$ physical state condition implies
\beq{5.17}
 (G_0-\beta)\ket{\phi_{1/3}}~=~0\quad
 \Longrightarrow\quad\zeta_\mu=\zeta^\dagger_\mu~~
 {\rm and}~~\beta={1\over2}~.
\eeq

Thus the physical state conditions (\ref{5.7}), (\ref{5.12}) and
(\ref{5.17}) reduce the original $2D$ components of the
$\zeta_\mu$ and $\zeta^\dagger_\mu$ polarization vectors to
only $D-1$ independent components.  In addition, there is a
critical value of the intercept $v$ for which an additional
degree of freedom is removed.  In particular, when
\beqq v={1\over3}~, \eeq
by the $L_0$ condition (\ref{5.7}), we find that $p^2=0$,
so that $\ket{\phi_{1/3}}$ describes a massless vector
particle.  Thus, the state with polarization $\zeta_\mu\propto p_\mu$
has zero norm, and there are only $D-2$ physical polarizations.

\setcounter{footnote}{0}
\section{Higher mass levels}

So far in this paper we have constructed the Fock space of states
for the (open) $K=4$ fractional superstring, and have shown that the
simplest guess for the physical state conditions provides the correct
equations of motion for the lowest-lying states in the fractional
superstring spectrum.  Also, demanding the presence of extra null
states (or, equivalently, demanding massless vector and spinor
particles with the same number of propagating degrees of freedom as
in the partition function), fixed the intercepts in the space-time
bosonic and fermionic sectors.  This is all in agreement with the
$K=4$ FSS partition function, whose derivation is reviewed in Appendix D.

The next step is, clearly, to examine higher mass levels in the FSS
Fock space.  However, at these levels we run into two separate problems,
which may or may not be related:  Lorentz non-invariance and the failure
of the physical state conditions to implement the internal projection.
We will describe below in some detail how these problems arise, and then
we will briefly outline a few possible ways in which they may be resolved.
We would like to point out at the outset that we do not have a clear
resolution of these problems.

\subsection{Lorentz non-invariance}

We constructed the FSS Fock space by tensoring together $D$ copies
of the $\bZ_4$ parafermion conformal field theory, with the aim
of describing the FSS spectrum in $D$ flat spacetime dimensions.
However, {\it a priori}, taking the tensor product only ensures a
permutation symmetry among the dimensions, and not the full rotational
Lorentz symmetry.

A concrete realization of this point is encountered at the first massive
level in the fermionic sector.  In particular, we will impose the physical
state conditions (\ref{Vpsc}) and (\ref{Gpsc}) on the
level $1$ states, and show that the
resulting equations of motion are not Lorentz covariant.

Consider the general state at level $1$:
\beqq
 \ket{\psi_1}~=~\alpha^\mu_{-1}\ket{\alpha,p} u^\alpha_\mu(p)+
 {\tilde\epsilon}^\mu_{-1} \ket{\alpha,p} w^\alpha_\mu(p)~,
\eeq
where $u^\alpha_\mu$ and $w^\alpha_\mu$ are spinor wave functions.
The $L_n$ and $G_r$ modes for $n,r > 1$ are easily seen to
identically annihilate $\vert \psi_1 \rangle$.  In addition, by
eq.~(\ref{LGalg}),
the $G_1$ condition is not independent of the $L_1$
and $G_0$ conditions, and the $G_{2/3}$ condition identically
annihilates $\vert \psi_1 \rangle$.  Thus, the only physical
state conditions that need to be checked are those corresponding
to $L_0$, $L_1$, $G_0$ and $G_{1/3}$.  The $L_0$ and $L_1$ conditions
are Lorentz covariant.  In particular, since $\vert \psi_1 \rangle$
has conformal dimension ${{p^2} / 2} + {D / 12}+1$ , the $L_0$
condition, with the value of the intercept derived in Sect.~4, gives
\beqq
 (L_0-{D/12})~\ket{\psi_1}=0~~\Longrightarrow~~p^2=-2~.
\eeq
Thus the level $1$ physical state corresponds to particles of mass
$m=\sqrt 2$.  Using the commutators (\ref{2.57}), one also finds that
\beqq
 L_1\ket{\psi}=0~~\Longrightarrow~~p\cdot u(p)+{1\over3}\wsl(p)=0~,
\eeq
where we have replaced ${\tilde\epsilon}^\mu_0$ with $\gamma^\mu$
when acting on the ground state $\ket{\alpha,p}$.

To evaluate the other physical state conditions, we must use the
generalized commutation relations (GCRs) of the ${\tilde\epsilon}^\mu$
fields.  The GCR of ${\tilde\epsilon}^{\mu}$ and ${\tilde\epsilon}^\nu$
is given by the single-component relation (\ref{epgcr1})
when $\mu=\nu$.  When $\mu \neq \nu$, however, all modes simply
anticommute.  In particular, when acting on the ground state
${\tilde\epsilon}^{\mu}_0$ and ${\tilde\epsilon}^{\nu}_{-1}$ satisfy
\beqa{noncovgcr}
 \left({\tilde\epsilon}^{\mu}_0 {\tilde\epsilon}^{\nu}_{-1}
 +{\tilde\epsilon}_{-1}^{\nu}{\tilde\epsilon}_{0}^{\mu}
 \right)\ket{\alpha,p}&=&0~,~~~~\mu \neq \nu~,\nonumber\\
 \left({\tilde\epsilon}^\mu_0{\tilde\epsilon}^\nu_{-1}
 +{1\over3} {\tilde\epsilon}_{-1}^\nu{\tilde\epsilon}_0^\mu
 \right)\ket{\alpha,p}&=&0~,~~~~\mu = \nu~.
\eeqa

The $G_0$ physical state condition can now be evaluated:
\beqqa
 G_0 \ket{\psi_1}&=&{1\over\sqrt2}
 [ \alpha_1 \cdot {\tilde\epsilon}_1+\alpha_0 \cdot {\tilde\epsilon}_0
 +{\tilde\epsilon}_{-1} \cdot \alpha_1 ] \ket{\psi_1} \nonumber\\
 &=&{1\over\sqrt2} \Bigl\lbrack \alpha_{-1}\cdot({\slp}u+{2\over3}w)
 +{\tilde\epsilon}_{-1} \cdot (u-{\slp}w)\\
 &~&~~~~+{2\over3}\sum_\mu {\tilde\epsilon}^\mu_{-1} p^\mu \gamma^\mu
 w^\mu \Bigr\rbrack \ket{\alpha,p}~,
\eeqa
or, $G_0 \vert \psi_1 \rangle=0$ implies that
\beqqa
 {\slp}u_{\mu}+{2 \over 3} w_{\mu}&=&0~,\nonumber\\
 u_{\mu}-{1 \over 3}{\slp}w_{\mu}-{2 \over 3}
 \left(\sum_{\nu \neq \mu} p^{\nu}\gamma^{\nu}\right)w_{\mu}&=&0~.
\eeqa
Compatibility of these equations with the $L_0$ condition implies the
Lorentz non-covariant constraint
\beq{noncov}
 \left(\sum_{\nu\neq\mu} p^\nu \gamma^\nu \right)w_\mu=0~.
\eeq
Note that even though the $G_1$ physical state condition is not
independent of the $G_0$ condition, since $[L_1,G_0]={1 \over 3}G_1$,
it does, however, give rise to a Lorentz covariant constraint:
\beqq
 G_1\ket{\psi_1}=0~~\Longrightarrow~~\usl+{2\over3}p\cdot w=0~.
\eeq

Similarly, the $G_{1/3}$ condition can be evaluated using the fact
that, on the ground state,
\beqqa
 {\tilde\epsilon}_{1/3}^{\mu}{\tilde\epsilon}_{-1}^{\nu}
 \ket{\alpha,p}&=&~0,~~~~\mu \neq \nu~,\nonumber\\
 \left({\tilde\epsilon}_{1/3}^{\mu}{\tilde\epsilon}_{-1}^{\nu}+
 {2 \over 3} {\tilde\epsilon}^{\nu}_{-2/3}{\tilde\epsilon}_0^{\mu}
 \right)\ket{\alpha,p}&=&~0,~~~~\mu = \nu~.
\eeqa
Thus
\beqa{onethirdcond}
 G_{1/3}\ket{\psi_1}&=&
 \left\{ {1\over\sqrt2}
  [{\tilde\epsilon}_{-2/3} \cdot \alpha_1 +
 \alpha_0 \cdot {\tilde\epsilon}_{1/3}] + {\tilde\epsilon}_{-2/3}
 \cdot {\tilde\epsilon}_1 \right\} \ket{\psi_1} \nonumber\\
 &=&{1\over\sqrt2}{\tilde\epsilon}^\mu_{-2/3}
 \left[u_\mu+\sqrt2 w_\mu-{2\over3}p^\mu \gamma^\mu w^\mu\right]
 \ket{\alpha,p}~,
 \eeqa
where we have used (\ref{4.19}) to rewrite the ${\tilde\epsilon}^{(2)}
_{1/3}$ term as an expression quadratic in ${\tilde\epsilon}^{(1)}$ modes.
It is clear from (\ref{onethirdcond}) that the $G_{1/3}$ condition
gives rise to the same Lorentz non-invariant piece found in the
$G_0$ condition.  Using (\ref{noncov}), the $G_{1/3}$ physical
state condition becomes
\beqq
G_{1/3}\vert \psi_1 \rangle=0~~ \Longrightarrow~~
u_{\mu}+ {\sqrt 2}w_{\mu}-{2 \over 3} {\slp}w_{\mu}=0~.
\eeq
The only solution to the physical state conditions
is $w_{\mu}=u_{\mu}=0$. So,
not only are the $L_n$ and $G_r$ physical state conditions Lorentz
non-covariant at the first massive level, but they are also
too strong; they allow no propagating states, even though the
partition function predicts $32$ such states at this level
(see Appendix D).

\subsection{Internal projection}

A second problem that arises at the higher mass levels concerns the
presence of extra cancellations between states of the FSS Fock space,
which have no analog in usual superstring.

Recall from the discussion in Appendix D that we can identify the form of
the vertex operators for space-time bosons (\ref{pfbos}) or fermions
(\ref{pfferm}) in the $A$ block of the partition function on the basis
of a statistics selection rule.  By matching $\bZ_4$ parafermion quantum
numbers, we can easily identify the bosonic and fermionic pieces of the $A$
block partition function, which we write separately as
\beqa{intproj}
A_b&=&4(c^0_0+c^4_0)^3(c_0^2)-4(c^2_0)^4~,\nonumber\\
A_f&=&4(c_2^2)^4-32(c^2_2)(c^4_2)^3~,
\eeqa
so that $A=A_b-A_f$.  The puzzling feature of these identifications is that
not all the terms contributing to space-time bosons have positive
coefficients, and likewise not all space-time fermions have negative
coefficients.  By the supersymmetric vanishing of $A$ (\ref{pfSUSY}),
as functions of the modular parameter $q$ we have $A_b=A_f$. It turns out
\cite{ADT} that the fermionic (or bosonic) piece satisfies the identity
\beq{6.14}
A_f=4\biggl(\prod_{n=1}^{\infty}{{1+q^n} \over {1-q^n}}\biggr)^4~,
\eeq
which gives the same counting of physical degrees of freedom as the Ramond
sector in $6$ space-time dimensions.
Since the coefficients in the $q$-expansion of the right-hand side are all
positive, we are led to view the minus signs in $A_b$ and $A_f$ as
``internal projections" (or cancellations) of degrees of freedom in the
fractional superstring.

The problem with implementing the internal projections in the framework
of this paper can be described as follows.  Note that the internal projection
occurs only at or above the first massive level in the bosonic sector and
the second massive level in the fermionic sector.  For example, consider the
Fock space description of the states at the first massive level in the bosonic
sector:
\beqa{masbos}
\vert \phi_1 \rangle = \bigl\lbrace A_{\mu}{\tilde\epsilon}^{\mu}_{-4/3}
 &+&B_{\mu\nu}\alpha^{\mu}_{-1}{\tilde\epsilon}^{\nu}_{-1/3}
 ~+~C_{\mu\nu}s^{\mu}_{-1}{\tilde\epsilon}^{\nu}_{-1/3}\\
 &+&D_{\mu\nu\rho\sigma}{\tilde\epsilon}^{\mu}_{-1/3}
 {\tilde\epsilon}^{\nu}_{-1/3} {\tilde\epsilon}^{\rho}_{-1/3}
 {\tilde\epsilon}^{\sigma}_{-1/3}\bigr\rbrace~,\nonumber
\eeqa
where ${\tilde\epsilon}$ can stand for either $\epsilon^{(+1)}$ or
$\epsilon^{(-1)}$, $\mu\neq\nu\neq\rho\neq\sigma$ in the last term,
and $A,B,C,D$ are polarization tensors.  The first three terms in
(\ref{masbos}) all correspond to contributions from terms with positive
coefficients in $A_b$ (\ref{intproj}), while the last term, since it involves
four ${\tilde\epsilon}$'s, corresponds to the projection term $-4(c^2_0)^4$
in $A_b$.  Thus, even though the last term adds more states to the FSS Fock
space, once the physical state conditions are implemented they must actually
subtract states.  This implies that the physical state conditions must, at
least, mix the states in the Fock space corresponding to the ``ground state"
and ``projection" sectors.

However, the general form of the physical state conditions, assumed in
Sect.~3, is, schematically, in terms of the $\alpha$ and $\epsilon$
modes,
\beqqa
L_n &\sim& \sum_m(\alpha_m \cdot \alpha_{n-m}+\epsilon_m \cdot
\epsilon_{n-m})~,\nonumber\\
G_r &\sim& \sum_m(\alpha_m \cdot \epsilon_{r-m}+\epsilon_m \cdot
\epsilon_{r-m})~.
\eeqa
It is easy to see that operators of this form can never mix the first three
terms with the last one in $\vert \phi_1 \rangle$.  This statement holds
generally for all the higher mass levels in the FSS Fock space as well.

Note that the counting of degrees of freedom $\it after$ the internal
projection implied by (\ref{6.14}) is precisely that of $D-2$ pairs of
world-sheet bosons and fermions, similar to that of the usual superstring.
Thus the asymptotic degeneracy of states in $A_b$ or $A_f$ corresponds to
the effective central charge
\beq{6.18} c_{\rm eff}~=~(D-2)\cdot{3\over2}~, \eeq
or $c_{\rm eff}=6$ for the $K=4$ FSS where $D=6$.  Since each dimension in
the FSS Fock space corresponds to a CFT with central charge $c_0=2$,
we see that, with the internal projection, the physical state conditions
must remove the equivalent of $\it three$ dimensions' worth of states, and
not just the timelike and longitudinal ones expected from a large critical
string gauge invariance.

\subsection{Possible resolutions}

We will now describe possible modifications of the framework presented
in this paper, which may solve the above-mentioned problems.
First, we will discuss some of the assumptions which underlie the discussions
of the last two subsections.  The most important is Lorentz invariance.
Clearly, if one were willing to give up Lorentz invariance, the occurence
of non-invariant physical state conditions would not be a problem.  Similarly,
the existence of the internal projection was based on an identification of
space-time bosons and fermions from their statistics selection rules.  Without
Lorentz invariance, this argument also has no force.  So, it is a logical
possibility that the above-mentioned problems are simply an indication that
we have to give up Lorentz invariance above the massless level of states.
(String interactions will presumably then give Lorentz non-invariant
contributions to the effective action for the massless states as well.)

However, the fact that there does exist a Lorentz-covariant description of
the massless state, that the A block of the partition function admits a
separation into pieces satisfying statistics selection rules, and that the
chiral counting argument described in Sect.~4.4 works, all hint that a
Lorentz-covariant interpretation of the $K=4$ FSS should exist.  Two possible
ways in which such an interpretation could be realized are to either change
the physical state conditions or to change the Fock space on which they act.
We will describe below how these proposals can be systematically explored.

The idea behind changing the physical state conditions is to add terms to the
$G_r$ that cancel the Lorentz non-covariant pieces in the equations of motion
they generate.  One could do this level by level in the FSS Fock space.
For example, we saw above that the $G_0$ condition gives rise to Lorentz
non-covariant terms in the equations of motion of the first massive fermion
states. An example of the kind of modification that could cancel those terms is
\beq{newG0}
G_0 ~ \longrightarrow {\tilde G}_0 = G_0 + \kappa G_{-1/3}G_{1/3}+...~,
\eeq
where $\kappa$ is a parameter to be fixed by the requirement of Lorentz
invariance.  Note that the new mode operator ${\tilde G}_0$ has the same
action on the massless states as did the old.
The ${\tilde G}_0$ proposed in (\ref{newG0}) is just meant to be
illustrative of this idea; in fact there are many more terms that contribute
at the first massive level and conceivably could contribute to (\ref{newG0}),
since there is no reason that ${\tilde G}_0$ must be manifestly covariant.
All we require is that the Lorentz non-covariance of the tensor-product
Fock space cancels against appropriate non-covariant physical state
conditions.

The modified ${\tilde G}_r$ must satisfy other conditions besides the
requirement that they yield covariant equations of motion.  In particular,
they must implement the internal projections described above, and must give
rise to extra towers of null states, indicative of the critical string gauge
invariance.  There is at least a hope of implementing the internal projection,
since the modified ${\tilde G}_r$'s can now include terms that mix the
projection and ground state sectors, and an extra condition (or perhaps
${\tilde G}_{1/3}$) could lead to the associated reduction in the effective
central charge.  The existence of towers of null states at the critical
dimension depends on the structure of the chiral algebra of the
${\tilde G}_r$'s.  The form of this
world-sheet symmetry algebra is, of course, one of the main mysteries of
fractional superstrings.  Some speculations on what this algebra may be will
be discussed in Sect.~7.1.

Note that the resolution described above gives up $\it manifest$ Lorentz
covariance, and only demands covariance in the final step---the equations
of motion.  It is possible that a manifestly covariant formulation of the
constraint algebra and states requres the introduction of the BRST ghost
system, as in the fermionic sector of the usual superstring.

Another possible resolution aims at preserving manifest Lorentz covariance
at all stages.  Referring back to the first massive fermionic level
calculation, we see that the non-covariance first appeared in the mode
GCRs (\ref{noncovgcr}).  One could try to maintain manifest covariance
by modifying the
commutation relations for $\mu\neq\nu$.  This means that we are no longer
simply taking the direct tensor product of $\bZ_4$ parafermion theories.
Instead, we are ``deforming" this tensor product to obtain a new CFT.  Since
we are changing the GCRs among the modes, we are effectively changing the
Fock space.

This program can be made systematic in the following way.  Working level
by level in the Fock space, one covariantizes the GCRs of the modes,
which leads in general to
the introduction of many free parameters.  These parameters are fixed by
then demanding that the resulting GCRs are associative (a necessary condition
for them to describe a consistent two-dimensional field theory).  One then
has to check that this new Fock space provides a representation of the
Virasoro algebra, as well as construct the new fractional
supercurrent $\tilde G$.

The deformed tensor product approach is in general much harder to implement
than the previous idea of changing the physical state conditions.  There
are many reasons for this.  One reason is that in the former approach one
has to solve the associativity conditions at each mass level.  Another is
that since the Fock space has been changed, a comparison to the partition
function can only be made after solving the physical state conditions and
removing the null states at each mass level.  Note, however, that the two
approaches outlined above may in principle be equivalent:  the extra
``physical state condition" which implements the internal projection on
the original Fock space could be a weak operator realization of the tensor
product deformation which reduces the Fock space to a manifestly covariant
one.  Indeed, one expects such a picture to hold if the deformed tensor
product approach is to reproduce the partition function derived in
Appendix D.

It is entirely possible that the use of the tensor-product Fock space
as a starting point for the understanding of the $K=4$ FSS is incorrect.
There are other possibilities, to which we will now turn.

\setcounter{footnote}{0}
\section{Remarks}

In this section we describe two features of the $K=4$ FSS which, though
not directly related to the open string physical state condition calculations
described above, may nevertheless play a role in the ultimate solution to the
problems raised in the previous section.

\subsection{Connection to the spin-$4/3$ string}

A crucial issue in the understanding of the FSS is the determination
of its critical space-time dimension $D$.  We described in \ref{sthree} an
argument that determines the $K=4$ FSS critical dimension to be
$D=6$.  However, this argument is indirect, in that it does not
determine $D$ from the consistency condition (anomaly cancellation)
for the world-sheet gauge invariance.  By solving the physical
state conditions we can in principle check this determination
of $D$---it should be the largest dimension in which a no-ghost
theorem holds.  This critical dimension is signalled by the presence
of extra towers of null states in the spectrum.  Because the generalized
commutator algebra (\ref{Jalg}) for the components of the fractional
supercurrent $G$ on the tensor-product Fock space cannot be combined
into a single algebra for the modes of $G$ itself, we can not construct
these null state towers.

In light of the discussion of Sect.~6.3, however, we should consider
modifying the algebra of physical state conditions.  A particularly
simple algebra is the one-component spin-$4/3$ algebra (\ref{ftalg}),
but with arbitrary central charge \cite{FZft},
\beqa{ft76}
  G^{+}(z)G^{+}(w)&= & {\lambda(c)\over(z-w)^{4/3}}
   \left\{G^{-}(w)+{1\over2}(z-w)\partial G^{-}(w)\right\},\nonumber\\
  G^{-}(z)G^{-}(w)&= & {\lambda(c)\over(z-w)^{4/3}}
   \left\{G^{+}(w)+{1\over2}(z-w)\partial G^{+}(w)\right\},\nonumber\\
  G^{+}(z)G^{-}(w)&= & {(1/2)\over(z-w)^{8/3}}\left\{{3c\over4}+
   2(z-w)^2 T(w)\right\},\\
  T(z)G^\pm(w)&= & {(4/3)G^\pm(w)\over(z-w)^2}+
   {\partial G^\pm(w)\over(z-w)}~,\nonumber\\
  T(z)T(w)&=& {(c/2)\over(z-w)^4}+{2T(w)\over(z-w)^2}+
   {\partial T(w)\over(z-w)}~.\nonumber
\eeqa
The structure constant $\lambda(c)$ is fixed by the condition of
associativity of this algebra (see Appendix C):
\beq{lcrel} \lambda^2(c)={{8-c} \over 6}~.  \eeq
We will call the hypothetical string theory with (\ref{ft76}) as
its constraint algebra a ``spin-$4/3$ string.''

A straightforward calculation, presented in Appendix E, shows that
the spin-$4/3$ string has towers of extra null states at the critical
central charge $c=10$.  This result was obtained earlier by an examination
of the Ka\v{c} determinant formula \cite{ALT}.  This value of $c$
is different from the central charge $c=2D=12$
of the $K=4$ FSS discussed above.  However, as mentioned
in Sect.~6.2, the effective central charge for the light-cone degrees
of freedom of the $K=4$ FSS is $c_{\rm eff}=6$ by the partition
function, instead of the expected $c_{\rm eff}=8$.  This suggests
that the correct covariant space of states might be a $c=10$
representation of the spin-$4/3$ string algebra (\ref{ft76}),
instead of the $c=12$ tensor product representation of the FSS
constraint algebra (\ref{truefss}).

Indeed, some numerical evidence can be adduced to support this
supposition.  As mentioned in the Introduction, the general-$K$
FSS has critical space-time dimension
\beqq D=2+{16\over K}~, \eeq
and the central charge $c_0$ per dimension is thought to
correspond to that of a $\bZ_K$ parafermion theory plus a free
coordinate boson; {\it i.e.}
\beqq c_0={2(K-1)\over K+2}+1={3K\over K+2}~.\eeq
On the other hand, the effective central charge for the light-cone
degrees of freedom (from the FSS partition functions) is \cite{ADT}
\beqq c_{\rm eff}=(D-2)\cdot{3\over2}~.\eeq
Thus we could expect the total critical central charge to be
\beq{ccritfz}
c~=~c_{\rm eff}+2c_0~=~{6K\over K+2}+{24\over K}~.
\eeq
But these $c$ are precisely the critical central charges obtained
in ref.~\cite{ALT} for the spin-$(K+4)/(K+2)$ strings, by demanding
towers of extra null states.  In particular, when $K=4$ we find
$c=10$, corresponding to the spin-$4/3$ case calculated in Appendix E.

Unfortunately, the representations of the spin-$4/3$ algebra at
$c=10$ are not well understood, and in particular no representation
with a flat target space-time interpretation is known.  Thus, a
direct construction of the spin-$4/3$ string and comparison to the
$K=4$ FSS partition function is not yet possible.  However, it is possible
that string theories with fractional world-sheet supersymmetry embedded
in curved backgrounds can be constructed, and may prove interesting to
study.  A suggestive example of an affine Lie algebra whose
Wess-Zumino-Witten theory has central charge $c=10$ and has the correct
field content to construct a dimension-$4/3$ group-invariant current
is $SO(5,1)_8$.  This conformal field theory is thus a candidate
representation of the spin-$4/3$ string constraint algebra (\ref{ft76})
at $c=10$.

We should point out that the $c=12$ FSS may be compatible with the
$c=10$ spin-$4/3$ string in the following sense.  Partial gauge-fixing
of the $c=12$ FSS may reduce to a $c=10$ theory with the spin-$4/3$
algebra as the remaining constraint algebra.

\subsection{\label{sseven}Chirality and anomalies in the
fractional superstring}

In the preceding sections, we discussed the massless spectrum of
the open $K=4$ FSS, which contains a minimal Yang-Mills supermultiplet
$(A_{\mu}, \psi^{L})$, where $A_{\mu}$ is a vector field and
$\psi^{L}$ is a left-handed Weyl spinor field in six dimensions.
The massless spectrum described by
the partition function of the closed FSS
follows in a simple way from that of the open string theory.
In the closed FSS there are supermultiplets
corresponding to both the left-moving and
righ-moving sectors, so to obtain the spectrum we must
take the tensor product of these two supermultiplets.  There are two
possibilities for doing this, which we will call Type IIA and Type IIB, in
analogy to the ten-dimensional superstring.  In the Type IIA model, the
two supermultiplets are chosen so that the corresponding spinor
fields are of opposite chirality.  Taking the tensor product,
we see that the Type IIA closed FSS contains an
$N=2$ supergravity multiplet:
\beqqa
 &&(A_{\mu}, \psi^{L}) \otimes (A_{\mu}, \psi^{R})=
 g_{\mu \nu}+B^{L}_{\mu \nu}+B^{R}_{\mu\nu} \nonumber\\
 &&~~~~+\psi^{L}_{\mu}+\psi^{R}_{\nu}+4A_{\mu}+\psi^{L}+\psi^{R}+\phi,
\eeqa
where $g_{\mu \nu}$ is the graviton, $B_{\mu\nu}$ are (anti-)self-dual
antisymmetric tensor fields, $\psi_{\mu}$ is the gravitino,
$\psi$ is a spinor field, and $\phi$ is a scalar field.
$L$ and $R$ denote left- and right-handedness, respectively.
This spectrum clearly shows that the Type IIA closed FSS is non-chiral,
exactly as in the ten-dimensional ($K=2$) case.

The second possibility, Type IIB, is realized by choosing spinors
of the same chirality in the two supermultiplets.
In the ten-dimensional superstring, the spectra of the
two types of closed superstrings are different, but the counting of states
in each model is the same and they therefore have the same partition
function.  In addition, both models yield low-energy effective theories
which are free of gravitational anomalies.

Let us naively try to obtain the chiral supergravity multiplet for
the Type IIB FSS:
\beqq
 (A_{\mu},\psi^{L}) \otimes (A_{\mu},\psi^{L})~=~
 g_{\mu\nu}+5 B^{L}_{\mu\nu}+B_{\mu\nu}^{R}+2\psi^{L}_{\mu}
 +2 \psi^{R}+5 \phi~.
\eeq
We will refer to this chiral $N=2$ multiplet as the $\alpha$ multiplet.
It is easy to check that the resulting six-dimensional effective
low-energy theory containing the $\alpha$ multiplet has gravitational
anomalies.  To cancel the anomalies, we may try to extend the theory
by adding additional multiplets to the $\alpha$ multiplet.  There are
some constraints that must be satisfied.  Since there is only one modular
invariant partition function for the $K=4$ closed FSS (without any
compactification), ${\cal Z}$ in eq.~(\ref{ZFSS}), the only possible way
to accommodate additional massless states in the spectrum is to multiply
${\cal Z}$ by an integer; this increases the number of massless states by
this integral factor.  We do not want to increase the number of gravitons
or gravitinos in the theory, however, so we cannot obtain these states from
tensoring extra vector supermultiplets $(A_{\mu}, \psi^{L(R)})$.  The only
other option is to form these extra multiplets, to be called $\beta$
multiplets, from tensoring chiral (scalar) supermultiplets
$(\psi^{L(R)},4\phi)$:
\beqq
 (\psi^{R},4\phi)\otimes(\psi^{R},4\phi)~=~
 4 B_{\mu\nu}^{R}+8 \psi^{R}+20 \phi~.
\eeq
One can easily verify that the only anomaly-free
chiral supergravity multiplet is given by adding five copies of the
$\beta$ multiplet to the $\alpha$ multiplet, which yields the particle
content
\beqq
 g_{\mu\nu}+5B^L_{\mu\nu}+2\psi^{L}_{\mu}+21B^{R}_{\mu\nu}
 +42\psi^R+105\phi~.
\eeq
This six-dimensional chiral supergravity multiplet was first constructed
in the $K_3$ compactification of the usual superstring \cite{GSWe}. The
number of massless states in this chiral model is exactly six times that
of the non-chiral model.  Thus this counting of states implies that the
partition function for the chiral model is
\beqq
 {\cal Z}({\rm chiral})~=~6\,{\cal Z}({\rm non}-{\rm chiral})~.
\eeq

To obtain five $\beta$ multiplets in this model, the left- and
right-moving sectors must each have at least three chiral multiplets.
How can this theory account for all these extra states?
One possibility can be found by recalling that there is more
than one possible realization of the $\bZ_4$ parafermion theory.
In the bosonization given in Sect.~2, we see that each parafermion
field is realized by two fields $\epsilon^{(\pm a)}$
with the same conformal dimension.  This splitting
mechanism can be carried out further with the introduction
of cocycle operators, thus increasing the number
of copies of a particular field.  In Appendix A, we discuss the
introduction of cocycle operators in the parafermion theory.

It is clear how this construction can increase the number of fermion
fields in the model, by choosing a cocycle subalgebra which increases
the number of spin fields $\sigma_2^{\mu}$, for example.  However, the
origin of the scalar fields which are the superpartners of these additional
fermions is harder to ascertain, since we did not have scalar fields
originally in the spectrum of the open string FSS.  One possible
solution to this problem is to reintroduce $\sigma_{\pm 1}$,
the $\bZ_4$ parafermion spin fields of dimension $1/16$, mentioned
in Sect.~2.  States in the sector generated by these fields
did not appear to contribute to the partition functions given in
Sect.~3.  However, if we consider the fusion rule
$[\sigma_1][\sigma_{-1}] \sim [\id]+[\epsilon]$, we see that the
$\epsilon$ field appears.
In the tensor product theory, the fields $\sigma_{\pm 1}$ acquire
a vector index $\mu$, and thus we may be able to regard some of the
$\epsilon$ fields as Lorentz scalar composites of these spin fields,
in a manner similar to the way that $\sum_\mu \epsilon^{(\pm2),\mu}$
is realized as a scalar composite of $\epsilon^{(\pm1),\mu}$'s in
eq.~(\ref{4.19}).  In this way we may be able to obtain the required scalars.

\vspace{2.5ex}
\begin{flushleft}
\large\bf Acknowledgments
\end{flushleft}

It is a pleasure to thank K.~Dienes, J.~Grochocinski and
J.~Schwarz for useful discussions and comments.  This work
was supported in part by the National Science Foundation.

\appendix

\setcounter{footnote}{0}
\Appendix{appA}

In this appendix we discuss representations of the
$\bZ_4$ parafermion theory (or $SU(2)_4/U(1)$ coset theory)
that can be constructed in terms of cocycles and the
free boson $\rho(z)$ introduced in Sect.~\ref{stwo}.
These alternative representations could, in principle,
be the correct building blocks for the $K=4$ FSS Fock
space, instead of the free boson representation
discussed in the body of this paper.  For this reason it
is important to explore the space of inequivalent
representations of the $\bZ_4$ parafermion theory.

Restricting ourselves just to those that can be constructed
with a free boson plus cocycles, there is already an infinite
number of representations.  By further
restricting our inquiries by {\it ad hoc}\/ simplifying
assumptions, we will construct a few inequivalent
representations, and point out their main properties.
In particular, some of these representations have structure
constants for their operator algebra different from those of
the free boson representation.  In other
representations the OPEs no longer satisfy abelian braiding
relations as (\ref{epope})--(\ref{epope2}) do.
This makes the physical state conditions technically
more difficult to implement.  It is for these reasons that
the discussion in the body of the paper has been limited to
the free boson representation without cocycles.

Since the free boson $\rho(z)$ CFT is associative by itself,
any cocycles that are attached to it must also be associative
if the combined theory is to be.  Thus, we expect the cocycles
to form a finite-dimensional associative algebra including the
identity.  Inequivalent examples of such cocycle algebras are
the algebras of $n\times n$ matrices with real, complex or
quaternionic entries.  By taking the direct product of these algebras
with the free boson OPE algebra, we obtain an infinite number of
inequivalent representations of the $\bZ_4$ parafermion theory.
In the construction that follows we will restrict ourselves to
the simplest cocycle algebras: $1\times1$ real matrices
({\it i.e.}~no cocycles) and $2\times2$ real matrices.

We start with the free boson representation without cocycles used
in the body of this paper.  By comparing the OPEs
(\ref{epope})--(\ref{epope2}) of the free boson primary fields
to the parafermion fusion rules
(\ref{fusionrules}) we can identify which primary fields
belong to each parafermion sector $[\phi^j_m]$:
\beq{Apfsect}\begin{array}{ll}
 \ [\phi^0_0]_0 ~=~\left\{\epsilon^{(3n)}+\epsilon^{(-3n)}\,,~
  n\geq0\right\}\,,&
 \ [\phi^2_0]_0 ~=~\left\{i\partial\rho~;~\epsilon^{(3n)}-
  \epsilon^{(-3n)}\,,~n>0\right\}\,,\nonumber\\
 \ [\phi^1_1]_0 ~=~\left\{\epsilon^{(3n+1/2)}~\right\}\,,&
 \ [\phi^1_1]^\prime_0~=~\left\{\epsilon^{(3n-1/2)}~\right\}\,,\\
 \ [\phi^1_0]_0 ~=~\left\{\epsilon^{(3n+1)}~\right\}\,,&
 \ [\phi^1_0]^\prime_0~=~\left\{\epsilon^{(3n-1)}~\right\}\,,\nonumber\\
 \ [\phi^0_1]_0+[\phi^0_{-1}]_0~=~\left\{\epsilon^{(3n+3/2)}~
  \right\}\,,&\nonumber
\end{array}\eeq
where $n$ runs over the integers in the specified ranges.
The subscript zero is to identify the representation.
Note that the $[\phi^1_0]$ and $[\phi^1_1]$ sectors each appear
twice in this representation
of the parafermion theory.  This is clear from the identifications
of the energy operator $\epsilon$ and the spin field
$\sigma_2$ in (\ref{pfid}), where they each have
multiplicity two:  $\epsilon$ can be represented by either
$\epsilon^{(+1)}$ or $\epsilon^{(-1)}$, and $\sigma_2$ by either
$\epsilon^{(+1/2)}$ or  $\epsilon^{(-1/2)}$.  Also, the parafermion
current sectors $[\phi^0_{\pm1}]$ cannot be separated in a
way consistent with the fusion rules (\ref{fusionrules}).
This property implies that this $\bZ_4$ parafermion representation
is not really a faithful representation of the $SU(2)_4$ fusion
rules (\ref{fusionrules}).  This is a reflection of the fact that in
(\ref{epope})--(\ref{epope2}) only a single cut occurs
on the right hand side of any given OPE.  This situation
is called abelian (or sometimes parafermionic) braiding.

We can form a representation with single multiplicity for the
$[\phi^1_0]$ and $[\phi^1_1]$ sectors by considering only the
symmetric subalgebra of the $0$-representation (\ref{Apfsect}):
\beqa{Apfsect1}
 [\phi^0_0]_1 &=&\left\{\epsilon^{(3n)}+\epsilon^{(-3n)}~,~~
  n\geq0\right\}~,\nonumber\\
 \ [\phi^1_1]_1 &=&\left\{\epsilon^{(3n+1/2)}+\epsilon^{(-3n-1/2)}
  \right\}~,\\
 \ [\phi^1_0]_1 &=&\left\{\epsilon^{(3n+1)}+\epsilon^{(-3n-1)}
  \right\}~,\nonumber\\
 \ [\phi^0_1]_1&=&\left\{\epsilon^{(3n+3/2)}+\epsilon^{(-3n-3/2)}~,~~
  n\geq0\right\}~.\nonumber
\eeqa
This representation satisfies non-abelian braiding relations.  In
particular, upon braiding the fields in this representation we
introduce new fields (on the next Riemann sheet).  The set of all
fields on all three sheets just reproduces the field content of
representation-0, though the field content on each sheet separately
is that given in (\ref{Apfsect1}).  Representation-1 has single energy
and spin-field multiplicity, but has zero spin-2 multiplicity
({\it i.e.}\ the $[\phi^2_0]$ sector does not appear).  This latter
property implies that representation-1 is no good for FSS building
since the arguments of \ref{sthree} show that the
spin-2 parafermion fields enter into the FSS spectrum.  Note that
this representation is a faithful representation of the $SU(2)_4$
fusion rules (with only spin-0 and spin-1 fields) but, since the
spin-2 fields decouple, cannot realize the $\bZ_4$ parafermion
current algebra, and therefore also can not realize the $SU(2)_4$
Kac-Moody current algebra {\it via}\/ eq.~(\ref{KMcurralg}).

The algebra of $2\times2$ real matrices is generated by the
four elements:
\beqq
 \a=\pmatrix{0&1\cr 0&0\cr}~,\quad
 \b=\pmatrix{0&0\cr 1&0\cr}~,\quad
 \c=\pmatrix{1&0\cr 0&-1\cr}~,\quad
 {\bf 1}=\pmatrix{1&0\cr 0&1\cr}~,
\eeq
which satisfy the algebra
\beqqa
 \a^2~=~\b^2~=0~,&\quad\quad
  &\c^2=~{\bf 1}~,\nonumber\\
 2\a\b~=~{\bf 1}+\c~,&\quad\quad
  & 2\b\a~=~{\bf 1}-\c~,\\
 \b\c~=~-\c\b~=~\b~,&\quad\quad
  &\c\a~=~-\a\c~=~\a~.\nonumber
\eeqa
If we take a direct tensor product of the $c=1$ free boson
theory with this set of cocycles, we simply obtain a new
$c=1$ theory with four times as many fields.  The point, however,
is that this new theory has closed subalgebras which could
not be realized in the free boson theory alone.  We will construct
two inequivalent such representations of the $\bZ_4$ parafermion
sectors with multiplicities less than or equal to those of the
$0$-representation.

The first is
\beqa{Apfsect2}
 [\phi^0_0]_2&=&\left\{{\bf 1}\left(\epsilon^{(3n)}+
  \epsilon^{(-3n)}\right)~,~~n\geq0\right\}~,\nonumber\\
 \ [\phi^2_0]_2&=&\left\{\c(i\partial\rho)~;~\c\left(\epsilon^{(3n)}-
  \epsilon^{(-3n)}\right)~,~~n>0\right\}~,\nonumber\\
 \ [\phi^1_1]_2&=&\left\{\b\epsilon^{(3n+1/2)}+
  \a\epsilon^{(-3n-1/2)}\right\}~,\\
 \ [\phi^1_0]_2&=&\left\{\a\epsilon^{(3n+1)}+
  \b\epsilon^{(-3n-1)}\right\}~,\nonumber\\
 \ [\phi^0_1]_2&=&\left\{{\bf 1}\left(\epsilon^{(3n+3/2)}+
  \epsilon^{(-3n-3/2)}\right)~,~~n\geq0\right\}~,\nonumber\\
 \ [\phi^0_{-1}]_2&=&\left\{\c\left(\epsilon^{(3n+3/2)}-
  \epsilon^{(-3n-3/2)}\right)~,~~n\geq0\right\}~.\nonumber
\eeqa
It is easily checked that this set of fields forms a closed
algebra under fusion.  This representation is also non-abelianly
braided, but has single
multiplicities, and is a faithful representation of the $SU(2)_4$
fusion rules (with all spins).  However, it has no
$\epsilon\sigma_2\sigma_2$=$c_{111}$ coupling ({\it i.e.}\ no coupling
between the $[\phi^1_0]$ and two $[\phi^1_1]$ sectors), which implies
that representation-2 is not suitable for FSS building, since, as will
be shown in Sect.~\ref{sfive}, this coupling is crucial for the
description of space-time spinors in the FSS Fock space.  Note that
the spin-2 sector OPEs (involving the $[\phi^2_0]$ and $[\phi^0_{\pm1}]$
fields) do not have a $\bZ_4$ symmetry and thus do not form a representation
of the parafermion current algebra.

It is easiest to write the second representation with cocycles
in terms of the basis $\{{\bf 1}, {\bf i},{\bf j}, {\bf k}\}$ where
${\bf i}\equiv-\a-\b$, ${\bf j}=\c$ and ${\bf k}=\a-\b$.  They
satisfy the relations
\beqq
  {\bf i}^2={\bf j}^2=-{\bf k}^2={\bf 1}~,~~
  {\bf i}{\bf j}=-{\bf j}{\bf i}={\bf k}~,~~
  {\bf j}{\bf k}=-{\bf k}{\bf j}=-{\bf i}~,~~
  {\bf k}{\bf i}=-{\bf i}{\bf k}=-{\bf j}~.
\eeq
A closed algebra is composed of the sectors
\beqa{Apfsect3}
 [\phi^0_0]_3&=&\left\{{\bf k}^n\epsilon^{(3n)}+
  {\bf k}^{-n}\epsilon^{(-3n)}~,~~n\geq0\right\}~,\nonumber\\
 \ [\phi^2_0]_3&=&\left\{{\bf k}(i\partial\rho)~;~
  {\bf k}^{n+1}\epsilon^{(3n)}+{\bf k}^{-n-1}\epsilon^{(-3n)}~,~~
  n>0\right\}~,\nonumber\\
 \ [\phi^1_1]_3&=&\left\{{\bf k}^n{\bf i}\epsilon^{(3n+1/2)}+
  {\bf k}^{-n}{\bf j}\epsilon^{(-3n-1/2)}\right\}~,\nonumber\\
 \ [\phi^1_1]_3^\prime&=&\left\{{\bf k}^n{\bf j}\epsilon^{(3n+1/2)}+
  {\bf k}^{-n}{\bf i}\epsilon^{(-3n-1/2)}\right\}~,\\
 \ [\phi^1_0]_3&=&\left\{{\bf k}^n\epsilon^{(3n+1)}+
  {\bf k}^{-n}\epsilon^{(-3n-1)}\right\}~,\nonumber\\
 \ [\phi^1_0]_3^\prime&=&\left\{{\bf k}^{n+1}\epsilon^{(3n+1)}+
  {\bf k}^{-n-1}\epsilon^{(-3n-1)}\right\}~,\nonumber\\
 \ [\phi^0_1]_3&=&\left\{{\bf k}^n{\bf i}\epsilon^{(3n+3/2)}+
  {\bf k}^{-n}{\bf j}\epsilon^{(-3n-3/2)}~,~~n\geq0\right\}~,\nonumber\\
 \ [\phi^0_{-1}]_3&=&\left\{{\bf k}^n{\bf j}\epsilon^{(3n+3/2)}+
  {\bf k}^{-n}{\bf i}\epsilon^{(-3n-3/2)}~,~~n\geq0\right\}~.\nonumber
\eeqa
Note that ${\bf k}^{-1}=-{\bf k}$.
This representation is non-abelianly braided in general but has a $\bZ_4$
symmetry in its parafermion current sector.  Thus the parafermion
fields $\phi^0_0$, $\phi^0_{\pm1}$ and $\phi^2_0$ form an
abelianly braided subalgebra.  Of the four $\bZ_4$ parafermion
representations presented in this Appendix, this is the only one
to have the defining $\bZ_4$ symmetry, and, consequently, the
only one which allows a realization of the $SU(2)_4$ current
algebra by adding a boson.  Note that this representation has
the same multiplicities
as the free boson representation (representation-0), and forms a
faithful representation of the $SU(2)_4$ fusion rules (\ref{fusionrules})
with double spin-1 multiplicities.
{\it A priori}, representation-3 is just as good as representation-0
for building the FSS Fock space.  In practice, it has the
technical disadvantage that the $\bZ_4$ symmetry only extends to
part of its spectrum, unlike the free boson case, where its $\bZ_3$
symmetry includes all its fields.  This $\bZ_3$ symmetry is a major
simplifing feature of the analysis presented in the body of this paper.

There exist many more representations that can be found by taking
larger cocycle spaces.  It is possible that some of these other
representations have subalgebras with low multiplicities which are
not equivalent to representations 0--3 constructed above.
Also, one can build other, essentially free, representations of the
$\bZ_4$ parafermions by orbifolding the free boson $\rho(z)$ of the above
representations.  It is found \cite{Yang} that the twist fields
introduced by a $\bZ_2$ orbifolding correspond to the half-odd integer
spin sectors of the parafermion theory,
which include the dimension-$1/16$ spin-fields.
We have not considered orbifolds in this paper
because the arguments of \ref{sthree} show that the half-odd
spin sectors do not contribute to the FSS partition function.
See however the discussion of chiral fermions in Sect.~\ref{sseven}
where these twist fields may play a role.

\setcounter{footnote}{0}
\Appendix{appB}

In this appendix we will review the argument of
Zamolodchikov and Fateev \cite{ZFpara} which leads to the
derivation of generalized commutation relations (GCRs)
for abelianly braided operators.  We will do this by an
example:  we derive the GCRs for the $\bZ_4$ parafermion fields
$\epsilon^{(+1)}$ and $\epsilon^{(-1)}$.  The general
procedure should be clear from this example.

The $\bZ_4$ parafermion Fock space falls into sectors
${\cal H}_q$ labelled by their $\bZ_3$ charge $q$.  The
dimension-$1/3$ energy operators $\epsilon^{(+1)}$ and $\epsilon^{(-1)}$
carry $\bZ_3$ charges $q=+1$ and $-1$, respectively.  To
simplify notation, we will denote $\epsilon^{(+1)}$ by
$\epsilon$ and $\epsilon^{(-1)}$ by $\epsdag$.  They act
on the Fock space by the rule
\beqq
 \epsilon: {\cal H}_q\rightarrow{\cal H}_{q+1}~,\quad
 \epsdag : {\cal H}_q\rightarrow{\cal H}_{q-1}~,
\eeq
where the $\bZ_3$ charge is defined mod 3.  Their mode
expansions are defined, as in eq.~(\ref{epmode}), by
\beqa{edomode}
 \epsilon(z)\chi_q(0)&=&\sum_n z^{n-{q\over3}}
  \epsilon_{-n+(q-1)/3}\chi_q(0)~,\nonumber\\
 \epsdag(z)\chi_q(0)&=&\sum_n z^{n+{q\over3}}
  \epsdag_{-n-(q+1)/3}\chi_q(0)~,
\eeqa
where $\chi_q$ is an arbitrary state in ${\cal H}_q$.  The
mode expansions can be inverted to give
\beqa{modedom}
 \epsilon_{n+(q-1)/3}\chi_q(0)&=&\oint_\gamma{{\rm d}z
  \over2\pi i} z^{n-1+q/3}\epsilon(z)\chi_q(0)~,\nonumber\\
 \epsdag_{n-(q+1)/3}\chi_q(0)&=&\oint_\gamma{{\rm d}z
  \over2\pi i} z^{n-1-q/3}\epsdag(z)\chi_q(0)~,
\eeqa
where $\gamma$ is a contour encircling the origin once,
where $\chi_q(0)$ is inserted.

The OPE of $\epsilon$ with $\epsdag$ is given by [see
eq.~(\ref{epope1})]
\beq{eedag}
 \epsilon(z)\epsdag(w)~=~{1\over(z-w)^{2/3}}+
 (z-w)^{1/3}\,i\partial\rho(w)+\ldots~.
\eeq
Now we will derive the generalized commutation relations
that the $\epsilon$ and $\epsdag$ modes satisfy as a result of
this OPE.  Consider the integral
\beq{II}
 {\cal I}~=~\oint_{\gamma}{{\rm d}z\over2\pi i}
  \oint_{\delta}{{\rm d}w\over2\pi i} z^{m-1+q/3}
  w^{n-1-q/3}(z-w)^{p+2/3}\epsilon(z)\epsdag(w)\chi_q(0)~,
\eeq
where $m$, $n$ and $p$ are arbitrary integers.  The contour
$\gamma$ encircles $\delta$, which in turn encircles the
origin.  The the fractional parts of the exponents in the integrand
are chosen so that the whole integrand is single-valued in both
the $z$- and $w$-planes.  This is possible only because of the
abelian nature of the $\epsilon\epsdag$ OPE (\ref{eedag}).
We can evaluate this integral by letting $\delta$ shrink down
to a small circle near to the origin.  In this limit we can expand
the $(z-w)^{p+2/3}$ factor as
\beq{cexp}
  (z-w)^{p+2/3}=~z^{p+2/3}\sum_{\ell=0}^\infty c^{(p+2/3)}_\ell
  \left({w\over z}\right)^\ell~,
\eeq
where $c^{(\alpha)}_\ell$ are the appropriate binomial
coefficients.  Inserting this expansion into eq.~(\ref{II}),
and using the mode definitions (\ref{modedom}), we find
\beqq
  {\cal I}~=~\sum_{\ell=0}^\infty c^{(p+2/3)}_\ell
  \epsilon_{m+p-\ell+{{q+1}\over3}}
  \epsdag_{n+\ell-{{q+1}\over3}}\chi_q(0)~.
\eeq

${\cal I}$ can also be evaluated in another way, by first
deforming the $\gamma$ contour so that it lies inside
$\delta$.  Upon performing this deformation, one picks
up in the usual way two contributions
\beq{deform}
  {\cal I}={\cal I}^\prime+{\cal I}_0~,
\eeq
corresponding to the same integral
${\cal I}^\prime$ with $\gamma$ and $\delta$
interchanged, and the new contribution ${\cal I}_0$ where the
$\gamma$ contour encircles the $\epsdag$ insertion
at the point $w$ on the $z$-plane.  ${\cal I}^\prime$ can
be evaluated in the same way as ${\cal I}$ was,
after interchanging $\epsilon(z)$ and $\epsdag(w)$
as well as $z$ and $w$ in the $(z-w)^{p+2/3}$ factor.
Taking care to perform these interchanges along equivalent
paths in the complex plane gives an overall phase
${\rm e}^{i\pi(-2/3)}\times{\rm e}^{i\pi(p+2/3)} = (-1)^p$.  Thus
\beqa{IIprime}
  {\cal I}^\prime&=&(-1)^p\oint_{\delta^\prime}{{\rm d}z\over2\pi i}
  \oint_{\gamma^\prime}{{\rm d}w\over2\pi i}
  z^{m-1+q/3}w^{n-1-q/3}(w-z)^{p+2/3}
  \epsdag(w)\epsilon(z)\chi_q(0)\nonumber\\
  &=&(-1)^p\sum_{\ell=0}^\infty c^{(p+2/3)}_\ell
   \epsdag_{n+p-\ell-{{q-1}\over3}}
   \epsilon_{m+\ell+{{q-1}\over3}}\chi_q(0)~,
\eeqa
where we have again used the mode definitions (\ref{modedom}).
Note that the abelian braid property of the $\epsilon\epsdag$ OPE
was important in performing the analytic continuation needed
to define the integrand of ${\cal I}^\prime$.  Because only one kind of
cut appears in  (\ref{eedag}), under the analytic continuation
which interchanges $z$ and $w$, $\epsilon(z)\epsdag(w)$
only gains a simple phase.  Indeed, this property can be taken as
the definition of abelian braiding.

The contribution ${\cal I}_0$ is the same as ${\cal I}$ except that
$\gamma$, instead of circling the origin, now only encircles
the point $w$ in the $z$-plane.  Letting this contour shrink
to a small circle around $w$, we can replace $\epsilon(z)
\epsdag(w)$ by their OPE (\ref{eedag}).  The value of the
integer $p$ in the integrand controls the number of terms
in the OPE that contribute.  For example, taking $p=-1$ gives
\beqqa
  {\cal I}_0&=& \oint_{0}{{\rm d}w\over2\pi i}w^{n-1-q/3}
  \oint_{w}{{\rm d}z\over2\pi i} z^{m-1+q/3}
  \left\{(z-w)^{-1}+\ldots\right\}\chi_q(0)\nonumber\\
  &=&\oint{{\rm d}w\over2\pi i}w^{n+m-2}\chi_q(0)~=~
  \delta_{n+m-1}~.
\eeqa

These expressions for ${\cal I}$, ${\cal I}^\prime$ and
${\cal I}_0$ can be combined according to
eq.~(\ref{deform}) to give a generalized commutation
relation for the $\epsilon$ and $\epsdag$ modes:
\beqq
 \sum_{\ell=0}^\infty c^{(-1/3)}_\ell\left[
 \epsilon_{m-1-\ell+{{q+1}\over3}}\epsdag_{n+\ell-{{q+1}\over3}}+
 \epsdag_{n-1-\ell-{{q-1}\over3}}\epsilon_{m+\ell+{{q-1}\over3}}\right]
 =\delta_{n+m-1},
\eeq
understood to be acting on any state $\chi_q\in{\cal H}_q$.
Alternatively we could have chosen $p=-2$, which would pick
up a contribution from the $\partial\rho(w)$ term in the
$\epsilon\epsdag$ OPE, to give the GCR
\beqqa
 &\displaystyle{ \sum_{\ell=0}^\infty c^{(-4/3)}_\ell\left[
 \epsilon_{m-2-\ell+{{q+1}\over3}}\epsdag_{n+\ell-{{q+1}\over3}}-
 \epsdag_{n-2-\ell-{{q-1}\over3}}\epsilon_{m+\ell+{{q-1}\over3}}
 \right] } &\nonumber\\
 &\displaystyle{ =~\left(m-1+{q\over3}\right)\delta_{n+m-1}+
 s_{n+m-2}~, } &
\eeqa
where we have used the mode expansion of the $\rho(z)$ field
in the form
\beqq
 s_m~=~\oint{{\rm d}z\over2\pi i}\,i\partial\rho(z)~z^m~.
\eeq
It is clear that by letting $p$ take more negative values, more
complicated GCRs involving more terms from the $\epsilon\epsdag$
OPE can be obtained.  By conformal invariance, this tower
of GCRs is consistent.  Indeed, the GCR obtained with
$p=p_0$ can be derived from the GCR with $p=p_0-1$ using
the binomial coefficient identity
\beqq
 c^{(\alpha)}_\ell-c^{(\alpha)}_{\ell-1}~=~c^{(\alpha+1)}_\ell~.
\eeq
It should also be clear that the argument reviewed here works
equally well for deriving GCRs from any abelianly braided OPE.

\setcounter{footnote}{0}
\Appendix{appC}

In this appendix, we discuss the representation theory and
associativity (or consistency) conditions of the fractional
superconformal algebras.  We begin with the split algebra (\ref{ftalg}).
The explicit form for the currents of this algebra in the
coordinate boson plus $\bZ_4$ parafermion CFT is given in
eq.~(\ref{curform}).  This CFT
has central charge $c_0=2$.  We can generalize the split algebra
(\ref{ftalg}) to the case of a CFT with arbitrary central
charge $c_0$ \cite{FZft}:
\beqa{ftalg1}
  G^{+}(z)G^{+}(w)&= & {\lambda(c_0)\over(z-w)^{4/3}}
   \left\{G^{-}(w)+{1\over2}(z-w)\partial G^{-}(w)\right\},\nonumber\\
  G^{-}(z)G^{-}(w)&= & {\lambda(c_0)\over(z-w)^{4/3}}
   \left\{G^{+}(w)+{1\over2}(z-w)\partial G^{+}(w)\right\},\\
  G^{+}(z)G^{-}(w)&= & {(1/2)\over(z-w)^{8/3}}\left\{{3c_0\over4}+
   2(z-w)^2{T}(w)\right\},\nonumber
\eeqa
where we have only included terms on the right hand side with
negative powers of $(z-w)$.  $\lambda(c_0)$ is the undetermined
structure constant of this algebra.  (The other coefficients are
determined by conformal invariance.)  We found by explicit construction
in Sect.~\ref{sfour} that $\lambda(2)=1$.

Because the operator algebra (\ref{ftalg1}) is abelianly braided
(that is, only a single cut appears on the right hand side of
each OPE), we can
derive a Ward identity relating correlators with
a $G^{+} G^{-}$ pair to ones with the pair removed.
Following Zamolodchikov and Fateev \cite{ZFpara,FZft},
we can then solve for the structure constant $\lambda$
by imposing the associativity condition on the four-point
function
\beq{4pnt}
  {\cal G}(z_i)=\langle G^{+}(z_{1}) G^{+}(z_{2})
  G^{-}(z_{3})G^{-}(z_{4})\rangle.
\eeq
Actually, we will derive a simpler Ward identity which
is valid only for the four-point function ${\cal G}$.
Consider the function ${\cal F}(z_i)$ defined by
  \beq{holof}
  {\cal F}(z_i)=(z_1-z_2)^{1/3}(z_1-z_3)^{2/3}
  (z_1-z_4)^{2/3}{\cal G}(z_i).
  \eeq
Because of the abelian nature of the spin-4/3 algebra,
${\cal F}$ is a holomorphic function of $z_1$ with a
first-order pole at $z_2$ and second-order poles at
$z_3$ and $z_4$.  Also, since by conformal invariance
${\cal G}\sim z_1^{-8/3}$ as $z_1\rightarrow\infty$, in the
same limit ${\cal F}\sim 1/z_1\rightarrow 0$.  Thus ${\cal F}$
is completely determined by the residues of its poles,
which are easily read off from eq.~(\ref{holof}) and
the OPEs (\ref{ftalg1}).\footnote{ This argument
must be changed for higher-point functions to make
${\cal F}$ vanish as $z_1\rightarrow\infty$.  In general,
this requires the presence of third-order poles in
${\cal F}$ as $z_1$ approaches a $G^{-}$ insertion.}
The resulting Ward identity for ${\cal G}$ is
\beqa{Widen}
  {\cal G} &=& (z_1-z_2)^{-1/3}(z_1-z_3)^{-2/3}
   (z_1-z_4)^{-2/3}\nonumber\\
  && \times\Biggl\{\lambda(z_1-z_2)^{-1}
   (z_2-z_3)^{2/3}(z_2-z_4)^{2/3}
   \langle G^{-}(z_2)G^{-}(z_3)G^{-}(z_4)\rangle\nonumber\\
  && \quad +{c_0\over8}(z_1-z_3)^{-1}\left[
   3(z_1-z_3)^{-1}-(z_2-z_3)^{-1}+2(z_3-z_4)^{-1}
   \right]\nonumber\\
  && \quad\quad\times(z_2-z_3)^{1/3}(z_3-z_4)^{2/3}
   \langle G^{+}(z_2)G^{-}(z_4)\rangle\nonumber\\
  && \quad +{c_0\over8}(z_1-z_4)^{-1}\left[
   3(z_1-z_4)^{-1}-(z_2-z_4)^{-1}-2(z_3-z_4)^{-1}
   \right]\nonumber\\
  && \quad\quad\times(z_2-z_4)^{1/3}(z_3-z_4)^{2/3}
   \langle G^{+}(z_2)G^{-}(z_3)\rangle\Biggr\}.
\eeqa
The remaining two- and three-point functions can be
directly evaluated from the spin-4/3 OPEs to give
a closed form expression for ${\cal G}$.  Associativity
can then be checked by taking, say, the limit as
$z_2\rightarrow z_3$ in this expression, and checking
that the residues of the poles match those determined
by the OPEs.  This fixes $\lambda$ as a function of $c_0$:
\beq{lamb}
  \lambda^2={8-c_0\over6}.
\eeq
A crucial assumption that enabled us to integrate the Ward
identity was that no fractional cuts not allowed by our
abelian braiding assumption occur on the right-hand side
of the OPEs (\ref{ftalg1}), even among the ``regular'' terms.
We present an example below of how the associativity
condition (\ref{lamb}) may be modified by the inclusion of
new cuts among the regular terms.

Much work has been done to understand the representation
theory of this algebra (and related non-local algebras)
\cite{FZft,KMQ,BNY,Rav,AGT1,AGT2,Pog,CLT}.  In particular, it is
known that the split algebra (\ref{ftalg1}), as well as the
algebra formed from it by adding together $G^{+}$ and $G^{-}$,
have a series of unitary minimal
representations.  These can be realized by the coset models
$SU(2)_4\otimes SU(2)_L/SU(2)_{L+4}$ with central charges
\beqq
 c_0(L)~=~2-{24\over(L+2)(L+6)}\quad\quad
 {\rm for}~~L=1,2,\ldots~.
\eeq
Note that the central charge of these minimal models has an
accumulation point at $c_0=2$, the central charge of the
free CFT we are using in the construction of the FSS.  This
is analogous to the role played by the free representation of
the superconformal algebra with central charge $3/2$ in the
superstring theory.  However, there are important differences
between the representation theory of the spin-$4/3$ algebras
and that of the spin-$3/2$ superVirasoro algebra of the superstring.
In particular, the superconformal algebra has no parameters
besides the central charge, whereas the split algebra has the
structure constant $\lambda$.

It is easy to see that if we have two representations of the split
algebra with currents and constants $\{G^\pm_i,T_i,c_i,\lambda_i\}$
for $i=1$ and $2$, then we can form a new representation of the split
algebra by tensoring them together only if $\lambda_1=\lambda_2$.  The
tensor-product algebra has currents and constants given by
$\{G^\pm=G^\pm_1+G^\pm_2, T=T_1+T_2, c=c_1+c_2, \lambda=\lambda_1=\lambda_2\}$.
If the $\lambda_i$ and $c_i$ are related by (\ref{lamb}), then the new
$\lambda$ and $c$ of the tensor-product algebra will satisfy an
appropriately modified relation.  The argument leading to (\ref{lamb})
breaks down because the tensor-product algebra is not abelianly braided.
In particular, new fractional powers appear among the regular terms of
the OPEs.  For example, the first regular term not shown in the
$G^{+}G^{+}$ OPE of eq.~(\ref{ftalg1}) is
\beq{regterm}
  G^{+}(z)G^{+}(w)~\sim~(z-w)^0 :\!G^{+}_1 G^{+}_2\!:(w)~.
\eeq
This term and its descendants all have integer powers of $(z-w)$.
Though this is not a ``cut,'' it nevertheless is a new fractional
power, not included among the allowed exponents necessary for
abelian braiding.

Using a simple {\it anzatz\/} for the
four-point functions of the tensored currents $G^{+}(z)$ and
$G^{-}(z)$, we can show explicitly how the appearance
of this new ``cut'' relaxes the abelian associativity
condition (\ref{lamb}).  First, we will assume this
``cut'' does not occur, and we will show how a simple
algebraic argument recovers (\ref{lamb}).

By $SL(2,{\bf R})$ invariance \cite{BPZ}, we can write the four-point
function (\ref{4pnt}) without any loss of generality as the sum of terms
  \beq{correxp}
  {\cal G}(z_i)=\sum_j k_j{\cal A}^{r_j}
  {\cal B}^{s_j}{\cal C}^{t_j}
  \eeq
where $k_j$ are coefficients to be determined, and
we have defined the combinations
  \beqa{ABC}
  {\cal A} &= & (z_1-z_2)(z_3-z_4),\nonumber\\
  {\cal B} &= & (z_1-z_3)(z_2-z_4),\\
  {\cal C} &= & (z_1-z_4)(z_2-z_3).\nonumber
  \eeqa
In order for ${\cal G}$ to have the right behavior
as its arguments approach infinity, the exponents
must satisfy, for each $j$,
  \beq{rsteq}
  r_j+s_j+t_j=-8/3.
  \eeq
Since only ${\cal A}$ vanishes as $z_1\rightarrow z_2$ or
$z_3\rightarrow z_4$, its exponent is determined by
the $G^{+} G^{+}$  or $G^{-}G^{-}$ OPE in eq.~(\ref{ftalg}), so
that the $r_j$ must belong to the set $\{-4/3+n\}$ where
$n$ is an integer.  Similarly, the exponents of ${\cal B}$
and ${\cal C}$ are determined by the $G^{+} G^{-}$ OPE, implying
$s_j,t_j\in\{-8/3+n\}$.  We make the {\em anzatz\/} that the
integers $n$ in the above sets satisfy $n\geq 0$.
This assumption is not necessarily true in general; in
the present case it can be verified using the Ward
identity (\ref{Widen}).  With this assumption, there
are 15 solutions for $\{r_j,s_j,t_j\}$ in these sets
satisfying (\ref{rsteq}).  Because ${\cal A}$, ${\cal B}$ and
${\cal C}$ are related by ${\cal A}-{\cal B}+{\cal C}=0$,
there are only five independent terms in our expansion for ${\cal G}$:
  \beqa{iiix}
  {\cal G}&=&{3c_0\over8}\Biggl[
   k\left({{\cal A}\over{\cal B}}\right)^{-4/3}
  +l\left({{\cal A}\over{\cal B}}\right)^{-1/3}
  +m\left({{\cal A}\over{\cal B}}\right)^{2/3}\nonumber\\
  &&\qquad\qquad\mbox{}
  +n\left({{\cal A}\over{\cal B}}\right)^{5/3}
  +p\left({{\cal A}\over{\cal B}}\right)^{8/3}
  \Biggr] {\cal C}^{-8/3}~.
  \eeqa
Now if we take the limit $z_3\rightarrow z_4$ and
compare the residues of the poles with those predicted
by the OPEs (\ref{ftalg1}), we find
  \beqa{res}
  \lambda^2 &= & k~,\nonumber\\
  2\lambda^2 &= & 8k+3l~.
  \eeqa
The limit $z_2\rightarrow z_4$ implies
  \beqa{resone}
  3c_0/8 &= & p~,\nonumber\\
  0 &= & 3n+8p~,\\
  12 &= & 9m+24n+44p~,\nonumber
  \eeqa
and the remaining limit $z_1\rightarrow z_4$ implies
  \beqa{restwo}
  3c_0/8 &= & k+l+m+n+p~,\nonumber\\
  0 &= & -4k-l+2m+5n+8p~,\\
  12 &= & 2k-l+5m+20n+44p~.\nonumber
  \eeqa
This overdetermined set of equations has a solution
  \beq{solone}\begin{array}{ll}
  k={1\over6}(8-c_0)~,& l=-{1\over3}(8-c_0)~,\nonumber\\
  m={1\over6}(8+5c_0)~,& n=-c_0~,\quad\quad p=3c_0/8~,
  \end{array}\eeq
only if $\lambda$ satisfies the abelian associativity condition
(\ref{lamb}).

The crucial assumption that enabled us to restrict
the exponents $\{r_j,s_j,t_j\}$ to certain sets was that
no fractional cuts besides those occurring in the singular
terms on the right-hand
side of the OPEs (\ref{ftalg1}), appear among the
``regular'' terms.  This, of course, is just the assumption of
abelian braiding.

Let us now consider the tensor product algebra with currents
$G^\pm=G^\pm_1+G^\pm_2$, discussed above.  New terms
like (\ref{regterm}) appear in the OPEs; these terms and their
descendants all have integer powers of $(z-w)$.  Though this is not
a ``cut,'' it nevertheless is a new fractional power, not included in
our earlier set of allowed values for the exponents
$\{r_j,s_j,t_j\}$.  Allowing $r_j$, $s_j$ and $t_j$ to take on
non-negative integer values (by our {\it anzatz\/}) implies
that one new term appears in the expansion of ${\cal G}$ in
addition to the five in eq.~(\ref{iiix}):  $(3c_0/8)q{\cal C}^{-8/3}$.
This results in a change in only the first residue equation in
(\ref{res}) to
\beqq
 \lambda^2~=~k+q~.
\eeq
With this change, the residue equations have the same solution
(\ref{solone}), but now with
\beqq
 \lambda^2~=~{8-c_0\over6}+q~.
\eeq
Thus by varying $q$, we can achieve any value of $\lambda$ and
still have an associative four-point function.

Similar observations hold for the full fractional superconformal
algebra
\beqa{fscalg1}
  G(z)G(w)&= & {1\over(z-w)^{8/3}}\left\{{3c_0\over4}+
    2(z-w)^2 {T}(w)\right\} \nonumber\\
  &&\mbox{}+ {\lambda\over(z-w)^{4/3}} \left\{
    G(w)+{1\over2}(z-w)\partial G(w) \right\} .
\eeqa
This algebra is non-abelianly braided since it has $(z-w)^{n-2/3}$
and $(z-w)^{n-1/3}$ cuts apparent among its singular terms.
Thus, the relation derived above between the structure constant
and central charge (\ref{lamb}) will not necessarily be valid.
Indeed, one of the alternative free boson plus cocycle representations
of the fractional superconformal algebra (\ref{fscalg1}) constructed
in \ref{appA} violates (\ref{lamb}), having $c_0=2$ but $\lambda=0$.
Also, when we tensor together copies of the algebra (\ref{fscalg1})
as in the discussion of the last paragraph, we also add in new
``cuts'' among the regular terms of the form $(z-w)^{0+n}$.

In summary, we have learned that
the data necessary to specify a non-local chiral
algebra consists not only of the singular pieces of
the OPEs, but also of a list of all the fractional
parts of the exponents of $(z-w)$ that may appear
in the ``regular'' terms, as well.  The abelian nature
of the split algebra (\ref{ftalg1}) allowed us
to perform the analytic continuation necessary to
derive a Ward identity, which in turn allowed us to
relate the structure constant and the central charge.
The non-split or tensored algebras have non-abelian braiding
properties and do not, in general, satisfy any particular
relation between $\lambda$ and $c$.  In principle, though,
if the exact form of the (non-abelian) braiding of the currents were
known, one could solve the associativity conditions to
find new relations between $\lambda$ and $c$.  (For
an example of this approach, see refs.~\cite{AGT1,AGT2}.)

\setcounter{footnote}{0}
\Appendix{sthree}

In this appendix we will build the modular-invariant partition function
for the closed $K=4$ FSS.  This will primarily be a review of results
obtained in refs.~\cite{AT,ADT}.  We will use the partition function
to predict the critical space-time dimension, to identify the bosonic
and fermionic states of the FSS, and to determine the analog of the GSO
projection.

We have argued in the body of this paper that the $K=4$ FSS consists of
one free coordinate boson theory and one $\bZ_4$ parafermion theory per
space-time dimension.  From this description we can readily deduce the
form that the partition function for the $K=4$ fractional superstring must
take.  Each propagating bosonic world-sheet field $X^\mu$ contributes to
the total one-loop partition function a factor
\beq{bosfac}
 {\rm each~boson}~\Longrightarrow~ {1\over{ \sqrt{\tau_2} \,
     \eta(q)\overline{\eta(q)} }} ~.
\eeq
Here $\eta(q)$ is the Dedekind $\eta$-function defined in (\ref{dede}),
$q={\rm e}^{2\pi i\tau}$
and $\tau\equiv \tau_1  + i \tau_2$ is the torus modular parameter.
Note that this factor includes the contributions from both holomorphic
and anti-holomorphic (or left- and right-moving) components.
The factor contributed by each world-sheet parafermion is
\beq{pffac}
 {\rm each~parafermion}~\Longrightarrow~ {\cal Z}^j_m(q)
\eeq
(the above is for left-moving parafermions;  right-moving
parafermions contribute the complex-conjugate).  Here
${\cal Z}^j_m(q)$ are the parafermion characters whose expressions
were derived earlier (\ref{pfchars}).

It is convenient to introduce
the string functions \cite{KP} $c^\ell_n(q)$ by
\beqq
  {\cal Z}^j_m(q)~\equiv~\eta(q)\,c^{2j}_{2m}(q)~.
\eeq
{}From the parafermion field identities (\ref{phid}) and the
expressions (\ref{pfchars}) for the parafermion characters,
we see that the string functions obey
\beqq
  c^{2j}_{2m}(q)~=~c^{2j}_{-2m}(q)~=~c^{4-2j}_{2m-4}(q)~.
\eeq
Thus we can take as a basis of string functions the
set $\{c^0_0, c^2_0, c^4_0, c^1_1, c^3_1, c^2_2, c^4_2\}$.
Also, the string functions have power series expansions
in $q$ starting with $c^{2j}_{2m}(q)=q^{h(j,m)}(1+\cdots)$
where
\beq{sjm}
  h(j,m)={2j(j+1)-1 \over 12}-{m^2\over 4}
  \quad\quad {\rm for}  ~|m|\leq j~.
\eeq
Under the group of modular transformations generated by
$T:\tau\rightarrow\tau+1$ and $S: \tau\rightarrow-1/\tau$,
the string functions mix among themselves, forming a closed
set.  From eq.~(\ref{sjm}) it follows that the string functions
are eigenfunctions under the $T$ modular transformation,
while the $S$ transformation is given by \cite{KP,GQ}
  \beq{Strans}
  c^\ell_n(-1/\tau)=\left(-24i\tau\right)^{-1/2}\sum^4_{L=0}
   \sum^4_{N=-3}{\rm e}^{i\pi nN/4}\ {\rm sin}
   \left[{{\pi(\ell+1)(L+1)}\over{6}}\right]c^L_N(\tau).
  \eeq

To construct the modular-invariant partition
function for the closed FSS, we impose the
conditions that there be a graviton in the closed
string spectrum, and that no tachyonic states appear
in the spectrum.  Since the timelike and longitudinal
polarizations of a massless graviton are not propagating
degrees of freedom, we expect the partition function for the
massless level to have contributions from only $D-2$
transverse dimensions.  In order for the FSS partition
function to have a chance of being modular invariant, it must
be composed of products of the string functions
which we have seen close among themselves under modular
transformations.  Thus, if only $D-2$ dimensions worth
of states contribute at the massless level, this must also
be true of the whole tower of massive states in the FSS.
Therefore, for the closed $K=4$ fractional superstring, the
propagating world-sheet field content consists
of $D-2$ coordinate bosons and $D-2$ each
of left- and right-moving parafermions.  We therefore
obtain the form for the total partition function:
\beq{kktheory}
 {\cal Z}~\sim~(\tau_2)^{1-D/2} ~\sum\,
 {\overline{c}}^{D-2}\,c^{D-2}~,
\eeq
where $c$ stands for any $K=4$ string function.  Note that the
$\eta$-functions have cancelled between the bosonic and
parafermionic contributions.

Consider for the moment the left-moving part of the
$D$-dimensional FSS.  The term in its
partition function that includes the ground state is
represented by the first term in the expansion of
\beq{ptach}
  (c^0_0)^{D-2}~\sim~ q^{-(D-2)/12}\,(1+\ldots)~,
\eeq
as follows from the string function expansions (\ref{sjm}).
In the usual way, the power of $q$ in the partition function
is interpreted as the mass squared of the state, while its
coefficient is the state's multiplicity ({\it i.e.}, the number
of bosons minus the number of fermions).  In this way we see
that for $D>2$ the ground state of the FSS is tachyonic.
Now let us consider the first excited state built from the
ground state by the action of the parafermion energy
operators $\epsilon^{(\pm1)}$.  This is the analog of the first
excited state of the Neveu-Schwarz sector of the superstring,
which describes a vector particle.
 Thus, the
vector particle in the left-moving FSS partition function
comes from the first term in the expansion of
\beq{pvect}
  (c^0_0)^{D-3}(c^2_0)~\sim~ q^{{1\over3}-
     {(D-2)\over12}}(1+\ldots)~.
\eeq
Since the graviton in closed string theory comes from
combining a left-moving and a right-moving
vector particle, the masslessness of the graviton implies
$D=6$.  This is the basis of our claim that the $K=4$ FSS
has critical dimension six.

Physically, we are only
interested in string theories that are tachyon-free.  This requires
that tachyons must be projected out of the physical spectrum.  Hence
we are interested in constructing a partition function
${\cal Z}$ for closed string theories which contain the massless term
(\ref{pvect}) but no tachyonic term like (\ref{ptach}).  We find only
one $D=6$ modular invariant partition function that satisfies these
conditions \cite{AT}:
\beq{ZFSS}
 (\tau_2)^{D/2-1} {\cal Z}(q)=|A|^2+3|B|^2,
\eeq
with
\beqa{ABxpr}
  A &= & 4(c^0_0+c^4_0)^3(c^2_0)-4(c^2_0)^4
   -4(c^2_2)^4+32(c^2_2)(c^4_2)^3~,\\
  B &= & 8(c^0_0+c^4_0)^2(c^2_2)(c^4_2)+16(c^0_0+c^4_0)
   (c^2_0)(c^4_2)^2-8(c^2_0)^2(c^2_2)^2~.\nonumber
\eeqa
Eq.~(\ref{sjm}) implies that the leading terms in a power
series expansion in $q$ are $A\sim q^0$ and $B\sim q^{1/2}$.
Thus there are indeed no tachyons in this theory.
The only contributions to the massless states are
from the terms $4(c^0_0)^3c^2_0-4(c^2_2)^4$ in $A$.  The first
term we have already interpreted as the massless vector
particle.  The number
of degrees of freedom of a massless vector particle
in six dimensions is four, fixing the normalization of
the partition function.  The second term, appearing with
a minus sign, must be interpreted as a space-time fermion.
It is composed of $j=1$ spin fields in the parafermion
theory, commonly denoted $\sigma_{\pm2}=\phi^1_{\pm1}$.
The normalization of this term suggests that it is a massless space-time
spin-$1/2$ Weyl fermion.  In Sect.~\ref{sfive} we will confirm this
identification.

It turns out that there is an additional
remarkable property shared by the expressions in
(\ref{ABxpr}).  It can be shown \cite{AT,ADT}
that each of these new parafermionic
string-function expressions vanishes as a function of $q$:
\beq{pfSUSY}
 A(q)~=~B(q)~=0~.
\eeq
This is interpreted as a sign of space-time supersymmetry
cancellations in the fractional superstring spectrum of states.

{}From the expressions (\ref{ZFSS}) and (\ref{ABxpr}) for the
partition function, we can immediately deduce a number of important
properties of the full $K=4$ FSS Fock space.

First, and most obviously, we notice that the set of string functions
$c^{2j}_{2m}$ which close on half-odd integer $j$ quantum numbers under
the $S$ transformation (\ref{Strans}) does not appear anywhere
in the partition function.  We deduce that the corresponding parafermion
sectors with half-odd integer $j$ are projected out of the full string
Fock space.  Indeed, we have already assumed this fact in the
discussion of Sect.~\ref{stwo}.

Next, we notice that the term $4(c^0_0)^3c^2_0$ in the $A$ block
of the partition function, identified as contributing to a space-time
boson state in the left-moving theory, has the form of a product of
parafermion sectors all with quantum numbers $m=0$.
Thus it is natural to guess that in the light-cone gauge space-time
bosons will have vertex operators proportional to
\beq{pfbos}
 {\cal B}~\sim~\prod_{i=1}^{D-2}\left(\phi^{j_i}_0\right)~,
\eeq
where $j_i=0$, $1$ or $2$.  Any of the parafermion primary fields
$\phi^j_m$ may, of course, be replaced by one of their descendant
fields.  Note that we are suppressing the
world-sheet boson ($X^\mu$) contributions to the vertex
operators.  Since the world-sheet bosons only give rise to
states with space-time bosonic statistics, this suppression
will not affect the identification of the statistics of
vertex operators.  By a similar argument following from the
$A$ block term $-4(c^2_2)^4$, the fermions
will have vertex operators proportional to
\beq{pfferm}
 {\cal F}~\sim~\prod_{i=1}^{D-2}\left(\phi^{j_i}_{\pm1}\right)~,
\eeq
where $j_i=1$ or $2$.  According to
the fusion rules (\ref{fusionrules}) the parafermion $m$ quantum
numbers add, so we see that (\ref{pfbos}) and
(\ref{pfferm}) are consistent with the spin statistics
connection.  Indeed, these vertex operators satisfy the selection
rules under fusion
\beqa{pstats}
    {\cal F}\ \cdot\ {\cal F}&\to&{\cal B}~\nonumber\\
    {\cal F}\ \cdot\ {\cal B}&\to&{\cal F}~\\
    {\cal B}\ \cdot\ {\cal B}&\to&{\cal B}~,\nonumber
\eeqa
which follow from $m$ quantum number conservation and the
parafermion field identifications (\ref{phid}).
These selection rules serve to confirm our guess of the form
of space-time boson and fermion states in the left-moving
half of the FSS.  In particular, we have identified the $m=0$
sector of the FSS Fock space as the analog of the superstring
Neveu-Schwarz sector, and the $m=\pm1$ sectors as the analog
of the Ramond sector.  In Sects.~\ref{sfive} and \ref{ssix} we
will confirm these identifications by finding the equations of
motion satisfied by the massless physical states in each sector.

Note that not every state built by the action of
$\epsilon^{(\pm1)}$ on the bosonic or fermionic ground
states (\ref{bfgrnd}) appears in the $A$ block partition
function.  For example, the bosonic ground state itself
would contribute to the term $(c^0_0)^4$, a tachyonic
state, which does not appear in (\ref{ABxpr}).  This
projection, and similar ones at higher mass-levels in
the FSS spectrum, are the analog of the GSO projection
\cite{GSO} in the superstring.  The $A$ block partition function
has the general expansion $A(q)\sim\sum_{N=0}^\infty a_N q^N$,
whereas if arbitrary states built by the action of the
$\epsilon^{(\pm1)}$ fields on the bosonic and fermionic
ground states appeared in the partition function, it would
have the expansion $A(q)\sim\sum_{n=-1}^\infty a_n q^{n/3}$.
Thus, in general, the analog of the GSO projection in the $K=4$
FSS removes $2/3$ of all the states from the bosonic and fermionic
sector Fock spaces.  The effect of this GSO-like projection
on the low-lying states is shown in Fig.~2.

\begin{figure}
 \[
 \begin{array}{lclcccl}
 &\vdots &&&\vdots &&\vdots\\
 &&&&&& \\
 &&(c^0_0)^{\phantom{2}}(c^4_0)(c^2_0)^2 & .............. & 4/3
  & .............. & (c^2_2)^2(c^4_2)^2 \\
 \phantom{(c^0_0)^2}(c^2_0)^4 &,&(c^0_0)^2(c^4_0)(c^2_0) & \hrulefill & 1
  & \hrulefill & \\
 (c^0_0)^{\phantom{2}}(c^2_0)^3 &,&(c^0_0)^3(c^4_0) & .............. & 2/3
  & .............. & (c^2_2)^3(c^4_2) \\
 (c^0_0)^2(c^2_0)^2 &&& .............. & 1/3 && \\
 (c^0_0)^3(c^2_0) &&& \hrulefill & 0 & \hrulefill &(c^2_2)^4 \\
 (c^0_0)^4 &&& .............. & -1/3 && \\
 &&&&&& \\
 \multicolumn{4}{c}{\rm Bosonic\ States}  & {\rm MASS}^2
  & \multicolumn{2}{c}{\rm Fermionic\ States}
 \end{array}
 \]
 \caption{The low-lying mass levels for the $A$ block states of the $K=4$
 FSS, along with the corresponding combinations of string functions which
 first get contributions from those levels.  The masses of states in each
 level are indicated in units of the Planck mass.  The mass levels removed
 by the GSO-like projection ({\it i.e.}, whose string functions do not
 appear in the partition function) are indicated by dotted lines. }
\end{figure}

In the $A$ block of the FSS partition function, we have noted
that the states fall into space-time bosonic and fermionic
sectors.  These sectors are built by the action of the energy
operators $\epsilon^{(\pm1)}\in[\phi^1_0]$ on the ground states
\beqa{bfgrnd}
 {\rm bosonic~ground~state}&\sim
  &\prod_{\mu=0}^{D-1}\,\id^\mu~,\nonumber\\
 {\rm fermionic~ground~state}&\sim
  &\prod_{\mu=0}^{D-1}\,\sigma_{\pm2}^\mu~,
\eeqa
where $\id\in[\phi^0_0]$ is the identity and
$\sigma_{\pm2}\in[\phi^1_1]$ are the dimension $1/12$ spin fields.
In the $A$ block, just as in the usual superstring,
no other ``mixed'' sectors appear, built on ground
states of the form
\beqq
 {\rm mixed~ground~state}~\sim~\prod_{\mu\in N}
 \prod_{\nu\in R}\,\id^\mu\sigma^\nu~,
\eeq
where $N$ and $R$ are disjoint sets of indices satisfying
$N\cup R=\{0,1,\ldots,D\}$.  This is another restriction
placed on the full FSS Fock space by modular invariance.
However, unlike the superstring, a certain ``mixed sector''
does appear in the $B$ block states of the FSS.
The possible implications of the presence of
$B$ block states are discussed in \cite{DT,ADT}.
This paper is restricted to an examination
of the $A$ block states only.

\Appendix{appD}

We compute the critical central charge of a string theory
with world-sheet fractional supersymmetry generated by
the spin-$4/3$ currents $G^+$ and $G^-$.  These currents
satisfy the split algebra with arbitrary central charge $c$
introduced in eqs.~(\ref{ftalg}) and (\ref{ftalg1}).
Instead of rescaling $G^+$ and $G^-$ so that the
structure constants are equal, as in eq.~(\ref{ftalg}), we will
leave them free in this appendix.  This will allow us to
discuss the hermiticity assignments of the currents more
easily.  Thus, the split algebra is
\beqa{ftalg2}
  G^{+}(z)G^{+}(w)&= & {\lambda^+\over(z-w)^{4/3}}
   \left\{G^{-}(w)+{1\over2}(z-w)\partial G^{-}(w)\right\},\nonumber\\
  G^{-}(z)G^{-}(w)&= & {\lambda^-\over(z-w)^{4/3}}
   \left\{G^{+}(w)+{1\over2}(z-w)\partial G^{+}(w)\right\},\\
  G^{+}(z)G^{-}(w)&= & {(1/2)\over(z-w)^{8/3}}\left\{{3c\over4}+
   2(z-w)^2{T}(w)\right\}.\nonumber
\eeqa
The associativity condition (\ref{lamb}) for the split algebra
then becomes
  \beq{nonlin}
  \lambda^+\lambda^-={8-c\over6}~.
  \eeq

The modes of the $G^\pm$ currents and the enegry-momentum tensor
$T$ satisfy the commutation relations
\beqa{Jalgcom}
 \left[L_m,L_n\right]&=&(m-n)L_{m+n}
 +{c\over12}(m^3-m)\delta_{m+n}~,\nonumber\\
 \left[L_m,G^\pm_r\right]&=&
 \left({m\over3}-r\right)G^\pm_{m+r}~,
\eeqa
and the GCRs
\beqa{JalgGCR1}
 &\displaystyle{ \sum_{\ell=0}^\infty c^{(-2/3)}_\ell\left[
  G^\pm_{\pm{q\over3}+n-\ell}G^\pm_{{2\pm q\over3}+m+\ell}-
  G^\pm_{\pm{q\over3}+m-\ell}G^\pm_{{2\pm q\over3}+n+\ell}\right]
  = } &\nonumber\\
  &\displaystyle{ {\lambda^\pm\over2}
  (n-m)G^\mp_{{2+2q\over3}+n+m}~, } &
\eeqa
and
\beqa{JalgGCR2}
 &\displaystyle{ \sum_{\ell=0}^\infty c^{(-1/3)}_\ell\left[
  G^{+}_{{1+q\over3}+n-\ell}G^{-}_{-{1+q\over3}+m+\ell}+
  G^{-}_{-{2+q\over3}+m-\ell}G^{+}_{{2+q\over3}+n+\ell}\right]
  = } &\nonumber\\
 &\displaystyle{ L_{n+m}+{3c\over16} \left(n+1+{q\over3}\right)
  \left(n+{q\over3}\right)\delta_{n+m}~, } &
\eeqa
when acting on a state with $\bZ_3$ charge $q$.

Following the discussion of Sect.~\ref{sfour}, the physical
state conditions derived from this constraint algebra are
\beqa{intercept}
   (L_0-v)~\ket{\phi_{\rm phys}} & = & 0~,\nonumber\\
   L_n\ket{\phi_{\rm phys}}&=&0~, \quad n>0~,\\
  G^\pm_r\ket{\phi_{\rm phys}}
   & = &0~, \quad r>0~.\nonumber
\eeqa
We can determine the critical values of the central
charge and intercept by demanding the appearance of
extra null states at those values of $c$ and $v$.
As in the bosonic string and the superstring, extra null
states can be taken as an indication of an enhanced
gauge symmetry in the string theory.

Consider the spurious state with $\bZ_3$ charge $q=1$
\beqq
  \ket{\phi}~=~G^+_{-1/3}\ket{\phi_0}~.
\eeq
It is built on a state with $\bZ_3$ charge $q=0$,
satisfying
\beqa{phi0}
  L_0\ket{\phi_0}&=&\left(v-{1\over3}\right)
   \ket{\phi_0}~,\nonumber\\
  L_{n+1}\ket{\phi_0}&=&G^\pm_{n+2/3}
   \ket{\phi_0}~=~0~,\quad n\geq0~.
\eeqa
If $\ket{\phi}$ is a physical state, it will be null,
since it is by construction spurious.  The conditions
(\ref{phi0}) imply that $\ket{\phi}$ obeys all the
physical state conditions (\ref{intercept}) except one:
\beqqa
  0&=&G^-_{1/3}\ket{\phi}~=~
   G^-_{1/3}G^+_{-1/3}\ket{\phi_0}\nonumber\\
  &=&L_0\ket{\phi_0}~=~
   \left(v-{1\over3}\right)\ket{\phi_0}~,
\eeqa
where the second line follows by use of the GCR
(\ref{JalgGCR2}).  Thus, we find a series of null states
if $v=1/3$.  Note that there is a second series with
$\bZ_3$ charge $q=-1$, found by exchanging $G^+$ and
$G^-$ modes.

To fix $c$, we consider the series
of spurious states with $q=1$, of the form
\beq{spur}
   \ket{\phi}~=~ \left\{\alpha G^+_{-4/3}+\beta G^+_{-1/3}L_{-1}
  +\gamma G^-_{-1}G^-_{-1/3}\right\}~\ket{\phi_0}~,
\eeq
where $\ket{\phi_0}$ now satisfies
\beqa{phi1}
  L_{0}~\ket{\phi_0}&=&-\ket{\phi_0}~,\nonumber\\
  L_{n+1}~\ket{\phi_0}&=&G^\pm_{n+2/3}~
    \ket{\phi_0}~=~0~,\quad n\geq0~.
\eeqa
Eq.~(\ref{spur}) is actually the most general spurious
state with $q=1$ built from $\ket{\phi_0}$, since it is easy to show
using the mode algebra (\ref{Jalgcom})--(\ref{JalgGCR2})
that other possible terms, like
\beqq
  \left\{\delta L_{-1}G^+_{-1/3}+ \epsilon G^+_{-1/3}
  G^-_{-2/3}G^+_{-1/3}\right\}\ket{\phi_0}
\eeq
can be expressed as linear combinations of the terms in
(\ref{spur}).  From the first condition in eq.~(\ref{phi1}),
$\ket{\phi}$ has intercept $v=1/3$.  It only remains to apply the
rest of the physical state conditions to $\ket{\phi}$ to
determine which values of $\alpha$, $\beta$, and $\gamma$
will give a new set of null states.

Using the commutation relations (\ref{Jalgcom})--(\ref{JalgGCR2}),
one can compute the action of the positive modes of the currents on
$\ket{\phi}$.  Imposing the physical state conditions
(\ref{intercept}) yields the four independent conditions:
\beqqa
  L_1~\ket{\phi}=0&\Longrightarrow&0=5\alpha
   -4\beta+2\sqrt2\lambda^-\gamma~,\nonumber\\
  G^-_{1/3}~\ket{\phi}=0&\Longrightarrow&0=\alpha
    -2\beta+4\sqrt2\lambda^-\gamma~,\nonumber\\
  G^-_{4/3}~\ket{\phi}=0&\Longrightarrow&0=
   \left({9c\over4}-4\right)\alpha-10\beta
   - 7\sqrt2\lambda^-\gamma~,\\
  G^+_1~\ket{\phi}=0&\Longrightarrow&0=7\sqrt2\lambda^+\alpha
   +4\sqrt2\lambda^+\beta+\left({5c\over2}-8
   -\lambda^+\lambda^-\right)\gamma~.\nonumber
\eeqa
This system of equations has the solution
\beqq
  \alpha=2\sqrt2\lambda^-\gamma~,~
  \beta=3\sqrt2\lambda^-\gamma~,~
  c=10~,~
  \lambda^+\lambda^-=-{1\over3}~.
\eeq
Thus, there are extra null states (since they are both physical
and spurious) at $c=10$.  Note that the associativity
condition, eq.~(\ref{nonlin}), is automatically satisfied by
our solution.

At $c=10$ there are extra null states with $\bZ_3$ charge
$q=0$, as well. Consider the state
\beqq
  \ket{\phi}~=~\alpha G^+_{-2/3}~\ket{\phi_0^-}+
  \beta G^-_{-2/3}~\ket{\phi_0^+}~,
\eeq
where $\ket{\phi_0^\pm}$ are primary states with $\bZ_3$
charges $q=\pm 1$, which satisfy
\beqqa
  L_0 ~\ket{\phi_0^\pm}&=&-{1\over3}
   \ket{\phi_0^\pm}~,\nonumber\\
  G^+_0~\ket{\phi_0^+} &=& b\ket{\phi_0^-}~,\nonumber\\
  G^-_{0}~\ket{\phi_0^-}&=&{1\over b}\left(L_0
   -{c\over24}\right)\ket{\phi_0^+}~=~
   -\left({c+8\over24b}\right)\ket{\phi_0^+}~,\\
  L_{n+1}~\ket{\phi_0^\pm}&=&G^+_{n+1/3}~\ket{\phi_0^-}~=~
  G^-_{n+1/3}~\ket{\phi_0^+}~=~0~,\quad n\geq 0~.\nonumber
\eeqa
The $G^+_0$ intercept $b$ is arbitrary.  The $G^-_0$ intercept
follows from the previous two equations and the mode algebra
(\ref{JalgGCR2}).  The physical state conditions then give
\beqqa
  G^+_{2/3}~\ket{\phi}=0 &\Longrightarrow&
  0=-(c+8){\lambda^+\over\sqrt2b}\alpha+(c-4)\beta~,\nonumber\\
  G^-_{2/3}~\ket{\phi}=0&\Longrightarrow&
  0=(c-4)\alpha+12{\lambda^- b\over\sqrt2}\beta~,
\eeqa
which when solved subject to the condition (\ref{nonlin}) again
give $c=10$.

So far this discussion has made no reference to the hermiticity
properties of the currents.  However, as discussed in Sect.~3.3
the hermiticity relations between mode operators play an
important role in translating the quantum constraint equations
into the physical state conditions (\ref{intercept}).  Also, it
is natural to require the constraint operators to be hermitian,
although there presumably is an interpretation of non-hermitian
constraints as corresponding to propagation in time-dependent
backgrounds.

Let us assume, then, that $G^+$ and $G^-$ are related by some
hermiticity relations.  It is not hard to show that, up to
rescalings of the currents, the algebra (\ref{ftalg2}) admits
only four inequivalent hermiticity assignments:
\beqq \begin{array}{rlll}
  (i):  & \quad(G^+)^\dagger=G^-  & \quad(G^-)^\dagger=G^+
   & \quad\lambda^+=\lambda^-~,\\
  (ii): & \quad(G^+)^\dagger=-G^- & \quad(G^-)^\dagger=-G^+
   & \quad\lambda^+=-\lambda^-~,\\
  (iii):& \quad(G^+)^\dagger=G^+    & \quad(G^-)^\dagger=G^-
   & \quad\lambda^+=\lambda^-~,\\
  (iv): & \quad(G^+)^\dagger=G^+    & \quad(G^-)^\dagger=G^-
   & \quad\lambda^+=-\lambda^-~,
  \end{array}\eeq
where in all cases $\lambda^+$ can be taken to be a postive
real number.  When $c>8$, eq.~(\ref{nonlin}) implies
$\lambda^+\lambda^-<0$, so only the hermiticity assignments $(ii)$ or
$(iv)$ are allowed.  In these cases we can still construct
non-split subalgebras, generated now by $G=G^+-G^-$.
In both cases $(ii)$ and $(iv)$,
$GG\sim -1+\ldots$, showing that such FSCAs are
necessarily non-unitary.  The FSS involves copies of the
FSCA at $c=2$, so cases $(i)$ and $(iii)$ apply.  Indeed,
the free field representation at $c=2$ constructed in
the body of the paper satisfies the hermiticity relations $(i)$.

We have determined the critical intercept and central charge
of a spin-4/3 string to be $v=1/3$ and $c=10$.  Note
that this whole calculation could have been performed using
the FSCA instead of the spin-4/3 algebra.  The physical
conditions would have been generated by the single
current $G=G^+ +G^-$.  The same intercept and
central charge emerge, using the sum of the $q=1$ and
$q=-1$ null states described above.  This result is in
agreement with that obtained by consideration of the Ka\v{c} determinant
formula for the FSCA \cite{ALT}.

\end{document}